\documentclass[acmsmall,authorversion, nonacm]{acmart}
\AtBeginDocument{%
  }

\usepackage{graphicx}
\usepackage{lscape}
\usepackage{longtable}
\usepackage{setspace}
\usepackage{booktabs}
\usepackage{pdflscape}
\usepackage{afterpage}
\usepackage{fancyhdr}
\usepackage{geometry}
\usepackage{multicol}
\usepackage{array}
\usepackage{pifont}
\usepackage{caption}
\usepackage{subcaption}

\begin{document}

\title{The Role of Generative AI in Facilitating Social Interactions: A Scoping Review}

\author{Teis T.J.E. Arets}
\email{t.t.j.e.arets@tue.nl}
\orcid{0009-0005-6325-7844}
\affiliation{%
  \institution{Eindhoven University of Technology}
  \city{Eindhoven}
  \country{The Netherlands}
}

\author{Giulia Perugia}
\orcid{0000-0003-1248-0526}
\affiliation{%
  \institution{Eindhoven University of Technology}
  \city{Eindhoven}
  \country{The Netherlands}
}

\author{Maarten Houben}
\orcid{0000-0002-6571-1925}
\affiliation{%
  \institution{Eindhoven University of Technology}
  \city{Eindhoven}
  \country{The Netherlands}
}

\author{Wijnand A. IJsselsteijn}
\orcid{0000-0001-6856-9269}
\affiliation{%
  \institution{Eindhoven University of Technology}
  \city{Eindhoven}
  \country{The Netherlands}
}

\renewcommand{\shortauthors}{Arets et al.}

\begin{abstract}
Reduced social connectedness increasingly poses a threat to mental health, life expectancy, and general well-being. Generative AI (GAI) technologies, such as large language models (LLMs) and image generation tools, are increasingly integrated into applications aimed at enhancing human social experiences. Despite their growing presence, little is known about how these technologies influence social interactions. This scoping review investigates how GAI-based applications are currently designed to facilitate social interaction, what forms of social engagement they target, and which design and evaluation methodologies designers use to create and evaluate them. Through an analysis of 30 studies published since 2020, we identify key trends in application domains including storytelling, socio-emotional skills training, reminiscence, collaborative learning, music making, and general conversation. We highlight the role of participatory and co-design approaches in fostering both effective technology use and social engagement, while also examining socio-ethical concerns such as cultural bias and accessibility. This review underscores the potential of GAI to support dynamic and personalized interactions, but calls for greater attention to equitable design practices and inclusive evaluation strategies.
\end{abstract}


\maketitle


\section{Introduction}
Social disconnectedness has been recognized as a `global epidemic' in recent years \cite{na_social_2023}, affecting approximately one in three adults in Europe and the US \cite{world_health_organization_social_2021}. Social disconnectedness has been associated with increased loneliness, social isolation, and reduced social integration \cite{hu_social_2022}, all conditions conducive to lower levels of mental and physical health, as well as diminished psychosocial well-being \cite{cornwell_social_2009,zhang_reconfiguring_2023,na_social_2023}. Social connectedness, or the `short-term experience of belonging and relatedness' between people \cite[p.67]{van_bel_social_2009}, instead, is a concept that has a demonstrated positive impact on mental health \cite{lamblin_social_2017}, life expectancy \cite{pachana_social_2015}, and general well-being \cite{jose_does_2012}. As a construct, social connectedness should not be confused with related concepts such as social integration, social isolation, or loneliness. Such constructs span longer periods of time and cannot be directly influenced by a single interaction \cite{flynn_social_2024}. Social connectedness, instead, is short-lived and can be shaped by powerful interactions. Addressing the global challenge of reduced social connectedness is not only urgent but also opens new avenues for exploring innovative solutions — including the role of emerging AI technologies in fostering human connection.

\subsection{GAI technology for social connectedness}
One domain with the potential to foster social connectedness that has been extensively investigated in recent years is technology \cite{barbosa_neves_can_2019}. 
Social Presence Theory explores the extent to which communication technology allows users to feel the presence of others in the interaction \cite{short_social_1976}. The experience of social presence is a crucial precedent for the experience of social connectedness \cite{oh_systematic_2018}. The fundamental components of social presence are intimacy and immediacy. 
Intimacy involves a co-regulated interpersonal state shaped by behavioral cues such as eye contact, proximity, and conversational depth, where changes in one cue are balanced by adjustments in others \cite{short_social_1976} (for example, when two people are placed in close proximity, eye contact is reduced).
Immediacy is the `measure of the psychological distance that a communicator places between himself and the object of his communication, his addressee, or his communication' \cite[p. 72]{short_social_1976}. 
If a technology can bring about the experience of both intimacy and immediacy among its users, it can be deemed an effective method to promote social connectedness.

A large proportion of research on the effectiveness of technology in fostering social connectedness focuses on older adults. In a scoping review on this topic, \citet{petersen_association_2023} found that Information and Communication Technologies (ICT) successfully promoted well-being and social connectedness in older people by facilitating conversations, and reducing the experience of loneliness.
In another review, \citet{you_sociocultural_2024} revealed the effectiveness of social robots and telepresence systems in reducing loneliness and stimulating social connectedness among older adults. This was mostly achieved by providing companionship, facilitating remote interactions with family and caregivers, and enhancing engagement through user-friendly, culturally relevant, and emotionally responsive technologies.

\subsubsection{Generative AI}
Although much research has been done on the role of technology in fostering social connectedness, little attention has been paid to the influence of emerging AI technologies on the experience of social connectedness. 
Generative AI (GAI) is a relatively novel type of technology that has the potential to support communication and foster social connectedness among users due to its near-human-like output \cite{feuerriegel_generative_2024,lyu_can_2024}. 
GAI refers to AI systems that create seemingly new content such as text, images, and videos based on user's request (i.e., prompt) leveraging large datasets \cite{feuerriegel_generative_2024,ooi_potential_2023}. 
There are two types of GAI technologies: \textit{unimodal} and \textit{multimodal}. Unimodal GAI generates one form of output (e.g., images only). Multimodal GAI is capable of generating multiple forms of output (e.g., ChatGPT: both text and images).
Since its launch in 2022, ChatGPT has become the most well-known and widely used GAI program, bringing both the application and the term `generative AI' into mainstream awareness 
\cite{eke_chatgpt_2023,openai_gpt-4_2023}. Where ChatGPT is able to generate text and images on request, other GAI applications can generate music (e.g., Suno \cite{suno_suno_2023}), videos (e.g., Runway ML \cite{runway_ml_runway_2025}), or avatars (e.g., D-ID\cite{d-id_d-id_2025}). 

GAI can be used to facilitate social interactions by generating personalized content and activities that continuously adapt to users' changing needs, behaviors, and capabilities. 
Its advanced natural language abilities enable rich, expressive communication and nuanced understanding across a wide range of topics, going beyond scripted or Q\&A-style interactions \cite{li_always_2025}. 
GAI can simulate social dynamics by adopting different roles or perspectives and incorporating elements such as politeness, playfulness, or intimacy, which can support parasocial experiences \cite{maeda_when_2024}. 
Furthermore, its personalization features allow it to be sensitive to individual differences in age, culture, and cognitive ability, especially when enhanced by memory functions that build on prior interactions \cite{ronit_enhancing_2024}. 
GAI’s multimodal capabilities — ranging from interpreting images or videos to producing text, speech, and visual content — enable complex, cross-modal exchanges (e.g., uploading an image and describing its content in text; \cite{feuerriegel_generative_2024}). 
Additionally, it integrates relatively easily with various platforms, including virtual and augmented reality, digital games, social robots, and smartphone applications \cite{joshi_review_2025}. 
Together, these features position GAI as uniquely capable of supporting and enhancing social connectedness. 

Despite this potential, though, the influence of GAI on social connectedness is relatively underexplored. 
This scoping review addresses this gap by mapping current GAI-based designs fostering social interaction. We explore the context in which such social interaction takes place, the role the GAI-based technology plays in the social interaction resulting from its use, and we examine the design methodologies employed to bring them to life. Finally, we address ethical considerations related to the use of GAI-based technology for this purpose.

\subsubsection{A typology for the role of Generative AI (GAI) in social interactions}

According to \citet{fogg_persuasive_2003}, technologies can play different roles in influencing human behavior. They can be \textit{tools}, \textit{media}, or \textit{agents}, the so-called `Functional Triad' \cite{fogg_persuasive_2003}. Technology as a \textit{tool} has the objective of making activities or tasks easier for the user (e.g., a fitness app tracking steps and calories, helping the user to achieve their health goals). Technology as a \textit{medium} functions as a channel through which experience can be facilitated (e.g., a flight simulator used for pilot training in a virtual environment). Finally, technology as an \textit{agent} requires the technology to embody a social actor with which communication occurs (e.g., a chatbot that provides mental health support).
This classification can be used to determine the degree to which technology can participate in social interactions and, consequently, influence social dynamics.

GAI can be considered a \textit{tool} when it supports individual actors in engaging in social interaction. For example, when it helps users prepare for a meeting by suggesting topics and summarizing previous meeting notes. 
It can be considered a \textit{medium} when it acts as a channel facilitating social interaction between people; take, for example, a language model that translates messages in real time between two people speaking different languages. Finally, it can be considered an \textit{agent} when it embodies an interactant in the social interaction, such as a chatbot responding to users' questions in real time. In all these declinations, GAI can foster experiences of intimacy and immediacy, thus promoting social connectedness. In this literature review, we use the `Functional Triad' framework to analyze the role that GAI-based technologies play in fostering social interactions, as described in the reviewed studies. We then draw conclusions about which role within the `Functional Triad' is the most appropriate for GAI technologies to adopt under different circumstances.

\subsubsection{A participatory approach to technology design}
While much attention has been given to the technological capabilities of GAI, how these technologies are designed is equally important to consider. Co-design — a participatory design approach that involves end-users throughout the design process — is particularly well-suited for fostering social connectedness \cite{sanders_co-creation_2008,schuler_principles_1998}. 
Unlike technology-oriented or user-centered methods, co-design creates opportunities for mutual understanding, shared creativity, and relational engagement. In doing so, not only it produces more relevant and personalized technologies, but also supports connection through the very act of designing together.
Co-design has been shown to improve the fit between user needs and design outcomes \cite{antonini_overview_2021,steen_benefits_2011}, yet it is often inconsistently applied. For instance, many studies claim to adopt co-design but involve users only in limited phases \cite{suijkerbuijk_active_2019,mannheim_ageism_2023}. Understanding the extent of end-user involvement in technology design is thus key to assessing a technology’s potential for fostering social connectedness.
In this review, we analyze how GAI-based technologies for social interaction are designed, with specific attention to whether and how co-design is used. Through this research endeavor, we hope to understand whether and how co-design practices are linked to design products that more effectively foster social connectedness.

\subsubsection{Socio-ethical perspectives on generative AI}
The widespread use of GAI raises a variety of ethical concerns, particularly regarding biases and the potential reinforcement of harmful stereotypes \cite{fui-hoon_nah_generative_2023,weidinger_taxonomy_2022}. One socio-ethical consideration we would like to highlight is that the training data of GAI models often fail to represent the full spectrum of identities and cultures, a phenomenon that can exacerbate biases and promote exclusionary practices \cite{weidinger_taxonomy_2022}. 
This risk is exacerbated when users rely heavily on models developed by a limited number of organizations, such as OpenAI, potentially leading to one-sided depictions of marginalized populations. Consequently, the potential of GAI to foster social interactions may be hindered if the technology inadvertently fosters exclusionary norms or stereotypes \cite{shelby_sociotechnical_2023}.
Based on an interdisciplinary analysis of the ethical challenges posed by GAI-based technologies, \citet{al-kfairy_ethical_2024} advocate for policies, guidelines, and frameworks that prioritize human rights, fairness, and transparency.
Effective bias mitigation in GAI involves cleaning and balancing datasets, ensuring model transparency, conducting regular audits, involving diverse teams, adhering to ethical guidelines, and implementing user feedback mechanisms \cite{ferrara_fairness_2023}.
In this literature review, we adopt a socio-ethical lens to examine how GAI is currently used to stimulate social interactions.

\subsection{This Literature Review}
This literature review aims to survey studies focusing on the design of GAI-based technologies to gain insight into the extent to which current GAI-based applications have the potential to stimulate social connectedness. 
To do so, it posits the following research questions: 
\begin{enumerate}
    \item \emph{What kind of GAI-based applications (e.g., type of GAI modality, software, and embodiment) are currently designed for facilitating social interaction?}
    \item \emph{What kind of social interactions (e.g., target groups, type of interaction, interaction setting, and equality of access to the system) does GAI currently attempt to facilitate?}
    \item \emph{Which methodologies (e.g., user-centered design or co-design) are applied to design and evaluate these GAI applications?}
\end{enumerate}

\section{Method}
\subsection{Review methodology}
Scoping reviews aim to explore the available research on a topic of interest. Systematic reviews pursue the same objective, but also aim to critically evaluate the scientific evidence generated by the reviewed studies \cite{munn_systematic_2018}.
Scoping reviews are particularly useful for emerging or evolving areas of research, where the evidence landscape is still developing. Given the young and fast developing nature of GAI-based design research and our research goal of exploring and characterizing the current landscape of GAI-based design research and design methodologies, we deemed a scoping review a more suitable review instrument than a systematic review \cite{munn_systematic_2018}. We report how we approached the literature review step-by-step following the guidelines of the Preferred Reporting Items for Systematic and Meta-Analysis Extension for Scoping Reviews (PRISMA-ScR) \cite{tricco_prisma_2018}. A copy of the PRISMA-ScR checklist can be found in the Supplementary Materials.

\subsection{Study retrieval}
We conducted a comprehensive search in four databases to identify studies to include: Scopus, the ACM Digital
Library, IEEE Xplore, and PubMed using slight variations of the following search query (the exact search queries we used for each database are reported in Table \ref{tab:appendixA_queries} in Appendix \ref{appendixA}): \emph{(Title/Abstract/Keywords: \textit{GAI keywords}) AND (Title/Abstract/Keywords: \textit{design keywords}) AND (Title/Abstract/Keywords: \textit{social interaction keywords})}. The exact GAI, design, and social interaction keywords are shown in Table \ref{tab:keywords}.

\begin{longtable}[c]{>{\raggedright\arraybackslash}p{.15\textwidth} >{\raggedright\arraybackslash}p{.3\textwidth} >{\raggedright\arraybackslash}p{.47\textwidth}}
    \caption{Overview of the exact keywords used for the database search.}
    \label{tab:keywords}\\
    \hline
    \textbf{Keywords} & \textbf{Search location} & \textbf{Search terms} \\ \hline
    \endfirsthead
    \hline
    \textbf{Keywords} & \textbf{Search location} & \textbf{Search terms} \\ \hline
    \endhead
    \hline
    \endfoot
\endlastfoot
GAI & Title, abstract, keywords & generative AI, generative-AI, generative artificial intelligence, ChatGPT, GPT*, large language model, LLM* \\
Design & Title, abstract, keywords & design*, creat* \\
Social interaction & Title, abstract, keywords & social* \\
\hline
\end{longtable}

On top of these four databases, we browsed the proceedings of the 2024 ACM Conference on Human Factors in Computing Systems (CHI). 
We did so as we noticed that many CHI papers fitting the scope of this literature review did not show up in our database search.
Due to the ubiquity of GAI technologies at the time of writing, we chose only to include the most recent CHI proceedings. We expected the majority of GAI-based designs to show up in the most recent proceedings. CHI proceedings were accessed through the ACM Digital Library by filtering on the CHI conference and using the same GAI-related search terms used for the database search.

The query in the databases yielded a list of 968 entries (July 2024), of which 623 were from Scopus, 154 from the ACM Digital Library, 191 from IEEE Xplore, and none from PubMed. The query in the ACM Digital Library yielded 274 entries (July 2024).

\subsection{Study selection}
The papers obtained from the electronic search were screened against the following eligibility criteria. We excluded entries that featured literature review studies and included studies: 
\begin{enumerate}
    \item Focusing on GAI-based applications offering an opportunity for verbal or non-verbal social interaction or the training skills necessary for users to engage socially with others (e.g., communication advice, face recognition aids).
    \item Describing the conceptualization, design, development, and / or evaluation of such GAI-based applications.
    \item Written in English.
    \item Published in 2020 or later.
\end{enumerate}

We selected 2020-onward as a time frame for the literature review as substantial advances in GAI, such as the advent of large language models (LLMs) like GPT-3 and GPT-4, arose from that year on. Since LLMs have transformed the field, they made studies performed before 2020 less representative of contemporary GAI capabilities and applications \cite{sengar_generative_2024}.

While screening the papers, we made a distinction between GAI-based technologies primarily designed to facilitate human-human interaction and GAI-based technologies designed to offer human-machine interaction. We included human-human interaction-facilitating designs by default. We only included human-machine interaction designs if the specific purpose of the technology was to stimulate social interaction with people (see Figure \ref{fig:appendixB_interaction_inclusion_pipeline} in Appendix \ref{appendixB}) or to help users develop the skills required to engage in social interaction with other people.
As an example, we included \citet{xygkou_mindtalker_2024}, who present a GPT-4 audio-based chatbot designed to interact with individuals with early-stage dementia, as the human-machine interaction presented in their paper served the purpose of fostering a meaningful conversation with users, but we also included \citet{tang_emoeden_2024}, who present an LLM- and text-to-image-based tool helping High-Functioning Autistic (HFA) children recognize and express emotions by providing personalized discussions points and human-like responses. 

\subsubsection{Selection pipeline}
\begin{figure}
    \centering
    \includegraphics[width=.65\textwidth]{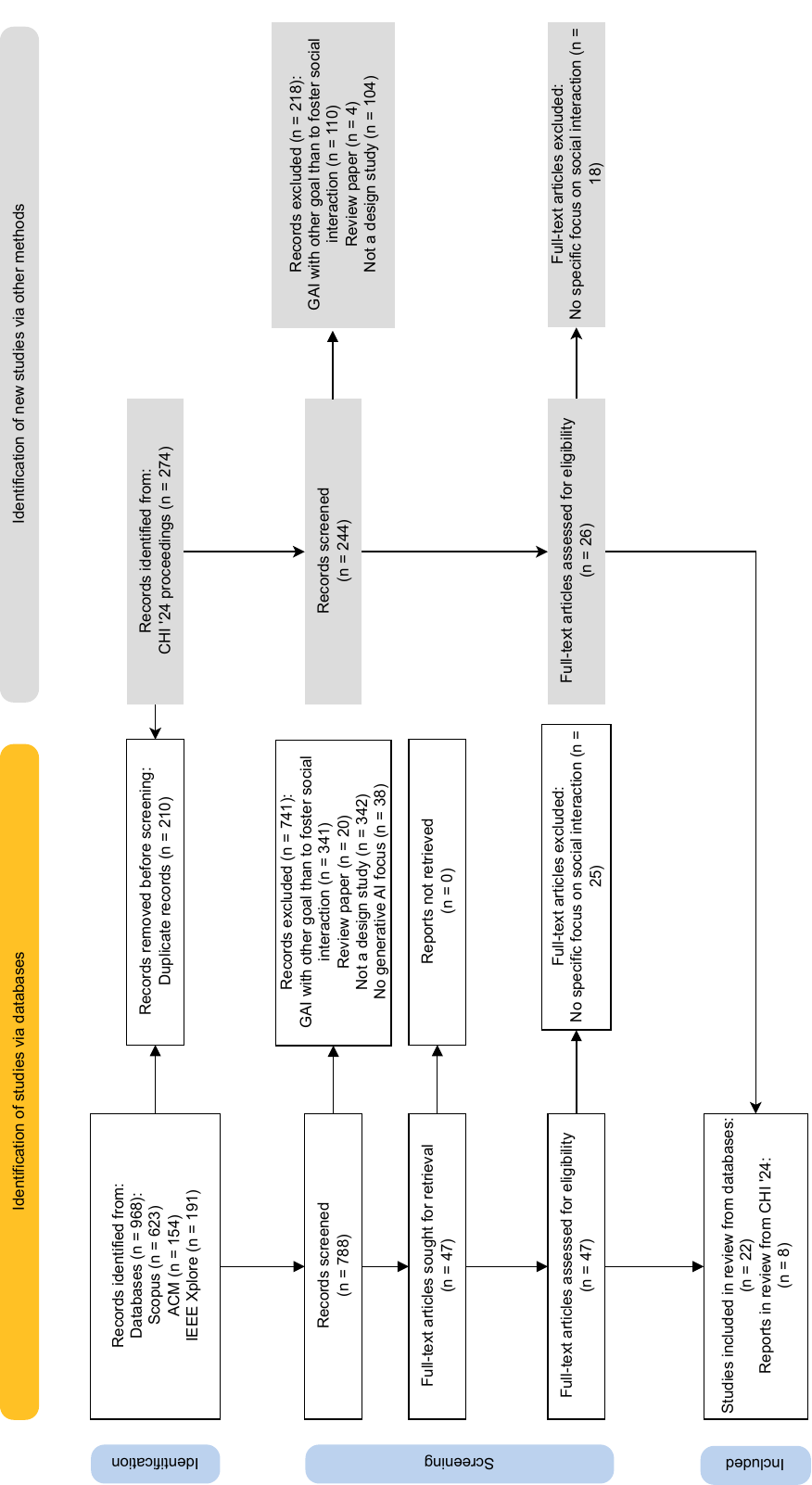}
    \caption{PRISMA flowchart. Studies were excluded if they did not meet the inclusion criteria. Most often, studies were excluded because they had another goal than to foster social interaction. Other studies were excluded because they were not design studies, had no focus on GAI, or because they were a review paper.}
    \label{fig:prisma}
\end{figure}

Figure \ref{fig:prisma} shows the PRISMA flowchart representing the study selection pipeline. As we went through the CHI 2024 proceedings separately, we decided to present this part of the selection in a separate column of the flowchart. 

As a first step, we checked all databases and CHI 2024 studies for duplicates. Overall, we found 210 duplicates of which 30 from the CHI 2024 proceedings results and 180 from the other database results. Once these were removed, we were left with 1032 records to screen, 788 from the database search, and 244 from the CHI 2024 proceedings.

As a second step, the first author screened the titles and abstracts of the remaining papers against the eligibility criteria using Rayyan\footnote{\url{https://www.rayyan.ai}}, an online tool for conducting literature reviews. The co-authors were consulted to reach a consensus on borderline studies. As a result of this screening process, we included 30 articles in this literature review, 22 from the database search and 8 from the CHI conference proceedings. 
The final set of 30 articles was read in its entirety by the first author for data extraction.

\subsection{Data extraction}
We extracted the following information from the papers:
\begin{enumerate}
    \item \textit{Study characteristics} (e.g., author and year of publication).
    \item \textit{General design concept} (e.g., type of GAI used, embodiment of the design, and intended end-user). We assembled an elaborate technical description of the presented GAI-based technologies to be able to map every technology on the `Functional Triad' of technologies as tools, media, or agents for social interaction \cite{fogg_persuasive_2003}. We did so to determine the degree to which GAI-based technology took part in social interactions and to gain insights into the circumstances in which technology can benefit from social participation or not.
    \item \textit{Type of social interaction} associated with the design (e.g., number of people involved in the social interaction and interaction setting). We recorded whether the social interaction resulting from using the GAI-based technology primarily brought about human-machine interaction, human-human interaction, or both. We also documented whether the technology was intended for an individual or a group, whether the interaction took place virtually or in a co-located space, and synchronously or asynchronously. These data were gathered to gain insights into the social contexts in which GAI-based technology, researchers, and designers are currently embedded.
    \item \textit{Methodological aspects} (e.g., design methodology and level of user involvement). We recorded the term authors used to describe their technology development process (e.g., user-centered design, participatory design, etc.), the different design phases they presented, as well as a description of how exactly participants were involved at each step of the design process. These data were crucial to determine how GAI-based technologies were thought of and brought to life.
    \item \textit{Outcomes of interaction} with the design (e.g., evaluation method and results). In particular, we logged whether and how GAI-based technologies were evaluated, and if representative end-users were involved in the evaluation process. For the papers that elaborated on the evaluation phase, we noted down the objective of the evaluation. In case of a social interaction-related evaluation, we recorded the particular evaluation focus and the metrics used.
\end{enumerate}

We used a pre-defined Microsoft Excel spreadsheet for systematically recording the relevant data from the set of included papers. The final data extraction file can be found in the Supplementary Materials of this article.

To summarize and analyze the data, we adopted the narrative synthesis approach proposed by \citet{popay_guidance_2006}, which allows to interpret data across diverse study designs. From the data, themes were identified which were then condensed into categories relevant to answer our research questions. 

\section{Results}
\subsection{Characteristics of included studies}
Of the 30 papers included in this scoping review, 2 were published in a journal \cite{schnitzer_prototyping_2024,said_design_2024}. The other 28 were part of conference proceedings, either as full papers \cite{xygkou_mindtalker_2024,jang_its_2024,jin_exploring_2024,samsonovich_registrar_2024,wei_improving_2024,degen_retromind_2024,alessa_towards_2023,bravo_creating_2023,bhati_bookmate_2023,suh_ai_2021,chen_chatgpt_2023,elgarf_creativebot_2022,shakeri_saga_2021,liu_when_2024,tang_emoeden_2024,fontana_de_vargas_co-designing_2024,seo_chacha_2024}, extended abstracts \cite{zhou_my_2024,wan_building_2024,chen_closer_2023,li_blibug_2023,billing_language_2023,li_exploring_2024,cai_advancing_2024,liu_peergpt_2024,jeung_unlocking_2024}, short papers \cite{wang_aint_2024}, or posters \cite{fang_socializechat_2023}.
Looking into their year of publication, 19 papers were published in 2024 \cite{zhou_my_2024,schnitzer_prototyping_2024,xygkou_mindtalker_2024,said_design_2024,jang_its_2024,jin_exploring_2024,wan_building_2024,samsonovich_registrar_2024,wang_aint_2024,wei_improving_2024,degen_retromind_2024,liu_when_2024,tang_emoeden_2024,li_exploring_2024,cai_advancing_2024,liu_peergpt_2024,jeung_unlocking_2024,fontana_de_vargas_co-designing_2024,seo_chacha_2024}, 8 in 2023 \cite{chen_chatgpt_2023,bhati_bookmate_2023,bravo_creating_2023,billing_language_2023,li_blibug_2023,chen_closer_2023,fang_socializechat_2023,alessa_towards_2023}, 1 in 2022 \cite{elgarf_creativebot_2022}, and 2 in 2021 \cite{shakeri_saga_2021,suh_ai_2021}.

\subsubsection{Intended end-users}
In total, 17 of the 30 studies designed GAI-based technologies for vulnerable user groups \cite{xygkou_mindtalker_2024,tang_emoeden_2024,zhou_my_2024,jin_exploring_2024,alessa_towards_2023,fang_socializechat_2023,liu_peergpt_2024,liu_when_2024,seo_chacha_2024,jang_its_2024,chen_chatgpt_2023,bhati_bookmate_2023,li_exploring_2024,fontana_de_vargas_co-designing_2024,schnitzer_prototyping_2024,degen_retromind_2024,elgarf_creativebot_2022}. We define vulnerable users as those who do not meet the social standard of self-reliance and social participation. \cite{numans_vulnerable_2021}. 
In the included studies, vulnerable users were children aged 4 to 15 years (for those studies that specified children's age \cite{zhou_my_2024,chen_chatgpt_2023,liu_peergpt_2024,liu_when_2024,seo_chacha_2024,elgarf_creativebot_2022}), people with dementia or mild cognitive impairment (MCI) \cite{xygkou_mindtalker_2024,schnitzer_prototyping_2024,degen_retromind_2024}, older adults \cite{jin_exploring_2024,alessa_towards_2023}, autistic people \cite{tang_emoeden_2024,jang_its_2024,li_exploring_2024} or their caregivers, in which case autistic people were secondary users \cite{bhati_bookmate_2023,fontana_de_vargas_co-designing_2024}, and people with speech and motor impairments \cite{fang_socializechat_2023}. 

The 13 studies that did not design GAI-based technologies for vulnerable users targeted students \cite{cai_advancing_2024,wei_improving_2024}, gamers \cite{wan_building_2024}, or no specific end-user group \cite{chen_closer_2023,bravo_creating_2023,wang_aint_2024,samsonovich_registrar_2024,jeung_unlocking_2024,suh_ai_2021,shakeri_saga_2021,li_blibug_2023,billing_language_2023,said_design_2024}. Personal correspondence with the authors of the latter papers provided more details on the users' characteristics. \citet{wang_aint_2024} explained that the conversation system for the robot Haru `was designed for general public use with a special emphasis on interaction with children' (Z. Wang, personal communication, January 31, 2025).
\citet{bravo_creating_2023} clarified that `the dialogue system has been designed for elderly people to entertain and allow them to talk about topics of interest' (T. Onorati, personal communication, January 31, 2025). \citet{jeung_unlocking_2024} and \citet{billing_language_2023} reported that there was no particular intended end-user group for their specific studies (J. L. Jeung, personal communication, February 4, 2025; E. Billing, personal communication, January 30, 2025). However, both expressed an interest in applying their design to older adults.
Finally, \citet{chen_closer_2023} replied that they intentionally built Closer Worlds without specification of the intended end user: `I think that two people interested in connection, with some time and willingness can enjoy this game' (C. Lee, personal communication, February 18, 2025).

Looking at other studies in which personal correspondence with the authors did not disclose details about the intended end-users, the authors often expressed that the design choices were made in particular contexts or application scenarios and were not limited to specific end-users \cite{samsonovich_registrar_2024,li_blibug_2023,shakeri_saga_2021,suh_ai_2021,said_design_2024}. 

\subsection{Technical specifications of GAI-based technologies (RQ1)}
\subsubsection{General technical details}
Some papers explicitly stated their motivation for using GAI in their designs. GAI was mainly used for its personalization potential for experience enrichment \cite{tang_emoeden_2024,jin_exploring_2024,wan_building_2024,fang_socializechat_2023,li_exploring_2024,fontana_de_vargas_co-designing_2024}, the creativity that it gave access to \cite{elgarf_creativebot_2022,chen_closer_2023}, and its potential to support communication in people with limited cognitive or physical abilities \cite{xygkou_mindtalker_2024,degen_retromind_2024}.

Twenty-six out of 30 designs used GAI for text generation, 7 for image generation, and 1 for music generation. 
For the 26 making use of text generation, the output was either written (13 cases; \cite{cai_advancing_2024,chen_chatgpt_2023,elgarf_creativebot_2022,fang_socializechat_2023,fontana_de_vargas_co-designing_2024,jang_its_2024,jin_exploring_2024, li_blibug_2023,seo_chacha_2024,shakeri_saga_2021,tang_emoeden_2024,degen_retromind_2024,wei_improving_2024}), spoken (8 cases; \cite{billing_language_2023,li_exploring_2024,liu_peergpt_2024,bravo_creating_2023,said_design_2024,schnitzer_prototyping_2024,wang_aint_2024,xygkou_mindtalker_2024}) or both written and spoken (5 cases; \cite{alessa_towards_2023,bhati_bookmate_2023,jeung_unlocking_2024,samsonovich_registrar_2024,wan_building_2024}). 
To provide some examples, \citet{li_blibug_2023} developed Blibug, a GAI-powered virtual assistant that responded to audience messages with chat messages (written output) in an online live streaming environment.
 \citet{li_exploring_2024} developed an application in which users interacted with an LLM-powered avatar in VR through speech-based dialog (spoken output) to practice job-related communication skills.
Finally, \citet{bhati_bookmate_2023} developed BookMate, a tool for generating social stories for autistic children. BookMate presented the stories in a slide-show format for the children to read (written output), but also allowed for audio narration (spoken output) for users with limited reading capabilities.

Of the 30 designs included in the review, 26 made use of unimodal GAI and 4 of multimodal GAI. Notably, the 4 multimodal GAI-based designs made use of both text and image generation \cite{jin_exploring_2024,tang_emoeden_2024,degen_retromind_2024,wei_improving_2024}.
\citet{jin_exploring_2024} and \citet{degen_retromind_2024} expected that the integration of different types of GAI output would increase the vividness and immersion of the experience, as well as the effectiveness of reminiscence therapy compared to traditional methods.
Since we looked at the types of GAI that were applied, we did not consider designs showing both written text and speech as multimodal, as they use the same underlying GAI-based modality. An elaborate overview of the GAI-related characteristics of the designs presented in the included papers can be found in Table \ref{tab:appendixC_tech_details} in Appendix \ref{appendixC}.

\subsubsection{Software use}
What stood out from our data extraction process was that 25 of the 26 designs that used a text generation model used GPT. That is, 4 used a generic version of ChatGPT (they did not specify the exact model that powered ChatGPT \cite{bhati_bookmate_2023,chen_chatgpt_2023,said_design_2024,wei_improving_2024}), 4 used GPT-3 \cite{billing_language_2023,elgarf_creativebot_2022,li_blibug_2023,shakeri_saga_2021}, 9 used GPT-3.5 \cite{alessa_towards_2023,cai_advancing_2024,fontana_de_vargas_co-designing_2024,jeung_unlocking_2024,samsonovich_registrar_2024,li_exploring_2024,liu_peergpt_2024,bravo_creating_2023,schnitzer_prototyping_2024}, and 8 used GPT-4 \cite{fang_socializechat_2023,jang_its_2024,wan_building_2024,xygkou_mindtalker_2024,jin_exploring_2024,tang_emoeden_2024,degen_retromind_2024,seo_chacha_2024}. The only paper not using a GPT model used Llama-2 \cite{wang_aint_2024}.

The employed image generation models were more varied. Four of the 7 studies that used image generation used Midjourney \cite{jin_exploring_2024,liu_when_2024,tang_emoeden_2024,zhou_my_2024}, 2 used DALL-E \cite{degen_retromind_2024,chen_closer_2023}, and 1 mentioned the use of ChatGPT for text and image generation \cite{wei_improving_2024}. Since GPT cannot be used for image generation but only for image interpretation \cite{openai_vision_nodate}, the authors most likely used DALL-E.

Finally, the only study focusing on music generation used Cococo (collaborative co-creation) \cite{suh_ai_2021}. Cococo, developed by \citet{louie_novice-ai_2020} in 2020, is a web-based music editor designed for human-AI co-creation, improving traditional generative music interfaces with AI-guided specialized tools for creative control.

\subsubsection{Transparency of GAI use}
In terms of the system prompts that were the driving force behind the designs, only 15 of the 30 reviewed papers were transparent about their prompting decisions \cite{alessa_towards_2023,cai_advancing_2024,chen_chatgpt_2023,fang_socializechat_2023,fontana_de_vargas_co-designing_2024,jang_its_2024,jeung_unlocking_2024,li_exploring_2024,bravo_creating_2023,schnitzer_prototyping_2024,seo_chacha_2024,liu_when_2024,zhou_my_2024,tang_emoeden_2024,billing_language_2023}. In Table \ref{tab:appendixC_tech_details}, we report the example prompts provided by these authors. Three of the 30 papers made the source code of their GAI-based design publicly available on a GitHub repository \cite{billing_language_2023,seo_chacha_2024,chen_closer_2023}.

\subsubsection{Interfaces of the GAI-based technologies}
The designs in the 30 studies could be classified into four different categories: 1) Chatbot, 2) Embodied conversational agent, 3) Mobile platform powered by an LLM, and 4) Standalone GAI application. Table \ref{tab:tech/social_details} contains descriptions of the technologies.

\textbf{Chatbot} — The chatbots found in the included papers can be captured by the following definition: a chatbot has no physical or virtual embodiment and engages in conversation with the user through natural language processing (NLP) of speech or text input. In total, 10 out of 30 designs used a chatbot as the primary technology interface \cite{alessa_towards_2023,cai_advancing_2024,elgarf_creativebot_2022,jang_its_2024,samsonovich_registrar_2024,liu_peergpt_2024,seo_chacha_2024,tang_emoeden_2024,wei_improving_2024,xygkou_mindtalker_2024}. For example, \citet{alessa_towards_2023} developed a ChatGPT-based chatbot designed to provide companionship to older adults through text-based conversations to reduce loneliness and social isolation and \citet{seo_chacha_2024} developed ChaCha, a chatbot with a text-based interface that encourages children to express and share emotions.

\textbf{Embodied conversational agent} — Unlike chatbots, designs were classified as embodied conversational agents if the agent had a physical or virtual instantiation, granting access to non-verbal dimensions of social interaction. This was the case for 9 out of 30 designs \cite{billing_language_2023,li_blibug_2023,li_exploring_2024,bravo_creating_2023,said_design_2024,schnitzer_prototyping_2024,degen_retromind_2024,wan_building_2024,wang_aint_2024}. 
Six of these embodied conversational agents had a physical embodiment \cite{billing_language_2023,bravo_creating_2023,said_design_2024,schnitzer_prototyping_2024,degen_retromind_2024,wang_aint_2024} and 3 a virtual one \cite{li_blibug_2023,li_exploring_2024,wan_building_2024}.
Physically embodied conversational agents were typically instantiated in social robots. 

Embodied conversational agents typically employed GAI with the aim of improving verbal communication. For example, \citet{billing_language_2023} explored using GPT-3 with the Nao and Pepper robots to create an open verbal dialogue between robot and human interaction partners, and \citet{li_exploring_2024} developed a VR avatar, powered by an LLM, which allowed autistic people to practice job-related communication skills. 

\textbf{Mobile platform powered by an LLM} — In total, 8 of 30 designs could be classified as mobile LLM-powered platforms \cite{bhati_bookmate_2023,chen_closer_2023,fang_socializechat_2023,fontana_de_vargas_co-designing_2024,jeung_unlocking_2024,shakeri_saga_2021, jin_exploring_2024,suh_ai_2021}. These could be divided into 1) education and support tools \cite{bhati_bookmate_2023,fang_socializechat_2023,fontana_de_vargas_co-designing_2024} used to teach communication skills or provide the tools necessary to express oneself, 2) creative and collaborative platforms \cite{chen_closer_2023,shakeri_saga_2021,suh_ai_2021} allowing collaborative storytelling, story creation, and music generation, and 3) memory tools \cite{jeung_unlocking_2024,jin_exploring_2024} supporting reminiscence.

\textbf{Standalone GAI application} — In total, 3 out of 30 designs could be classified as a standalone GAI application. These did not envision any integration into a custom platform or system, and utilized the original interface or API for direct user interaction \cite{zhou_my_2024,chen_chatgpt_2023,liu_when_2024}. In other words, they executed the GAI task in its original form. \citet{zhou_my_2024} conducted co-design sessions with children in which the latter hand-crafted avatars. The moderators subsequently used images of the avatars and verbal descriptions of them to create AI-generated images of the avatars. These avatars could later be used as the main characters in a storytelling activity. \citet{liu_when_2024} used Midjourney to generate AI images from visual arts created by families and verbal descriptions provided by parents, children, and therapists, highlighting the importance of both verbal and non-verbal forms in expressive arts therapy. \citet{chen_chatgpt_2023} had participants use ChatGPT to create mindmaps to organize topics and information to include in storytelling.

\begin{longtable}[c]{>{\raggedright\arraybackslash}p{.18\textwidth} >{\raggedright\arraybackslash}p{.15\textwidth} >{\raggedright\arraybackslash}p{.48\textwidth} >{\raggedright\arraybackslash}p{.1\textwidth}}
\caption{Overview of the activity contexts for which the GAI-based technologies were designed and the embodiment of the technologies in those contexts.}
\label{tab:tech/social_details}\\
\hline
\textbf{Embodiment} & \textbf{Activity} & \textbf{Technology description} & \textbf{Reference} \\ \hline
\endfirsthead
\hline
\textbf{Embodiment} & \textbf{Activity} & \textbf{Technology description} & \textbf{Reference} \\ \hline
\endhead
\hline
\endfoot
\endlastfoot

Chatbot & Storytelling & CreativeBot is an autonomous chatbot used in a collaborative storytelling context. It generates creative versus non-creative behavior in storytelling scenarios. & \citet{elgarf_creativebot_2022}\\
Chatbot & Storytelling \& Socio-emotional skills training & ChaCha is a chatbot that combines a state machine and LLMs to guide children in sharing personal events and associated emotions. & \citet{seo_chacha_2024}\\
Chatbot & Socio-emotional skills training & The system uses an LLM-based chatbot to provide social communication advice for autistic individuals facing workplace-related social difficulties. & \citet{jang_its_2024}\\
Chatbot & Socio-emotional skills training & EmoEden is an AI-driven training tool designed to assist high-functioning autistic children in identifying and expressing their emotions. It integrates LLMs and text-to-image models to provide personalized assistance in emotional learning. & \citet{tang_emoeden_2024}\\
Chatbot & Reminiscence & MindTalker is an audio-based chatbot created using GPT-4 to support reminiscence therapy through meaningful conversations about personal photos. & \citet{xygkou_mindtalker_2024}\\
Chatbot & Collaborative learning activity & The system is an LLM-powered chatbot designed to enhance small group learning through advanced language understanding and generation capabilities. It facilitates transitions through phases of task completion, asks follow-up questions, challenges ideas, and promotes inclusivity. & \citet{cai_advancing_2024}\\
Chatbot & Collaborative learning activity & The system uses LLM-based peer agents to facilitate children's collaborative learning by acting as team moderators or participants. & \citet{liu_peergpt_2024}\\
Chatbot & Collaborative learning activity & The system integrates an LLM virtual assistant into a WeChat group to aid collaborative learning by providing problem feedback, resource gathering, and inspiration. & \citet{wei_improving_2024}\\
Chatbot & General conversation & The system is a ChatGPT-based conversational companion designed to provide companionship and reduce loneliness among older adults. It uses personalized prompts based on user information to generate relevant and engaging conversations. & \citet{alessa_towards_2023}\\
Chatbot & General conversation & Registrar is a virtual chatbot designed to assist with hotel registration tasks, integrating a cognitive model with neural network models. & \citet{samsonovich_registrar_2024}\\
\hline
Embodied conversational agent & Socio-emotional skills training & The system uses LLM-driven chatbots in VR environments to help autistic individuals practice job communication skills through speech-based interactions. & \citet{li_exploring_2024}\\
Embodied conversational agent & Reminiscence & RetroMind is a framework that combines LLMs and social robots to improve the effectiveness of reminiscence therapy by creating visual material based on memories. & \citet{degen_retromind_2024}\\
Embodied conversational agent & General conversation & The system integrates the OpenAI GPT-3 language model with the Aldebaran Pepper and Nao robots to create an open verbal dialogue with the robots. & \citet{billing_language_2023}\\
Embodied conversational agent & General conversation & Blibug is an AI-Vtuber on the Chinese Bilibili platform that interacts with viewers through "Danmu" messages (real-time comments overlaid on the video playback). & \citet{li_blibug_2023}\\
Embodied conversational agent & General conversation & The system uses LLMs to generate personalized verbal interactions based on users' interests and preferences. & \citet{bravo_creating_2023}\\
Embodied conversational agent & General conversation & The system is a humanoid robotic platform named Adam that engages in vocal conversations using ChatGPT and exhibits human-like movements. & \citet{said_design_2024}\\
Embodied conversational agent & General conversation & The system is a zoomorphic interactive robot companion that combines intelligent conversational abilities and emotion recognition to support older adults. & \citet{schnitzer_prototyping_2024}\\
Embodied conversational agent & General conversation & The system uses an LLM-based AI agent deployed in VRChat as a non-playable character (NPC) that responds to player interactions with context-relevant responses, facial expressions, and body gestures. & \citet{wan_building_2024}\\
Embodied conversational agent & General conversation & The study leverages LLMs to generate expressive behaviors and responses during conversations with tabletop robot Haru. & \citet{wang_aint_2024}\\
\hline
Mobile platform with LLM suggestions/input & Storytelling & Closer Worlds is a two-player world-building game that uses GAI to foster intimate conversation through imaginative play and reflection. Participants take turns describing an imaginative world through text input, which the AI converts to an image. & \citet{chen_closer_2023}\\
Mobile platform with LLM suggestions/input & Storytelling & SAGA is a web application that allows friends to collaboratively write stories with the help of a system that leverages GPT as a co-writing component. & \citet{shakeri_saga_2021}\\
Mobile platform with LLM suggestions/input & Socio-emotional skills training & BookMate is an AI-based interactive mobile application that helps caregivers of people with ASD generate social stories. It integrates AI processing and AI-based audio transcription for story generation, effective audio data extraction, and data processing. & \citet{bhati_bookmate_2023}\\
Mobile platform with LLM suggestions/input & Socio-emotional skills training & QuickPic is a mobile AAC application co-designed with Speech-Language Pathologists (SLPs) and special educators to enhance language learning for non-speaking individuals. It generates topic-specific communication boards automatically from photographs. & \citet{fontana_de_vargas_co-designing_2024}\\
Mobile platform with LLM suggestions/input & Reminiscence & Treasurefinder is a device powered by an LLM that generates open-ended questions based on stories stored in NFC-tagged physical objects or cards. & \citet{jeung_unlocking_2024}\\
Mobile platform with LLM suggestions/input & Reminiscence & The system uses GAI to support music-based reminiscence by generating conversations and images based on users' memories. & \citet{jin_exploring_2024}\\
Mobile platform with LLM suggestions/input & Music making & Cococo is an AI tool that supports collaborative music composition by automatically generating musical phrases based on user input. & \citet{suh_ai_2021}\\
Mobile platform with LLM suggestions/input & General conversation & SocializeChat is an Augmentative and Alternative Communication (AAC) tool that employs LLM technology to boost social chat with gaze inputs. It generates multiple sentences of conversation in real-time, offers suggestions tailored to users' preferences, and phrases sentences in a style that matches the relationship of people in conversation. & \citet{fang_socializechat_2023}\\
\hline
Standalone GAI application & Storytelling & The system uses ChatGPT to generate stories and mind maps to enhance children's storytelling and comprehension skills. It provides personalized feedback and recommendations. & \citet{chen_chatgpt_2023}\\
Standalone GAI application & Storytelling \& Socio-emotional skills training & The system integrates GAI with traditional art-making materials to support family expressive arts therapy. Families use both traditional materials and image-based GAI to create and evolve their family stories. & \citet{liu_when_2024}\\
Standalone GAI application & Storytelling \& Socio-emotional skills training & The system involves crafting agents by children from physical materials, which are then digitalized by using Midjourney. The digital agents act as storytelling agents designed to support socio-emotional learning by co-creating characters and social-emotional stories through collaborative activities. & \citet{zhou_my_2024}\\

\hline
\end{longtable}

\subsection{Social interaction dynamics of GAI-based technology use (RQ2)}

\subsubsection{Social activities}
In the papers included in the review, we identified six different social activities fostered by GAI-based designs. These are all shown in Table \ref{tab:tech/social_details}. 

\textbf{Storytelling or storymaking} — In total, 7 GAI-based designs were intended to support users in storytelling or storymaking \cite{chen_closer_2023,chen_chatgpt_2023,elgarf_creativebot_2022,shakeri_saga_2021,zhou_my_2024,liu_when_2024,seo_chacha_2024}. 
These included all 3 standalone GAI applications \cite{chen_chatgpt_2023,zhou_my_2024,liu_when_2024}.
\citet{chen_closer_2023} show how collaboratively creating and describing imaginative worlds, with GAI generated corresponding images based on users' descriptions, can facilitate intimate conversations and self-reflection in a world-building storytelling game. They did not focus on any particular end-user group, but rather acknowledged that anyone looking for a connection could benefit from using their design, Closer Worlds.
Children seemed to be the intended end-users of the other GAI-based storytelling and storymaking designs. Here, storytelling was used as a method for training socio-emotional skills \cite{zhou_my_2024,liu_when_2024,seo_chacha_2024}. Given the goal of training socio-emotional skills, we discuss these examples in the eponymous category.

\textbf{Socio-emotional skills training} — Eight GAI-based designs were presented in socio-emotional skills training settings \cite{liu_when_2024,seo_chacha_2024,bhati_bookmate_2023,fontana_de_vargas_co-designing_2024,jang_its_2024,li_exploring_2024,tang_emoeden_2024}, 3 of which also addressed storytelling \cite{liu_when_2024,seo_chacha_2024,zhou_my_2024}. 
The latter 3 illustrate how storytelling was perceived an effective method to support socio-emotional skills training for children. For example, in \citet{zhou_my_2024}, children crafted and digitalized (using Midjourney) storytelling agents, using AI-generated characters to foster socio-emotional learning by collaborating on social-emotional stories. Similarly, \citet{liu_when_2024} integrated AI-generated images with traditional art materials, enabling families to collaboratively craft and evolve stories that support emotional expression and empathy. In \citet{seo_chacha_2024}, the ChaCha chatbot guided children in sharing personal stories and emotions, promoting emotional reflection and understanding. These approaches allowed children to practice emotion recognition, empathy, and communication skills through co-created narratives.

\textbf{Reminiscence} — In 4 cases, GAI was applied to reminiscence activities \cite{jeung_unlocking_2024,jin_exploring_2024,degen_retromind_2024,xygkou_mindtalker_2024}. 
Reminiscence applications tended to be aimed at people with dementia or mild cognitive impairment \cite{degen_retromind_2024,xygkou_mindtalker_2024} or older adults in general \cite{jin_exploring_2024}.
An approach that different papers had in common was that of generating new text and images while reciting memories to foster recollection and conversations about those memories \cite{jin_exploring_2024,degen_retromind_2024}. For example, \citet{degen_retromind_2024} built the RetroMind framework that leverages LLMs and the Pepper robot to help people with demenetia engage in personalized sessions, where AI-generated images are created based on people with dementia's recollections. These visual representations of memories were used to stimulate memory recall and reinforce the narrative of the person's own life story. Similarly, \citet{jin_exploring_2024} addressed music-related memories through AI-generated images and conversation suggestions.

\textbf{Collaborative learning activity} — Three studies addressed the use of GAI to foster collaboration between people \cite{cai_advancing_2024,liu_peergpt_2024,wei_improving_2024}.
In all cases, a chatbot was used in the collaboration context. The aim of using the chatbot was to make the collaboration between human actors more efficient. \citet{cai_advancing_2024} built a text-based chatbot that was present with a group in real-time to enhance small group collaborative learning by facilitating smooth transitions through stages of discussion. Similarly, \citet{liu_peergpt_2024} introduced a virtually present, GAI-based peer agent with either the role of moderator or participant in children's collaborative workshops. Depending on the role, the agent either participated directly or managed the group as a whole.

\textbf{Music making} — Of all 30 papers included in the review, only one study focused on collaborative music composition. \citet{suh_ai_2021} used the GAI-based music composition tool Cococo to automatically generate musical phrases based on user inputs with a particular focus on how the social dynamics were affected by the presence of the AI.

\textbf{General conversation} — Finally, 10 papers did not address a specific social activity but rather targeted engagement in general conversation \cite{alessa_towards_2023,billing_language_2023,fang_socializechat_2023,samsonovich_registrar_2024,li_blibug_2023,bravo_creating_2023,said_design_2024,schnitzer_prototyping_2024,wan_building_2024,wang_aint_2024}. 
Many of these papers focused on using GAI in embodied conversational agents to establish natural human-machine interaction \cite{billing_language_2023,said_design_2024,schnitzer_prototyping_2024,wang_aint_2024}. For example, \citet{wang_aint_2024} leveraged LLMs to generate expressive behaviors and responses during interactions with the robot Haru without mentioning a particular application context.
In Table \ref{tab:tech/social_details}, we provide more details on the general conversation contexts of the papers in question.

\subsubsection{Social contexts}
A detailed description of the social context in which the designed GAI-based applications were used can be found in Table \ref{tab:tech/social_context}. The results revealed that the different embodiments adopted by the GAI applications were linked to different types of sociality (i.e., whether a person or thing can present itself as a social actor or not). 
Based on the technology embodiments presented to the end-user in Table \ref{tab:tech/social_details}, we map these technologies onto the adapted functional triad framework \cite{fogg_persuasive_2003}, which describes the roles technology can play in social interactions (i.e., as a tool, medium, or agent). 
Essentially, every chatbot and embodied conversational agent is an interactant in social interaction. Thus, in 19 cases, the GAI-based technology took up the role of an \textbf{agent} in the interaction \cite{elgarf_creativebot_2022,seo_chacha_2024,jang_its_2024,tang_emoeden_2024,xygkou_mindtalker_2024,cai_advancing_2024,liu_peergpt_2024,wei_improving_2024,alessa_towards_2023,samsonovich_registrar_2024,li_exploring_2024,degen_retromind_2024,billing_language_2023,li_blibug_2023,bravo_creating_2023,said_design_2024,schnitzer_prototyping_2024,wan_building_2024,wang_aint_2024}. 
Mobile platforms with LLM suggestions/input and cases of standalone GAI application typically supported people in engaging in social interaction or acted as a channel facilitating social interaction between people, respectively taking up the role of tool or medium in the interaction. In 9 cases, the technology functioned as a \textbf{tool} \cite{bhati_bookmate_2023,chen_chatgpt_2023,chen_closer_2023,fontana_de_vargas_co-designing_2024,jeung_unlocking_2024,jin_exploring_2024,liu_when_2024,suh_ai_2021,zhou_my_2024}; in 2 cases, the role of \textbf{medium} \cite{fang_socializechat_2023,shakeri_saga_2021}.

\textbf{GAI as an agent to facilitate social participation} — We found that \textit{chatbots} or \textit{embodied conversational agents} actively participated in the social interactions as social actors \cite{alessa_towards_2023,elgarf_creativebot_2022,jang_its_2024,samsonovich_registrar_2024,seo_chacha_2024,tang_emoeden_2024,xygkou_mindtalker_2024,cai_advancing_2024,liu_peergpt_2024,wei_improving_2024,billing_language_2023,bravo_creating_2023,said_design_2024,schnitzer_prototyping_2024,wang_aint_2024,li_blibug_2023,li_exploring_2024,wan_building_2024,degen_retromind_2024}. 
They primarily engaged in one-to-one interactions with individual users. In some cases though, they fostered human-machine and human-human social interaction simultaneously. This was the case of group-based social interactions with the technology \cite{cai_advancing_2024,liu_peergpt_2024,wei_improving_2024}.
In these instances, they were embedded in collaborative learning scenarios, in which the GAI-based chatbot took on the role of guiding and contributing to the collaboration process \cite{cai_advancing_2024,liu_peergpt_2024,wei_improving_2024}. The task of the GAI chatbot in these contexts was to help maintain real-time interaction, offer personalized, dynamic support, and promote a collaborative, inclusive learning environment.

\textbf{GAI as a medium through which social interaction takes place} — GAI-based technologies we classified as a medium acted as a communication channel through which a social interaction between humans took place \cite{fang_socializechat_2023,shakeri_saga_2021}. SocializeChat, developed by \citet{fang_socializechat_2023}, demonstrates how GAI can be applied to build a communication channel. It is an application designed for people with speech and motor impairments that uses eye-gaze technology to facilitate real-time conversations. Using GPT-4 and eye-gaze technology, it helps individuals with speech and motor impairments communicate by offering sentence suggestions and adapting communication styles according to the social context. Thus, GAI-based technology did not take on an participatory role, but rather acted as a communication channel.

\textbf{GAI as a tool to facilitate social interaction} — Technologies that we classified as a tool showed the opposite to those acting as an agent in terms of the social interaction contexts addressed \cite{jin_exploring_2024,bhati_bookmate_2023,chen_closer_2023,fontana_de_vargas_co-designing_2024,suh_ai_2021,jeung_unlocking_2024,chen_chatgpt_2023,liu_effectiveness_2021,zhou_my_2024}. Most importantly, these types of technology were intended to facilitate interaction between two or more humans without taking an active role in the social interaction itself (i.e., they did not act as social actors). 
When the GAI-based application was a tool, it acted as a source for the social interaction between two people \cite{bhati_bookmate_2023,chen_closer_2023,fang_socializechat_2023,fontana_de_vargas_co-designing_2024,suh_ai_2021,shakeri_saga_2021,jeung_unlocking_2024}. An example is Treasurefinder, a small device that uses an LLM to create open-ended questions for users to discuss \cite{jeung_unlocking_2024}. The device is activated when users scan an NFC-tagged item at home, after which it plays voice stories that were previously recorded by the object's owners, enabling users to engage with the objects and their related narratives. This demonstrates how LLMs can be used to foster conversation between users. 
Another example is Closer Worlds, designed by \citet{chen_closer_2023}. It is a two-player game focused on creating worlds, aiming to cultivate deep conversations via imaginative play and reflection using GAI. Players alternately input text to describe a fantasy world, which is then translated into an image by the GAI. This cycle is interwoven with reflective questions to support personal sharing.

A type of social interaction which deserves to be highlighted is the use of GAI to stimulate communication between pairs of people. This type of social interaction mainly occurred in therapy setting to enhance communication between therapists and vulnerable people like people with dementia \cite{degen_retromind_2024} or autistic people \cite{bhati_bookmate_2023,fontana_de_vargas_co-designing_2024}. Within this context, therapists used GAI-based technologies to prepare and support the therapy sessions with their clients. For example, \citet{fontana_de_vargas_co-designing_2024} co-designed QuickPic, a mobile application, with Speech-Language Pathologists and special education teachers to assist in the creation of topic-specific communication boards for autistic individuals. Such application allows therapists to generate communication boards automatically from photographs, providing "just-in-time" language support during therapy sessions, while also guiding the interaction by helping users, such as autistic children, to formulate sentences using the displayed symbols. 
Here, we would like to highlight a certain asymmetry of the therapy-setting in terms of access to GAI-based technology. Therapists are the only ones deciding how the communication boards are generated, autistic individuals and people with dementia do not have a say in this process and are not given the possibility to adapt the design to their needs. 

All pair-based technology interactions but one \cite{shakeri_saga_2021} took place in a co-located environment \cite{bhati_bookmate_2023,chen_closer_2023,fang_socializechat_2023,fontana_de_vargas_co-designing_2024,jeung_unlocking_2024,suh_ai_2021}. \citet{shakeri_saga_2021} dared to design an online pair-based interaction. They built SAGA, an asynchronous collaborative storytelling system that allows friends separated by distance to write stories together with the help of a system that leverages GPT as a co-writing component. Each user is assigned a unique color to denote their contributions, and takes turns entering prompts that guide the narrative, with the GAI adding to the story between turns. Users found themselves to be integrating their own personality into the characters in the storyline, this way expressing actions and feelings they would not have as easily in real life, creating a sense of connectedness. Even at a distance, SAGA acted as a medium through which social interaction could be fostered.

\begin{longtable}[c]{>{\raggedright\arraybackslash}p{.13\textwidth} >{\raggedright\arraybackslash}p{.14\textwidth} >{\raggedright\arraybackslash}p{.11\textwidth} >{\raggedright\arraybackslash}p{.1\textwidth} >{\raggedright\arraybackslash}p{.1\textwidth} >{\raggedright\arraybackslash}p{.16\textwidth} >{\raggedright\arraybackslash}p{.1\textwidth}}
\caption{Overview of the contexts of the social interactions brought about by using the GAI-based designs presented in the papers.}
\label{tab:tech/social_context}\\
\hline
\textbf{Embodiment} & \textbf{Type of interaction} & \textbf{Setting} & \textbf{Interaction space} & \textbf{Synchrony} & \textbf{Equality of access to GAI} & \textbf{Reference} \\ \hline
\endfirsthead
\hline
\textbf{Embodiment} & \textbf{Type of interaction} & \textbf{Setting} & \textbf{Interaction space} & \textbf{Synchrony} & \textbf{Equality of access to GAI} & \textbf{Reference} \\ \hline
\endhead
\hline
\endfoot
\endlastfoot
Chatbot & Human-Machine & Individual interaction & Virtual & Real-time & N/A & \citet{alessa_towards_2023}\\
Chatbot & Human-Machine & Individual interaction & Virtual & Real-time & N/A & \citet{elgarf_creativebot_2022}\\
Chatbot & Human-Machine & Individual interaction & Virtual & Real-time & N/A & \citet{jang_its_2024}\\
Chatbot & Human-Machine & Individual interaction & Virtual & Real-time & N/A & \citet{samsonovich_registrar_2024}\\
Chatbot & Human-Machine & Individual interaction & Virtual & Real-time & N/A & \citet{seo_chacha_2024}\\
Chatbot & Human-Machine & Individual interaction & Virtual & Real-time & N/A & \citet{tang_emoeden_2024}\\
Chatbot & Human-Machine & Individual interaction & Virtual & Real-time & N/A & \citet{xygkou_mindtalker_2024}\\
Chatbot & Human-Machine \& Human-Human & Group-based interaction & Virtual & Real-time & Equal access & \citet{cai_advancing_2024}\\
Chatbot & Human-Machine \& Human-Human & Group-based interaction & Co-located & Real-time & Equal access & \citet{liu_peergpt_2024}\\
Chatbot & Human-Machine \& Human-Human & Group-based interaction & Virtual & Real-time & Equal access & \citet{wei_improving_2024}\\ \hline
Embodied conversational agent & Human-Machine & Individual interaction & Co-located & Real-time & N/A & \citet{billing_language_2023}\\
Embodied conversational agent & Human-Machine & Individual interaction & Co-located & Real-time & N/A & \citet{bravo_creating_2023}\\
Embodied conversational agent & Human-Machine & Individual interaction & Co-located & Real-time & N/A & \citet{said_design_2024}\\
Embodied conversational agent & Human-Machine & Individual interaction & Co-located & Real-time & N/A & \citet{schnitzer_prototyping_2024}\\
Embodied conversational agent & Human-Machine & Individual interaction & Co-located & Real-time & N/A & \citet{wang_aint_2024}\\
Embodied conversational agent & Human-Machine & Individual interaction & Virtual & Real-time & N/A & \citet{li_blibug_2023}\\
Embodied conversational agent & Human-Machine & Individual interaction & Virtual & Real-time & N/A & \citet{li_exploring_2024}\\
Embodied conversational agent & Human-Machine & Individual interaction & Virtual & Real-time & N/A & \citet{wan_building_2024}\\
Embodied conversational agent & Human-Machine \& Human-Human & Pair-based interaction & Co-located & Real-time & Unequal — Therapist is in control of what the GAI model does rather than the person with dementia. Also, only memories of the person with dementia are used as input. & \citet{degen_retromind_2024}\\ \hline
Mobile platform with LLM suggestions/input & Human-Machine & Individual interaction & Virtual & Real-time & N/A & \citet{jin_exploring_2024}\\
Mobile platform with LLM suggestions/input & Human-Human & Pair-based interaction & Co-located & Asynchronous — The GAI output is shown to the interaction partner, the autistic person, later & Unequal access — caregiver uses GAI, autistic person is exposed to the result & \citet{bhati_bookmate_2023}\\
Mobile platform with LLM suggestions/input & Human-Human & Pair-based interaction & Co-located & Real-time & Equal access & \citet{chen_closer_2023}\\
Mobile platform with LLM suggestions/input & Human-Human & Pair-based interaction & Co-located & Real-time & Unequal access — person with impairment has access to four options; the interaction partner to every possible response & \citet{fang_socializechat_2023}\\
Mobile platform with LLM suggestions/input & Human-Human & Pair-based interaction & Co-located & Asynchronous — The GAI output is shown to the interaction partner, the non-speaking individual, later & Unequal access — caregiver uses GAI, non-speaking individual is exposed to the result & \citet{fontana_de_vargas_co-designing_2024}\\
Mobile platform with LLM suggestions/input & Human-Human & Pair-based interaction & Co-located & Real-time & Equal access & \citet{suh_ai_2021}\\
Mobile platform with LLM suggestions/input & Human-Human & Pair-based interaction & Virtual & Asynchronous — Users can work on their prompts at different times & Equal access & \citet{shakeri_saga_2021}\\
Mobile platform with LLM suggestions/input & Human-Machine \& Human-Human & Individual interaction \& Pair-based interaction & Co-located & Real-time & Equal access & \citet{jeung_unlocking_2024}\\ \hline
Standalone GAI application & Human-Human & Group-based interaction & Co-located & Real-time & Equal access & \citet{chen_chatgpt_2023}\\
Standalone GAI application & Human-Human & Group-based interaction & Co-located & Real-time & Equal access & \citet{liu_when_2024}\\
Standalone GAI application & Human-Human & Group-based interaction & Co-located & Real-time & Equal access & \citet{zhou_my_2024}\\
\hline
\end{longtable}
\begin{flushleft}
\textit{Note:} The type of interaction refers to the social interaction that takes place, which can be exclusively between the user and the machine, between multiple users, or both. The setting addresses how the technology is used — alone, in pairs, or in group. The interaction space denotes whether the interaction between the parties involved takes place virtually or physically (co-located). Synchrony refers to whether the social interaction takes place at the same time for each party involved or whether interaction may take place asynchronously. Equality of access is an indication of whether every party involved in the interaction had access to the same part of the GAI-based technology. This part was disregarded for exclusive Human-Machine interactions, as there is an inherent inequality present.
\end{flushleft}

\subsection{Design and evaluation methodologies (RQ3)}
\subsubsection{User involvement}
\begin{figure}[ht]
    \centering
    \includegraphics[width=\textwidth]{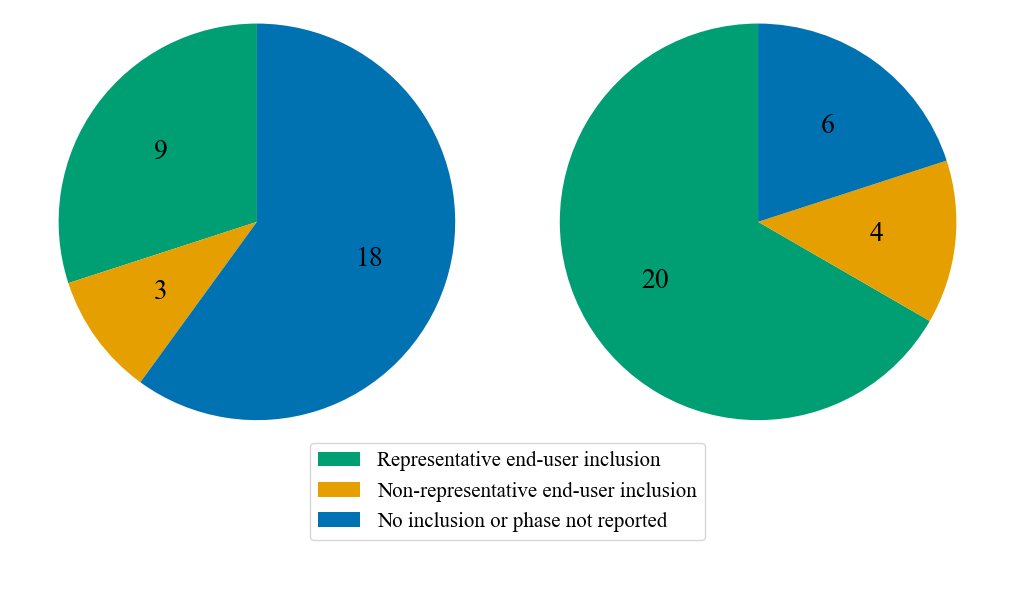}  
    \caption{An overview of intended end-user inclusion in the conceptualization phase (left graph) and the evaluation phase (right graph).}  
    \label{fig:inclusion_pie}
\end{figure}

In terms of end-user involvement in the design process, we differentiated between participant involvement in the \emph{conceptualization phase} and \emph{evaluation phase}. 
The conceptualization phase involves any activity that leads to the creation and envisioning of the application. 
In contrast, the evaluation phase refers to the evaluation and testing of the application to ensure it meets the desired requirements and works effectively. 
Figure \ref{fig:inclusion_pie} shows the level of involvement of the intended end-user in different stages of the design process.
We found three possible ways of participant inclusion: 1) representative end-user inclusion, 2) non-representative end-user inclusion, and 3) no inclusion or no mention of the conceptualization phase. 
By \textit{representative end-user}, we mean the user to whom the GAI-based design is intended. For example, \citet{xygkou_mindtalker_2024} built MindTalker, an audio-based conversational agent created using the state-of-the-art GPT-4 LLM to carry out meaningful conversations with people with early-stage dementia. In this case, we consider people with early-stage dementia as representative end-users. \citet{xygkou_mindtalker_2024} consulted dementia experts and therapists during the conceptualization phase of the technologies, and people with dementia only in their evaluation phase. Therefore, in Figure \ref{fig:inclusion_pie}, we classified as non-representative end-user inclusion for the conceptualization phase and representative end-user inclusion for the evaluation phase.
Although we positively value the inclusion of all relevant stakeholders (in this case, also dementia experts) in the design and evaluation processes, in this review, we mainly highlight whether and how the actual end-users of the GAI-based technologies have been involved in the design process.

\textbf{End-user inclusion in the conceptualization phase} (Figure \ref{fig:inclusion_pie} — left graph) — Of the 30 articles, 12 reported the inclusion of participants (either representative or non-representative) in the conceptualization phase \cite{fontana_de_vargas_co-designing_2024,fang_socializechat_2023,elgarf_creativebot_2022,xygkou_mindtalker_2024,liu_when_2024,jin_exploring_2024,zhou_my_2024,tang_emoeden_2024,schnitzer_prototyping_2024,seo_chacha_2024,cai_advancing_2024,chen_closer_2023}, as shown in green and yellow in the left graph Figure \ref{fig:inclusion_pie}. 
In 9 cases, the representative end-user was involved in conceptualization \cite{fontana_de_vargas_co-designing_2024,elgarf_creativebot_2022,jin_exploring_2024,zhou_my_2024,tang_emoeden_2024,schnitzer_prototyping_2024,seo_chacha_2024,cai_advancing_2024,chen_closer_2023}. Six of the articles demonstrated that including vulnerable end-users in the conceptualization phase is possible (i.e., children \cite{zhou_my_2024,seo_chacha_2024,elgarf_creativebot_2022}, including children with HFA \cite{tang_emoeden_2024}; older adults \cite{jin_exploring_2024}; people with dementia or MCI \cite{schnitzer_prototyping_2024}).

Of the 12 papers that included a conceptualization phase, 3 reported consulting non-representative end-users exclusively during that phase
\cite{fang_socializechat_2023,xygkou_mindtalker_2024,liu_when_2024}. For all 3 cases, the intended end-user was vulnerable (i.e., people with speech and motor impairments \cite{fang_socializechat_2023}; children \cite{liu_when_2024}; people with dementia \cite{xygkou_mindtalker_2024}). 
All 3 reported the use of experts to ensure that their designs were tailored to the intended end-users.
A more elaborate reasoning for choosing to include only non-representative participants in the conceptualization phase was offered by \citet{xygkou_mindtalker_2024}. They designed their chatbot to stimulate social connectedness among people with early-stage dementia, but only involved dementia care experts and therapists in the co-creation process. Their reason for not including people with dementia in the conceptualization phase was that they prioritized expert knowledge to ensure the app was aligned to therapeutic needs. In addition, they avoided involving people with dementia in the early stages of the process due to ethical concerns about the cognitive strain of long-term participation and safety concerns, as GPT-4's dialogues had not been extensively evaluated for their suitability or potential impact on people with dementia. During the evaluation phase, people with early-stage dementia were involved.

Most of the 12 studies that reported including end-users in the conceptualization phase did so through interviews to determine the technology requirements and context of use \cite{fang_socializechat_2023,jin_exploring_2024,liu_when_2024,seo_chacha_2024,tang_emoeden_2024,xygkou_mindtalker_2024}. 
A limited number of studies included participants in co-design sessions to establish design choices \cite{cai_advancing_2024,zhou_my_2024,fontana_de_vargas_co-designing_2024,chen_closer_2023}. For instance, \citet{zhou_my_2024} allowed children to craft physical objects and use image generation to create virtual agents and explore how they could be used to guide storytelling. 
\citet{cai_advancing_2024} used co-design sessions to understand the desired behaviors and optimal timing of chatbots. They did so by employing a Wizard-of-Oz technique where researchers simulated the chatbot, and undergraduate students crafted and refined the chatbot through a participatory design approach.
\citet{elgarf_creativebot_2022} had an alternative approach to end-user inclusion. They used a dataset with previously collected storytelling data of children telling a story to or collaboratively with a wizarded-robot. These data were used to extract statements to label them as creative or non-creative, which informed the training of storytelling robots.

The other 18 articles did not report participant inclusion in their conceptualization phase or did not present the conceptualization phase. The majority of them did not elaborate on how their applications came to the current stage \cite{alessa_towards_2023,bhati_bookmate_2023,chen_chatgpt_2023,jang_its_2024,jeung_unlocking_2024,samsonovich_registrar_2024,li_exploring_2024,bravo_creating_2023,said_design_2024,shakeri_saga_2021,suh_ai_2021,degen_retromind_2024,wan_building_2024,wang_aint_2024,wei_improving_2024,liu_peergpt_2024}. Two studies focused primarily on the technical description and feasibility of their application, without going into the design process or involving participants \cite{billing_language_2023,li_blibug_2023}. 

\textbf{End-user involvement in the evaluation phase} (Figure \ref{fig:inclusion_pie} — right graph) — Of the 30 articles, 24 reported the inclusion of participants (either representative or non-representative) in the evaluation phase \cite{fontana_de_vargas_co-designing_2024,fang_socializechat_2023,elgarf_creativebot_2022,xygkou_mindtalker_2024,seo_chacha_2024,tang_emoeden_2024,bhati_bookmate_2023,jang_its_2024,alessa_towards_2023,li_exploring_2024,liu_peergpt_2024,chen_chatgpt_2023,cai_advancing_2024,chen_closer_2023,bravo_creating_2023,wang_aint_2024,samsonovich_registrar_2024,said_design_2024,wan_building_2024,jeung_unlocking_2024,suh_ai_2021,shakeri_saga_2021,wei_improving_2024} (see Figure \ref{fig:inclusion_pie} — right graph).
\citet{degen_retromind_2024} also presented an evaluation step but used data from a documented case study of a person with dementia found in existing literature \cite{shenk_narratives_2002}. They tested the performance of RetroMind, a framework that combines LLMs and social robots to enhance reminiscence therapy by creating visual materials based on memories of people with dementia. The goal was to assess system performance in terms of the effectiveness of generating images that correspond with verbal descriptions of memories from the documented case study of a person with dementia. This one is also included in the right graph in Figure \ref{fig:inclusion_pie} as \textit{Representative end-user inclusion}.

In 4 cases, the authors reported that they had consulted only non-representative end-users during the conceptualization phase \cite{fang_socializechat_2023,bhati_bookmate_2023,elgarf_creativebot_2022,alessa_towards_2023}. In all 4 cases, the intended end-user could be considered vulnerable (i.e., older adults \cite{alessa_towards_2023}; autistic people \cite{bhati_bookmate_2023}; children \cite{elgarf_creativebot_2022}; people with speech and motor impairments \cite{fang_socializechat_2023}). 
Part of the reason for this is the preliminary nature of many of the designs presented in the papers. Eleven of the designs that were evaluated were explicitly reported to be in a `preliminary' \cite{alessa_towards_2023,bhati_bookmate_2023,cai_advancing_2024,chen_chatgpt_2023,fang_socializechat_2023,degen_retromind_2024,wan_building_2024}, `early-stage' \cite{li_exploring_2024}, or `pilot' stage \cite{liu_peergpt_2024,shakeri_saga_2021,wang_aint_2024}. 
The evaluation was often aimed at determining system parameters or optimizing performance to obtain desired application behavior \cite{elgarf_creativebot_2022,wan_building_2024,wang_aint_2024,cai_advancing_2024,degen_retromind_2024}, and evaluating the feasibility of using GAI-based technology in the intended use cases \cite{alessa_towards_2023,bravo_creating_2023}. This means that for many of the papers, the applications were not yet ready for longitudinal use and evaluation.
For example, \citet{elgarf_creativebot_2022} designed a chatbot to be used in collaborative storytelling with children. However, the focus of their evaluation was on whether the output was perceived as creative by adults. They expressed the intention to test with children in the future.
\citet{fang_socializechat_2023} designed a gaze-based communication interface for people with speech and motor impairments; they involved caregivers of people in the intended target group. However, they also mentioned that their study was in a preliminary testing phase, acknowledging the need for additional testing with the intended end-user as they might have unexpected demands for social communication. Similar intentions were expressed by \citet{alessa_towards_2023} and \citet{bhati_bookmate_2023}.

While a limited number of studies included end-users in the design of the GAI-based technologies, the majority of them (20 out of 30) reported representative end-user inclusion in the evaluation phase \cite{cai_advancing_2024,chen_closer_2023,fontana_de_vargas_co-designing_2024,seo_chacha_2024,tang_emoeden_2024,xygkou_mindtalker_2024,chen_chatgpt_2023,jang_its_2024,jeung_unlocking_2024,samsonovich_registrar_2024,li_exploring_2024,liu_peergpt_2024,bravo_creating_2023,said_design_2024,shakeri_saga_2021,suh_ai_2021,degen_retromind_2024,wan_building_2024,wang_aint_2024,wei_improving_2024}.
For example, \citet{tang_emoeden_2024} developed EmoEden, a tool that integrates LLMs and text-to-image models to support emotional learning for HFA children. They adopted a longitudinal approach and tested the system's effectiveness in engaging 6 HFA children and improving their emotional recognition and expression abilities for 22 days. The longitudinal approach allowed for observing changes and developments in the children's emotional learning over an extended period of time. 

\subsubsection{Evaluation metrics}
\begin{figure}[h!]
    \centering
    \includegraphics[width=.8\textwidth]{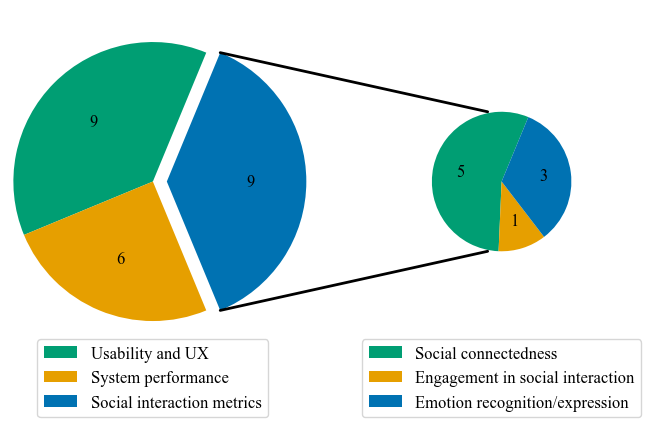}
    \caption{An overview of the focus of the evaluation phase of the 24 studies that reported one. The left pie chart shows the different focus areas: Usability and UX, System performance, and Social interaction metrics. The right pie charts gives more detail about the different social interaction metrics.}
    \label{fig:evaluation_metrics_pie}
\end{figure}

Delving deeper into the metrics used in the 24 studies that reported an evaluation phase, Figure \ref{fig:evaluation_metrics_pie} shows an overview of the evaluation focus. 
Nine studies focused on assessing the usability or user experience of their designs \cite{bhati_bookmate_2023,chen_chatgpt_2023,bravo_creating_2023,wang_aint_2024,samsonovich_registrar_2024,jang_its_2024,li_exploring_2024,cai_advancing_2024,fontana_de_vargas_co-designing_2024}. 
Typical evaluation angles were system usability and engagement, feedback and suggestions for improvement, and ease of use.
For 6 studies, the focus of the evaluation was system performance \cite{elgarf_creativebot_2022,alessa_towards_2023,said_design_2024,degen_retromind_2024,fang_socializechat_2023,wan_building_2024}. Here, the quality of generated content or a comparative evaluation between systems was addressed. 
This was mostly assessed through comparative user ratings, such as creativity ratings of storytelling agents to validate system behavior \cite{elgarf_creativebot_2022}, and task-based evaluations like measuring semantic similarity between user narratives and GAI-based content in reminiscence therapy \cite{degen_retromind_2024}.

The other 9 studies reported results of evaluations related to social interaction, both quantitative \cite{chen_closer_2023} and qualitative \cite{jeung_unlocking_2024,suh_ai_2021,shakeri_saga_2021,xygkou_mindtalker_2024,seo_chacha_2024,liu_peergpt_2024,wei_improving_2024,tang_emoeden_2024}. In general, a distinction could be made between social connectedness metrics, engagement in social interaction metrics, and emotion recognition and expression metrics (see Figure \ref{fig:evaluation_metrics_pie}).

\subsubsection{Social interaction metrics}
Interestingly, 8 out of 9 studies that focused on social interaction in their evaluation phase used qualitative data to assess this \cite{jeung_unlocking_2024,suh_ai_2021,shakeri_saga_2021,xygkou_mindtalker_2024,seo_chacha_2024,liu_peergpt_2024,wei_improving_2024,tang_emoeden_2024}. Most of them based their conclusions on post-interaction interviews \cite{tang_emoeden_2024, xygkou_mindtalker_2024,seo_chacha_2024}, observational data \cite{jeung_unlocking_2024,suh_ai_2021}, or session transcripts of data \cite{liu_peergpt_2024,seo_chacha_2024}.
Only one used quantitative measures \cite{chen_closer_2023}. 
Below, we will elaborate on the dimensions of social interaction that were evaluated and qualitative and quantitative methodologies what were applied to do so.

\textbf{Social connectedness metrics} — In total, 5 studies reported evaluation results related to the degree to which the participant experienced social connectedness in some way \cite{chen_closer_2023,jeung_unlocking_2024,suh_ai_2021,shakeri_saga_2021,xygkou_mindtalker_2024}. \citet{chen_closer_2023} quantitatively measured the extent to which users of their collaborative imaginative world-building game experienced intimacy, closeness with the study partner, and quality of conversation by administering the corresponding questionnaires before and after use. In the other 4 studies, a qualitative approach was used to assess the influence of the use of the application on experienced social connectedness \cite{jeung_unlocking_2024,suh_ai_2021,shakeri_saga_2021,xygkou_mindtalker_2024}. For example, 
\citet{suh_ai_2021} gathered participants' thoughts on the extent to which Cococo use for AI-enhanced collaborative music making fostered social connectedness through semi-structured interviews.
The main reason they reported for the experienced closeness was that some participants felt close to their partner in their shared humanness due to the unpredictable presence of the GAI.

\textbf{Measuring engagement in social interaction} — Three studies evaluated the extent to which participants engaged in social interaction \cite{liu_peergpt_2024,seo_chacha_2024,wei_improving_2024}. \citet{liu_peergpt_2024} and \citet{wei_improving_2024} specifically focused on using GAI-based technology to aid in effective group collaboration. The qualitative focus of these studies was on how much participants engaged in conversations with other participants and how much they contributed to the process. In both studies, the qualitative results showed contradictory findings: \citet{liu_peergpt_2024} found that agent-moderated discussions resulted in fewer opportunities for conversation among participants, while \citet{wei_improving_2024} found that the introduction of a virtual assistant improved communication and collaborative dynamics.
Finally, \citet{seo_chacha_2024} assessed to what extent participants shared emotional stories with a dedicated chatbot. The focus here was on how much personal information the participants shared with the robot. They found that participants shared personally relevant life events and concerns with the chatbot, and most saw the chatbot even as a close friend.

\textbf{Emotion recognition/expression metrics} — \citet{tang_emoeden_2024} developed a GAI-based training tool to train autistic children to recognize and express emotions effectively, skills that empower them to engage in social interaction in the long run. Through post-test interviews with the participants' parents, they evaluated the extent to which these children showed improvements in their emotion recognition and expression abilities. For the six participating children, they found improvements.

\section{Discussion}
This scoping review set out to investigate how GAI is currently being designed and evaluated in applications intended to support social interaction — as a proxy for social connectedness. Given the increasing interest in GAI technologies such as LLMs and multimodal generation systems, it is crucial to understand not only their technical affordances but also how they are being harnessed to promote social connectedness — a core human need increasingly threatened by isolation and marginalization. By systematically reviewing 30 studies published since 2020, this review provides a landscape-level view of emerging design methodologies, social interaction contexts, and evaluation metrics. The scoping review methodology, rooted in PRISMA-ScR guidelines \cite{tricco_prisma_2018} and supported by a structured database and conference search, was chosen to accommodate the rapid and diverse evolution of this field. In this section, we summarize the main findings, identify gaps in the literature, and provide recommendations for researchers and designers working with GAI for social connectedness.

\subsection{Summary of the main results}
GAI-based technologies for social interaction were mostly unimodal, making use of text-generating AI (most often a GPT-model), usually embedded in a chatbot or embodied conversational agent (RQ1). In the reviewed studies, GAI was often used to support social interaction in vulnerable user groups (e.g., children, autistic people, people with dementia) by facilitating storytelling, storymaking, socio-emotional skills training, reminiscence, collaborative learning, music making, and general conversation. When GAI was designed as an agent, it was embedded in one-to-one human-machine interactions, often taking the form of a chatbot or an embodied conversational agent, who proactively engaged with users. 
When GAI was designed as a medium, it functioned as a facilitative structure to support interaction between individuals — either when one person was unable to verbally express themselves or when participants were physically separated.
When GAI was designed as a tool, it usually fostered interactions between two or more users, and took the form of a mobile or standalone application aimed at enhancing users’ communication (RQ2). Only about one-third of the reviewed studies adopted a participatory design approach to GAI-based technology design. In this context, while studies rarely involved end-users in the conceptualization phase of the GAI-based designs (9/30), they were often included in their evaluation (20/30; RQ3). However, the preliminary nature of most of the technologies developed in the included studies affected the choice of evaluation metrics. The majority of studies focused on system performance, usability and user experience metrics (15/24), and a limited number of studies determined the extent to which the designed technologies effectively fostered social interactions (9/24).

\subsection{Potential beyond text generation}
The dominance of unimodal, text-based applications — especially those built on GPT models — illustrates a prevailing design orientation: conversational interfaces are currently the \textit{de facto} standard for social GAI applications. This aligns with broader trends in GAI deployment and user familiarity \cite{rebholz_conversational_2024}, but also raises questions about modality bias. That is, the reviewed literature demonstrates only limited engagement with other GAI modalities — such as image or music generation — despite their potential to support non-verbal expression, aesthetic co-creation, and multisensory engagement \cite{bland_enhancing_2025}, especially for users with communicative or cognitive differences. The ability of GAI to tailor output to individual preferences, memories, and abilities gives it unique potential in storytelling, reminiscence, and other social interaction contexts \cite{brankaert_assistive_2020,feuerriegel_generative_2024}. The example of \citet{chen_closer_2023}, who developed Closer Worlds, demonstrates how GAI can support a socially rewarding process.  Users are encouraged to share personal details, which are then transformed into co-created visual scenes that deepen connection and emotional resonance.

Multimodal GAI applications included in this review were limited in numbers, all integrating text- and image-based output. The incidence of text and image integration and the lack of multimodal use of GAI in general may partially be explained by the difficulty of seamlessly aligning multimodal outputs \cite{balkrishna_rasiklal_exploring_2024}. We see that the combination of multiple types of generated output potentially enhances the user experience by providing richer and more immersive interactions, depending on what user needs, activities, and contexts one aims to serve. Thus, opportunities for designing richer, embodied, and emotionally resonant interactions remain largely untapped. 

\subsection{The entanglement between the role of the technology and sociality}
The reviewed studies highlight a diversity in application domains with a shared ambition to facilitate moments of social connection. Particularly, researchers and designers tap into opportunities for GAI to foster social interaction in users who might have limited chances to engage socially, such as users belonging to vulnerable populations. The critical distinctive element that we identify in the proposed GAI-based technologies is the \textit{sociality} of the GAI-based technology, referring to the extent to which GAI has something to gain by presenting itself as a social entity in the interaction. 

Agents have either the purpose of fostering human-machine interaction exclusively or human-machine interaction alongside human-human interaction. According to the Social Presence Theory \cite{short_social_1976,gunawardena_social_1995}, technologies that present themselves as social entities that bring about a sense of intimacy and immediacy can enhance social presence, making interactions feel more personal and engaging. Additionally, the Media Equation \cite{reeves_media_1996}, particularly the Computers are Social Actors theory (CASA) \cite{nass_computers_1994}, supports the idea that agents designed to appear social (like chatbots or embodied conversational agents) are effective not just because of their functions, but because humans are predisposed to treat them as social partners. Social cues in agents are not merely enhancing presence, but are tapping into a deeper, automatic social response rooted in human psychology. These theories support the idea that chatbots and embodied conversational agents benefit from sociality by fostering a sense of presence and connection.

While technologies categorized as tools or media are not typically framed as social entities themselves, this does not mean they are socially neutral or passive. On the contrary, drawing on Actor-Network Theory (ANT) \cite{muniesa_actor-network_2015}, we recognize that these technologies can play an active role in shaping the dynamics of human interaction — even if they do not explicitly participate in the interaction as agents. ANT invites us to move beyond a dichotomy of active human users and passive technical instruments. Instead, it conceptualizes both humans and non-humans — including technologies — as part of networks of action, where agency is distributed across a web of interdependent actors. In this light, GAI technologies that function as tools or media — for example, a storytelling app like SAGA \cite{shakeri_saga_2021} that prompts conversation between two users, or a communication aid in a therapeutic setting like QuickPic \cite{fontana_de_vargas_co-designing_2024} — are not just neutral conduits. These technologies mediate, configure, and sometimes constrain the forms of engagement that are possible. We recognize that such technologies can subtly influence who speaks when, what topics are surfaced, how emotional cues are expressed, and what modes of interaction are seen as valid or desirable. Rather than being passive backdrops, these technologies co-produce the social realities they are embedded in. This perspective is essential when evaluating the social impact of GAI: even when it is not personified as an "agent," it can still be a powerful shaper of interactional norms, enabler of new communicative practices, and amplifier of certain voices or roles over others. Recognizing this influence of such technologies on the dynamics of human interaction allows us to better attend to the sociotechnical assemblages through which connectedness is fostered — and to the ethical responsibilities involved in designing them.

We would like to stress that while the role distinctions of tool, medium, and agent that we identified for the designs introduced in the studies are useful, the boundaries often blur in practice: a chatbot framed as an “agent” may be repurposed as a co-authoring “tool,” while a reminiscence platform may simultaneously serve as a channel and an interlocutor. This fluidity is one of GAI’s distinctive traits — but also introduces new complexities in design, interaction, and evaluation. In any case, the role that technology plays in social interaction should be contingent upon the specific interaction it is intended to facilitate, rather than being predefined by the technology itself. In other words, the design and use of technology should be guided by the desired outcomes of the interaction, ensuring that its role is contextually appropriate and responsive to the goals of the interaction.

\subsection{User involvement in design and post-design control}
Only a minority of studies involved representative end-users in the early design phases. Aligning with earlier findings \cite{suijkerbuijk_active_2019,mannheim_ageism_2023}, most included vulnerable users only in evaluation, if at all — often citing ethical or cognitive strain concerns. While these concerns are valid, they should not foreclose the possibility of inclusive, adapted design methods. Allowing end-users to have a say on what the technology does and looks like, thus including them in decision making, can result in technologies that more effectively cater to the social needs of the vulnerable populations they are aimed at \cite{sanders_co-creation_2008,moll_are_2020}. Encouragingly, a few cases (e.g., \cite{cai_advancing_2024, fontana_de_vargas_co-designing_2024, zhou_my_2024}) demonstrated that co-design is both feasible and enriching, even with children or older adults, when approached with care and creativity. These cases suggest that relational engagement can begin in the design process itself, not just in the final interaction.

A recurring insight is the uneven distribution of control and authorship within GAI-mediated social interaction. In many reviewed designs, particularly those in therapeutic or support contexts, GAI functions as a tool deployed by care professionals or educators, rather than one co-controlled by end-users themselves. In these cases, generative outputs (e.g., image boards, conversation prompts, or emotion cues) are configured by professionals, while the intended users — often autistic individuals, people with dementia, or children — respond within a predefined interaction structure. This model can yield practical benefits but also foregrounds an ethical tension: who gets to shape the interaction, and whose voice is amplified?

This limited involvement often carries through into post-design contexts, where end-user control over GAI systems remains constrained. A recurring insight across studies is the uneven distribution of authorship in GAI-mediated social interaction. Particularly in therapeutic or educational settings, GAI is frequently implemented as a tool used by professionals, rather than as a co-controlled medium for end-users themselves. In these cases, generative outputs (e.g., image boards, conversation prompts, or emotion cues) are configured by professionals, while the intended users — often autistic individuals, people with dementia, or children — respond within a predefined interaction structure. This approach may yield practical benefits, especially in structured settings, but also raises ethical questions: who gets to shape the interaction, and whose voice is amplified? 

\subsection{From system metrics to social meaning}
Another observation concerns the evaluation focus of the reviewed studies. While many applications aimed to foster social connectedness, relatively few measured this outcome directly. 
While the majority of studies prioritized usability, performance, or feasibility — all important foundations for responsible and accessible design — these evaluations offer only a partial view when it comes to the affective and relational dimensions of social interaction. Even when social interaction was assessed, it was often limited to short-term qualitative impressions or surface-level metrics (e.g., number of conversational turns, thematic engagement). This was likely the case because of the preliminary nature of most of the designs. These initial efforts provide a valuable starting point; future research could build on this foundation by incorporating deeper assessments of connectedness, reciprocity, or emotional resonance.
Notably, some studies adopted creative qualitative approaches, such as observing storytelling dynamics, analyzing collaborative dialogues, or conducting semi-structured interviews to explore subjective experiences of closeness and engagement. These methods offer a promising foundation, but more sustained, theory-informed, and longitudinal evaluation strategies are needed. This is particularly important for capturing GAI’s evolving role over time — as a social facilitator, co-creative partner, or relational proxy — especially as users’ trust, familiarity, and dependency may shift with prolonged use.

\subsection{Ethical reflection}
The widespread use of GPT-based models in GAI applications brings notable benefits but also raises ethical challenges — particularly regarding bias, representation, and inclusivity. Over-reliance on models developed by a few dominant players, such as OpenAI, risks amplifying unrepresentative training data and cultural blind spots, potentially leading to stereotyping, marginalization, and exclusion \cite{hagendorff_mapping_2024, fui-hoon_nah_generative_2023, zhuo_red_2023, weidinger_ethical_2021, dwivedi_opinion_2023}. While many reviewed papers briefly acknowledged these risks, few provided concrete strategies to mitigate them, reflecting broader uncertainty about how to engage with the “black box” nature of GAI in socially responsible ways. 

To move forward, researchers and designers need more than abstract ethical principles — they need practical, hands-on guidance. Existing frameworks on fairness and transparency are important, but often remain too high-level \cite{jacobs_bridging_2021}. What is needed are actionable design prompts, prototyping heuristics, and content guidelines that address issues such as cultural sensitivity, output curation, and bias mitigation at the micro-level of GAI interaction. These tools should empower — not burden — developers to integrate ethical reflection into their workflows. We expect that a lack of practical, hands-on guidance plays a role in the lack of involvement of vulnerable user groups in the studies included in this review. 

This is especially critical when designing for vulnerable populations \cite{ dennis_confronting_2022}, who were the primary focus in many of the reviewed studies. Exclusion from design processes — often justified by concerns over cognitive strain — risks reinforcing ableist assumptions and undermines the relational goals these technologies aim to support. Yet several papers, such as \citet{ zhou_my_2024}, illustrate that meaningful co-design is not only possible but socially and creatively enriching, especially when adapted to participants' capabilities. 

We advocate for an approach that balances care and inclusion, offering support structures and flexible design methods that enable representative end-users to contribute meaningfully from the earliest stages. Co-design should be seen not only as a path to better outcomes, but as a relational act that supports connection, dignity, and shared authorship. By fostering such inclusive practices — and by diversifying both design teams and training datasets — we can ensure that GAI technologies for social connectedness are themselves designed in socially connected ways.

\section{Conclusion}
This scoping review provides insights into the role GAI-based technology can play in fostering social interaction for human users. Emerging GAI-based technology has the potential to enhance social interaction by generating personalized content and activities that continuously adapt to the ever-changing user needs, behaviors, and capabilities. 
Based on this review, we found that most GAI-based applications to foster social interaction were meant for end-users that could be considered vulnerable in some way, showing an interesting tendency in the field. 
Currently, text-based GAI approaches are the most common, dominated by GPT. Multimodal GAI approaches are showing a careful increase, and have the potential to provide richer and more immersive interactions. 
Different types of social interactions (i.e., human-machine or human-human) can be brought about by GAI-based technology intervention, and the role the technology should play in the interaction (i.e., tool, medium, or agent) should depend on the kind of interaction and the type of social activity researchers and designers want to bring about.
However, most approaches are still in early stages of development, with limited research conducted in real-world settings.
Furthermore, effective inclusion of representative end-users, vulnerable populations in particular, at every stage of the design process remains a challenge.
By addressing current gaps and encouraging collaboration between researchers, designers, representative end-users, and other stakeholders, the future of GAI-based technology holds promise for effectively fostering social interactions for its users, thereby proactively supporting social connectedness.

\section*{CRediT authorship contribution statement}
\textbf{Teis Arets:} Conceptualization, methodology, validation, formal analysis, writing — original draft, writing — review and editing, visualization, project administration.
\textbf{Giulia Perugia:} Conceptualization, methodology, validation, formal analysis, writing — review and editing.
\textbf{Maarten Houben:} Conceptualization, methodology, validation, formal analysis, writing — review and editing.
\textbf{Wijnand IJsselsteijn:} Conceptualization, methodology, validation, writing — review and editing.


\begin{acks}
This research was funded by the Dutch Research Council (NWO) through the QoLEAD (Quality of Life by use of Enabling AI in Dementia) Project (project number: KICH1.GZ02.20.008). Additional support from Alzheimer Nederland is gratefully acknowledged.

This research was also funded by the research programme Ethics of Socially Disruptive Technologies (ESDiT), under the Gravitation programme of the Dutch Ministry of Education, Culture, and Science and the Netherlands Organization for Scientific Research (NWO grant number 024.004.031).

ChatGPT (OpenAI) was used to support language editing (grammar, spelling, and writing style) of author-written text. All conceptual and analytical contributions were made by the authors.
\end{acks}

\bibliographystyle{ACM-Reference-Format}
\bibliography{bibliography}


\begin{thebibliography}{93}


\ifx \showCODEN    \undefined \def \showCODEN     #1{\unskip}     \fi
\ifx \showISBNx    \undefined \def \showISBNx     #1{\unskip}     \fi
\ifx \showISBNxiii \undefined \def \showISBNxiii  #1{\unskip}     \fi
\ifx \showISSN     \undefined \def \showISSN      #1{\unskip}     \fi
\ifx \showLCCN     \undefined \def \showLCCN      #1{\unskip}     \fi
\ifx \shownote     \undefined \def \shownote      #1{#1}          \fi
\ifx \showarticletitle \undefined \def \showarticletitle #1{#1}   \fi
\ifx \showURL      \undefined \def \showURL       {\relax}        \fi
\providecommand\bibfield[2]{#2}
\providecommand\bibinfo[2]{#2}
\providecommand\natexlab[1]{#1}
\providecommand\showeprint[2][]{arXiv:#2}

\bibitem[Al-kfairy et~al\mbox{.}(2024)]%
        {al-kfairy_ethical_2024}
\bibfield{author}{\bibinfo{person}{Mousa Al-kfairy}, \bibinfo{person}{Dheya Mustafa}, \bibinfo{person}{Nir Kshetri}, \bibinfo{person}{Mazen Insiew}, {and} \bibinfo{person}{Omar Alfandi}.} \bibinfo{year}{2024}\natexlab{}.
\newblock \showarticletitle{Ethical {Challenges} and {Solutions} of {Generative} {AI}: {An} {Interdisciplinary} {Perspective}}.
\newblock \bibinfo{journal}{\emph{Informatics}} \bibinfo{volume}{11}, \bibinfo{number}{3} (\bibinfo{date}{Aug.} \bibinfo{year}{2024}), \bibinfo{pages}{58}.
\newblock
\showISSN{2227-9709}
\href{https://doi.org/10.3390/informatics11030058}{doi:\nolinkurl{10.3390/informatics11030058}}


\bibitem[Alessa and Al-Khalifa(2023)]%
        {alessa_towards_2023}
\bibfield{author}{\bibinfo{person}{Abeer Alessa} {and} \bibinfo{person}{Hend Al-Khalifa}.} \bibinfo{year}{2023}\natexlab{}.
\newblock \showarticletitle{Towards {Designing} a {ChatGPT} {Conversational} {Companion} for {Elderly} {People}}. In \bibinfo{booktitle}{\emph{Proceedings of the 16th {International} {Conference} on {PErvasive} {Technologies} {Related} to {Assistive} {Environments}}}. \bibinfo{publisher}{ACM}, \bibinfo{address}{Corfu Greece}, \bibinfo{pages}{667--674}.
\newblock
\showISBNx{9798400700699}
\href{https://doi.org/10.1145/3594806.3596572}{doi:\nolinkurl{10.1145/3594806.3596572}}


\bibitem[Antonini(2021)]%
        {antonini_overview_2021}
\bibfield{author}{\bibinfo{person}{Matteo Antonini}.} \bibinfo{year}{2021}\natexlab{}.
\newblock \showarticletitle{An {Overview} of {Co}-{Design}: {Advantages}, {Challenges} and {Perspectives} of {Users}’ {Involvement} in the {Design} {Process}}.
\newblock \bibinfo{journal}{\emph{Journal of Design Thinking}} \bibinfo{volume}{2}, \bibinfo{number}{1} (\bibinfo{date}{June} \bibinfo{year}{2021}).
\newblock
\href{https://doi.org/10.22059/jdt.2020.272513.1018}{doi:\nolinkurl{10.22059/jdt.2020.272513.1018}}


\bibitem[Balkrishna~Rasiklal(2024)]%
        {balkrishna_rasiklal_exploring_2024}
\bibfield{author}{\bibinfo{person}{Yadav Balkrishna~Rasiklal}.} \bibinfo{year}{2024}\natexlab{}.
\newblock \bibinfo{title}{Exploring {Multimodal} {Generative} {AI}: {A} {Comprehensive} {Review} of {Image}, {Text}, and {Audio} {Integration}}.
\newblock
\href{https://doi.org/10.5281/ZENODO.14288596}{doi:\nolinkurl{10.5281/ZENODO.14288596}}
\newblock
\shownote{ISSN: 2995-486X}.


\bibitem[Barbosa~Neves et~al\mbox{.}(2019)]%
        {barbosa_neves_can_2019}
\bibfield{author}{\bibinfo{person}{Barbara Barbosa~Neves}, \bibinfo{person}{Rachel Franz}, \bibinfo{person}{Rebecca Judges}, \bibinfo{person}{Christian Beermann}, {and} \bibinfo{person}{Ron Baecker}.} \bibinfo{year}{2019}\natexlab{}.
\newblock \showarticletitle{Can {Digital} {Technology} {Enhance} {Social} {Connectedness} {Among} {Older} {Adults}? {A} {Feasibility} {Study}}.
\newblock \bibinfo{journal}{\emph{Journal of Applied Gerontology}} \bibinfo{volume}{38}, \bibinfo{number}{1} (\bibinfo{date}{Jan.} \bibinfo{year}{2019}), \bibinfo{pages}{49--72}.
\newblock
\showISSN{0733-4648, 1552-4523}
\href{https://doi.org/10.1177/0733464817741369}{doi:\nolinkurl{10.1177/0733464817741369}}


\bibitem[Bhati et~al\mbox{.}(2023)]%
        {bhati_bookmate_2023}
\bibfield{author}{\bibinfo{person}{Deepshikha Bhati}, \bibinfo{person}{Angela Guercio}, \bibinfo{person}{Veronica Rossano}, {and} \bibinfo{person}{Rita Francese}.} \bibinfo{year}{2023}\natexlab{}.
\newblock \showarticletitle{{BookMate}: {Leveraging} {Deep} {Learning} to {Empower} {Caregivers} of {People} with {ASD} in {Generation} of {Social} {Stories}}. In \bibinfo{booktitle}{\emph{2023 27th {International} {Conference} {Information} {Visualisation} ({IV})}}. \bibinfo{publisher}{IEEE}, \bibinfo{address}{Tampere, Finland}, \bibinfo{pages}{403--408}.
\newblock
\showISBNx{9798350341614}
\href{https://doi.org/10.1109/IV60283.2023.00074}{doi:\nolinkurl{10.1109/IV60283.2023.00074}}


\bibitem[Billing et~al\mbox{.}(2023)]%
        {billing_language_2023}
\bibfield{author}{\bibinfo{person}{Erik Billing}, \bibinfo{person}{Julia Rosén}, {and} \bibinfo{person}{Maurice Lamb}.} \bibinfo{year}{2023}\natexlab{}.
\newblock \showarticletitle{Language {Models} for {Human}-{Robot} {Interaction}}. In \bibinfo{booktitle}{\emph{Companion of the 2023 {ACM}/{IEEE} {International} {Conference} on {Human}-{Robot} {Interaction}}}. \bibinfo{publisher}{ACM}, \bibinfo{address}{Stockholm Sweden}, \bibinfo{pages}{905--906}.
\newblock
\showISBNx{978-1-4503-9970-8}
\href{https://doi.org/10.1145/3568294.3580040}{doi:\nolinkurl{10.1145/3568294.3580040}}


\bibitem[Bland(2025)]%
        {bland_enhancing_2025}
\bibfield{author}{\bibinfo{person}{Tyler Bland}.} \bibinfo{year}{2025}\natexlab{}.
\newblock \showarticletitle{Enhancing {Medical} {Student} {Engagement} {Through} {Cinematic} {Clinical} {Narratives}: {Multimodal} {Generative} {AI}–{Based} {Mixed} {Methods} {Study}}.
\newblock \bibinfo{journal}{\emph{JMIR Medical Education}}  \bibinfo{volume}{11} (\bibinfo{date}{Jan.} \bibinfo{year}{2025}), \bibinfo{pages}{e63865--e63865}.
\newblock
\showISSN{2369-3762}
\href{https://doi.org/10.2196/63865}{doi:\nolinkurl{10.2196/63865}}


\bibitem[Cai et~al\mbox{.}(2024)]%
        {cai_advancing_2024}
\bibfield{author}{\bibinfo{person}{Zhenyao Cai}, \bibinfo{person}{Seehee Park}, \bibinfo{person}{Nia Nixon}, {and} \bibinfo{person}{Shayan Doroudi}.} \bibinfo{year}{2024}\natexlab{}.
\newblock \showarticletitle{Advancing {Knowledge} {Together}: {Integrating} {Large} {Language} {Model}-based {Conversational} {AI} in {Small} {Group} {Collaborative} {Learning}}. In \bibinfo{booktitle}{\emph{Extended {Abstracts} of the {CHI} {Conference} on {Human} {Factors} in {Computing} {Systems}}}. \bibinfo{publisher}{ACM}, \bibinfo{address}{Honolulu HI USA}, \bibinfo{pages}{1--9}.
\newblock
\showISBNx{9798400703317}
\href{https://doi.org/10.1145/3613905.3650868}{doi:\nolinkurl{10.1145/3613905.3650868}}


\bibitem[Chen et~al\mbox{.}(2023a)]%
        {chen_closer_2023}
\bibfield{author}{\bibinfo{person}{Tiffany Chen}, \bibinfo{person}{Cassandra Lee}, \bibinfo{person}{Jessica~R Mindel}, \bibinfo{person}{Neska Elhaouij}, {and} \bibinfo{person}{Rosalind Picard}.} \bibinfo{year}{2023}\natexlab{a}.
\newblock \showarticletitle{Closer {Worlds}: {Using} {Generative} {AI} to {Facilitate} {Intimate} {Conversations}}. In \bibinfo{booktitle}{\emph{Extended {Abstracts} of the 2023 {CHI} {Conference} on {Human} {Factors} in {Computing} {Systems}}}. \bibinfo{publisher}{ACM}, \bibinfo{address}{Hamburg Germany}, \bibinfo{pages}{1--15}.
\newblock
\showISBNx{978-1-4503-9422-2}
\href{https://doi.org/10.1145/3544549.3585651}{doi:\nolinkurl{10.1145/3544549.3585651}}


\bibitem[Chen et~al\mbox{.}(2023b)]%
        {chen_chatgpt_2023}
\bibfield{author}{\bibinfo{person}{Xieling Chen}, \bibinfo{person}{Haoran Xie}, \bibinfo{person}{Di Zou}, {and} \bibinfo{person}{Fu~Lee Wang}.} \bibinfo{year}{2023}\natexlab{b}.
\newblock \showarticletitle{{ChatGPT} for generating stories and mind-maps in storytelling}. In \bibinfo{booktitle}{\emph{2023 10th {International} {Conference} on {Behavioural} and {Social} {Computing} ({BESC})}}. \bibinfo{publisher}{IEEE}, \bibinfo{address}{Larnaca, Cyprus}, \bibinfo{pages}{1--8}.
\newblock
\showISBNx{9798350395884}
\href{https://doi.org/10.1109/BESC59560.2023.10386441}{doi:\nolinkurl{10.1109/BESC59560.2023.10386441}}


\bibitem[Cornwell and Waite(2009)]%
        {cornwell_social_2009}
\bibfield{author}{\bibinfo{person}{Erin~York Cornwell} {and} \bibinfo{person}{Linda~J. Waite}.} \bibinfo{year}{2009}\natexlab{}.
\newblock \showarticletitle{Social {Disconnectedness}, {Perceived} {Isolation}, and {Health} among {Older} {Adults}}.
\newblock \bibinfo{journal}{\emph{Journal of Health and Social Behavior}} \bibinfo{volume}{50}, \bibinfo{number}{1} (\bibinfo{date}{March} \bibinfo{year}{2009}), \bibinfo{pages}{31--48}.
\newblock
\showISSN{0022-1465, 2150-6000}
\href{https://doi.org/10.1177/002214650905000103}{doi:\nolinkurl{10.1177/002214650905000103}}


\bibitem[{D-ID}(2025)]%
        {d-id_d-id_2025}
\bibfield{author}{\bibinfo{person}{{D-ID}}.} \bibinfo{year}{2025}\natexlab{}.
\newblock \bibinfo{title}{D-{ID} [{AI} avatar generator]}.
\newblock
\urldef\tempurl%
\url{https://www.d-id.com}
\showURL{%
\tempurl}


\bibitem[Dwivedi et~al\mbox{.}(2023)]%
        {dwivedi_opinion_2023}
\bibfield{author}{\bibinfo{person}{Yogesh~K. Dwivedi}, \bibinfo{person}{Nir Kshetri}, \bibinfo{person}{Laurie Hughes}, \bibinfo{person}{Emma~Louise Slade}, \bibinfo{person}{Anand Jeyaraj}, \bibinfo{person}{Arpan~Kumar Kar}, \bibinfo{person}{Abdullah~M. Baabdullah}, \bibinfo{person}{Alex Koohang}, \bibinfo{person}{Vishnupriya Raghavan}, \bibinfo{person}{Manju Ahuja}, \bibinfo{person}{Hanaa Albanna}, \bibinfo{person}{Mousa~Ahmad Albashrawi}, \bibinfo{person}{Adil~S. Al-Busaidi}, \bibinfo{person}{Janarthanan Balakrishnan}, \bibinfo{person}{Yves Barlette}, \bibinfo{person}{Sriparna Basu}, \bibinfo{person}{Indranil Bose}, \bibinfo{person}{Laurence Brooks}, \bibinfo{person}{Dimitrios Buhalis}, \bibinfo{person}{Lemuria Carter}, \bibinfo{person}{Soumyadeb Chowdhury}, \bibinfo{person}{Tom Crick}, \bibinfo{person}{Scott~W. Cunningham}, \bibinfo{person}{Gareth~H. Davies}, \bibinfo{person}{Robert~M. Davison}, \bibinfo{person}{Rahul Dé}, \bibinfo{person}{Denis Dennehy}, \bibinfo{person}{Yanqing Duan},
  \bibinfo{person}{Rameshwar Dubey}, \bibinfo{person}{Rohita Dwivedi}, \bibinfo{person}{John~S. Edwards}, \bibinfo{person}{Carlos Flavián}, \bibinfo{person}{Robin Gauld}, \bibinfo{person}{Varun Grover}, \bibinfo{person}{Mei-Chih Hu}, \bibinfo{person}{Marijn Janssen}, \bibinfo{person}{Paul Jones}, \bibinfo{person}{Iris Junglas}, \bibinfo{person}{Sangeeta Khorana}, \bibinfo{person}{Sascha Kraus}, \bibinfo{person}{Kai~R. Larsen}, \bibinfo{person}{Paul Latreille}, \bibinfo{person}{Sven Laumer}, \bibinfo{person}{F.~Tegwen Malik}, \bibinfo{person}{Abbas Mardani}, \bibinfo{person}{Marcello Mariani}, \bibinfo{person}{Sunil Mithas}, \bibinfo{person}{Emmanuel Mogaji}, \bibinfo{person}{Jeretta~Horn Nord}, \bibinfo{person}{Siobhan O’Connor}, \bibinfo{person}{Fevzi Okumus}, \bibinfo{person}{Margherita Pagani}, \bibinfo{person}{Neeraj Pandey}, \bibinfo{person}{Savvas Papagiannidis}, \bibinfo{person}{Ilias~O. Pappas}, \bibinfo{person}{Nishith Pathak}, \bibinfo{person}{Jan Pries-Heje}, \bibinfo{person}{Ramakrishnan
  Raman}, \bibinfo{person}{Nripendra~P. Rana}, \bibinfo{person}{Sven-Volker Rehm}, \bibinfo{person}{Samuel Ribeiro-Navarrete}, \bibinfo{person}{Alexander Richter}, \bibinfo{person}{Frantz Rowe}, \bibinfo{person}{Suprateek Sarker}, \bibinfo{person}{Bernd~Carsten Stahl}, \bibinfo{person}{Manoj~Kumar Tiwari}, \bibinfo{person}{Wil Van Der~Aalst}, \bibinfo{person}{Viswanath Venkatesh}, \bibinfo{person}{Giampaolo Viglia}, \bibinfo{person}{Michael Wade}, \bibinfo{person}{Paul Walton}, \bibinfo{person}{Jochen Wirtz}, {and} \bibinfo{person}{Ryan Wright}.} \bibinfo{year}{2023}\natexlab{}.
\newblock \showarticletitle{Opinion {Paper}: “{So} what if {ChatGPT} wrote it?” {Multidisciplinary} perspectives on opportunities, challenges and implications of generative conversational {AI} for research, practice and policy}.
\newblock \bibinfo{journal}{\emph{International Journal of Information Management}}  \bibinfo{volume}{71} (\bibinfo{date}{Aug.} \bibinfo{year}{2023}), \bibinfo{pages}{102642}.
\newblock
\showISSN{02684012}
\href{https://doi.org/10.1016/j.ijinfomgt.2023.102642}{doi:\nolinkurl{10.1016/j.ijinfomgt.2023.102642}}


\bibitem[Eke(2023)]%
        {eke_chatgpt_2023}
\bibfield{author}{\bibinfo{person}{Damian~Okaibedi Eke}.} \bibinfo{year}{2023}\natexlab{}.
\newblock \showarticletitle{{ChatGPT} and the rise of generative {AI}: {Threat} to academic integrity?}
\newblock \bibinfo{journal}{\emph{Journal of Responsible Technology}}  \bibinfo{volume}{13} (\bibinfo{date}{April} \bibinfo{year}{2023}), \bibinfo{pages}{100060}.
\newblock
\showISSN{26666596}
\href{https://doi.org/10.1016/j.jrt.2023.100060}{doi:\nolinkurl{10.1016/j.jrt.2023.100060}}


\bibitem[Elgarf and Peters(2022)]%
        {elgarf_creativebot_2022}
\bibfield{author}{\bibinfo{person}{Maha Elgarf} {and} \bibinfo{person}{Christopher Peters}.} \bibinfo{year}{2022}\natexlab{}.
\newblock \showarticletitle{{CreativeBot}: a {Creative} {Storyteller} {Agent} {Developed} by {Leveraging} {Pre}-trained {Language} {Models}}. In \bibinfo{booktitle}{\emph{2022 {IEEE}/{RSJ} {International} {Conference} on {Intelligent} {Robots} and {Systems} ({IROS})}}. \bibinfo{publisher}{IEEE}, \bibinfo{address}{Kyoto, Japan}, \bibinfo{pages}{13438--13444}.
\newblock
\showISBNx{978-1-66547-927-1}
\href{https://doi.org/10.1109/IROS47612.2022.9981033}{doi:\nolinkurl{10.1109/IROS47612.2022.9981033}}


\bibitem[Fang et~al\mbox{.}(2023)]%
        {fang_socializechat_2023}
\bibfield{author}{\bibinfo{person}{Yuyang Fang}, \bibinfo{person}{Yunkai Xu}, \bibinfo{person}{Zhuyu Teng}, \bibinfo{person}{Zhaoqu Jiang}, {and} \bibinfo{person}{Wei Xiang}.} \bibinfo{year}{2023}\natexlab{}.
\newblock \showarticletitle{{SocializeChat}: a {GPT}-based {AAC} {Tool} for {Social} {Communication} {Through} {Eye} {Gazing}}. In \bibinfo{booktitle}{\emph{Adjunct {Proceedings} of the 2023 {ACM} {International} {Joint} {Conference} on {Pervasive} and {Ubiquitous} {Computing} \& the 2023 {ACM} {International} {Symposium} on {Wearable} {Computing}}}. \bibinfo{publisher}{ACM}, \bibinfo{address}{Cancun, Quintana Roo Mexico}, \bibinfo{pages}{128--132}.
\newblock
\showISBNx{9798400702006}
\href{https://doi.org/10.1145/3594739.3610705}{doi:\nolinkurl{10.1145/3594739.3610705}}


\bibitem[Ferrara(2023)]%
        {ferrara_fairness_2023}
\bibfield{author}{\bibinfo{person}{Emilio Ferrara}.} \bibinfo{year}{2023}\natexlab{}.
\newblock \showarticletitle{Fairness and {Bias} in {Artificial} {Intelligence}: {A} {Brief} {Survey} of {Sources}, {Impacts}, and {Mitigation} {Strategies}}.
\newblock \bibinfo{journal}{\emph{Sci}} \bibinfo{volume}{6}, \bibinfo{number}{1} (\bibinfo{date}{Dec.} \bibinfo{year}{2023}), \bibinfo{pages}{3}.
\newblock
\showISSN{2413-4155}
\href{https://doi.org/10.3390/sci6010003}{doi:\nolinkurl{10.3390/sci6010003}}


\bibitem[Feuerriegel et~al\mbox{.}(2024)]%
        {feuerriegel_generative_2024}
\bibfield{author}{\bibinfo{person}{Stefan Feuerriegel}, \bibinfo{person}{Jochen Hartmann}, \bibinfo{person}{Christian Janiesch}, {and} \bibinfo{person}{Patrick Zschech}.} \bibinfo{year}{2024}\natexlab{}.
\newblock \showarticletitle{Generative {AI}}.
\newblock \bibinfo{journal}{\emph{Business \& Information Systems Engineering}} \bibinfo{volume}{66}, \bibinfo{number}{1} (\bibinfo{date}{Feb.} \bibinfo{year}{2024}), \bibinfo{pages}{111--126}.
\newblock
\showISSN{2363-7005, 1867-0202}
\href{https://doi.org/10.1007/s12599-023-00834-7}{doi:\nolinkurl{10.1007/s12599-023-00834-7}}


\bibitem[Flynn et~al\mbox{.}(2024)]%
        {flynn_social_2024}
\bibfield{author}{\bibinfo{person}{Aisling Flynn}, \bibinfo{person}{Attracta Brennan}, \bibinfo{person}{Marguerite Barry}, \bibinfo{person}{Sam Redfern}, {and} \bibinfo{person}{Dympna Casey}.} \bibinfo{year}{2024}\natexlab{}.
\newblock \showarticletitle{Social connectedness and the role of virtual reality: experiences and perceptions of people living with dementia and their caregivers}.
\newblock \bibinfo{journal}{\emph{Disability and Rehabilitation: Assistive Technology}} \bibinfo{volume}{19}, \bibinfo{number}{7} (\bibinfo{date}{Oct.} \bibinfo{year}{2024}), \bibinfo{pages}{2615--2629}.
\newblock
\showISSN{1748-3107, 1748-3115}
\href{https://doi.org/10.1080/17483107.2024.2310262}{doi:\nolinkurl{10.1080/17483107.2024.2310262}}


\bibitem[Fogg(2003)]%
        {fogg_persuasive_2003}
\bibfield{author}{\bibinfo{person}{B.~J. Fogg}.} \bibinfo{year}{2003}\natexlab{}.
\newblock \bibinfo{booktitle}{\emph{Persuasive technology: {Using} computers to change what we think and do}}.
\newblock \bibinfo{publisher}{Morgan Kaufmann}, \bibinfo{address}{San Francisco}.
\newblock
\showISBNx{978-1-55860-643-2}


\bibitem[Fontana De~Vargas et~al\mbox{.}(2024)]%
        {fontana_de_vargas_co-designing_2024}
\bibfield{author}{\bibinfo{person}{Mauricio Fontana De~Vargas}, \bibinfo{person}{Christina Yu}, \bibinfo{person}{Howard~C. Shane}, {and} \bibinfo{person}{Karyn Moffatt}.} \bibinfo{year}{2024}\natexlab{}.
\newblock \showarticletitle{Co-{Designing} {QuickPic}: {Automated} {Topic}-{Specific} {Communication} {Boards} from {Photographs} for {AAC}-{Based} {Language} {Instruction}}. In \bibinfo{booktitle}{\emph{Proceedings of the {CHI} {Conference} on {Human} {Factors} in {Computing} {Systems}}}. \bibinfo{publisher}{ACM}, \bibinfo{address}{Honolulu HI USA}, \bibinfo{pages}{1--16}.
\newblock
\showISBNx{9798400703300}
\href{https://doi.org/10.1145/3613904.3642080}{doi:\nolinkurl{10.1145/3613904.3642080}}


\bibitem[Frohlich et~al\mbox{.}(2020)]%
        {brankaert_assistive_2020}
\bibfield{author}{\bibinfo{person}{David~M. Frohlich}, \bibinfo{person}{Emily Corrigan-Kavanagh}, \bibinfo{person}{Sarah Campbell}, \bibinfo{person}{Theopisti Chrysanthaki}, \bibinfo{person}{Paula Castro}, \bibinfo{person}{Isabela Zaine}, {and} \bibinfo{person}{Maria Da~Graça Campos~Pimentel}.} \bibinfo{year}{2020}\natexlab{}.
\newblock \showarticletitle{Assistive {Media} for {Well}-being}.
\newblock In \bibinfo{booktitle}{\emph{{HCI} and {Design} in the {Context} of {Dementia}}}, \bibfield{editor}{\bibinfo{person}{Rens Brankaert} {and} \bibinfo{person}{Gail Kenning}} (Eds.). \bibinfo{publisher}{Springer International Publishing}, \bibinfo{address}{Cham}, \bibinfo{pages}{189--205}.
\newblock
\showISBNx{978-3-030-32834-4 978-3-030-32835-1}
\href{https://doi.org/10.1007/978-3-030-32835-1_12}{doi:\nolinkurl{10.1007/978-3-030-32835-1_12}}
\newblock
\shownote{Series Title: Human–Computer Interaction Series}.


\bibitem[Fui-Hoon~Nah et~al\mbox{.}(2023)]%
        {fui-hoon_nah_generative_2023}
\bibfield{author}{\bibinfo{person}{Fiona Fui-Hoon~Nah}, \bibinfo{person}{Ruilin Zheng}, \bibinfo{person}{Jingyuan Cai}, \bibinfo{person}{Keng Siau}, {and} \bibinfo{person}{Langtao Chen}.} \bibinfo{year}{2023}\natexlab{}.
\newblock \showarticletitle{Generative {AI} and {ChatGPT}: {Applications}, challenges, and {AI}-human collaboration}.
\newblock \bibinfo{journal}{\emph{Journal of Information Technology Case and Application Research}} \bibinfo{volume}{25}, \bibinfo{number}{3} (\bibinfo{date}{July} \bibinfo{year}{2023}), \bibinfo{pages}{277--304}.
\newblock
\showISSN{1522-8053, 2333-6897}
\href{https://doi.org/10.1080/15228053.2023.2233814}{doi:\nolinkurl{10.1080/15228053.2023.2233814}}


\bibitem[Gunawardena(1995)]%
        {gunawardena_social_1995}
\bibfield{author}{\bibinfo{person}{Charlotte~N Gunawardena}.} \bibinfo{year}{1995}\natexlab{}.
\newblock \showarticletitle{Social {Presence} {Theory} and {Implications} for {Interaction} and {Collaborative} {Learning} in {Computer} {Conferences}}.
\newblock \bibinfo{journal}{\emph{International Journal of Educational Telecommunications}} \bibinfo{volume}{1}, \bibinfo{number}{2} (\bibinfo{year}{1995}), \bibinfo{pages}{147--166}.
\newblock


\bibitem[Hagendorff(2024)]%
        {hagendorff_mapping_2024}
\bibfield{author}{\bibinfo{person}{Thilo Hagendorff}.} \bibinfo{year}{2024}\natexlab{}.
\newblock \showarticletitle{Mapping the {Ethics} of {Generative} {AI}: {A} {Comprehensive} {Scoping} {Review}}.
\newblock \bibinfo{journal}{\emph{Minds and Machines}} \bibinfo{volume}{34}, \bibinfo{number}{4} (\bibinfo{date}{Sept.} \bibinfo{year}{2024}), \bibinfo{pages}{39}.
\newblock
\showISSN{1572-8641}
\href{https://doi.org/10.1007/s11023-024-09694-w}{doi:\nolinkurl{10.1007/s11023-024-09694-w}}


\bibitem[Haslam et~al\mbox{.}(2015)]%
        {pachana_social_2015}
\bibfield{author}{\bibinfo{person}{Catherine Haslam}, \bibinfo{person}{Tegan Cruwys}, \bibinfo{person}{S.~Alexander Haslam}, {and} \bibinfo{person}{Jolanda Jetten}.} \bibinfo{year}{2015}\natexlab{}.
\newblock \showarticletitle{Social {Connectedness} and {Health}}.
\newblock In \bibinfo{booktitle}{\emph{Encyclopedia of {Geropsychology}}}, \bibfield{editor}{\bibinfo{person}{Nancy~A. Pachana}} (Ed.). \bibinfo{publisher}{Springer Singapore}, \bibinfo{address}{Singapore}, \bibinfo{pages}{1--10}.
\newblock
\showISBNx{978-981-287-080-3}
\href{https://doi.org/10.1007/978-981-287-080-3_46-1}{doi:\nolinkurl{10.1007/978-981-287-080-3_46-1}}


\bibitem[Holtzblatt and Jones(1998)]%
        {schuler_principles_1998}
\bibfield{author}{\bibinfo{person}{Karen Holtzblatt} {and} \bibinfo{person}{Sandra Jones}.} \bibinfo{year}{1998}\natexlab{}.
\newblock \showarticletitle{Principles of {Contextual} {Inquiry}}.
\newblock In \bibinfo{booktitle}{\emph{Contextual {Design}: {Defining} {Customer}-{Centered} {Systems}} (\bibinfo{edition}{1} ed.)}, \bibfield{editor}{\bibinfo{person}{Douglas Schuler} {and} \bibinfo{person}{Aki Namioka}} (Eds.). \bibinfo{publisher}{Morgan Kaufmann}, \bibinfo{pages}{177--210}.
\newblock
\showISBNx{978-0-203-74433-8}
\href{https://doi.org/10.1201/9780203744338-9}{doi:\nolinkurl{10.1201/9780203744338-9}}


\bibitem[Hu and Li(2022)]%
        {hu_social_2022}
\bibfield{author}{\bibinfo{person}{Rita~Xiaochen Hu} {and} \bibinfo{person}{Lydia~W Li}.} \bibinfo{year}{2022}\natexlab{}.
\newblock \showarticletitle{Social {Disconnectedness} and {Loneliness}: {Do} {Self}-{Perceptions} of {Aging} {Play} a {Role}?}
\newblock \bibinfo{journal}{\emph{The Journals of Gerontology: Series B}} \bibinfo{volume}{77}, \bibinfo{number}{5} (\bibinfo{date}{May} \bibinfo{year}{2022}), \bibinfo{pages}{936--945}.
\newblock
\showISSN{1079-5014, 1758-5368}
\href{https://doi.org/10.1093/geronb/gbac008}{doi:\nolinkurl{10.1093/geronb/gbac008}}


\bibitem[Jacobs and {Wijnand IJsselsteijn}(2021)]%
        {jacobs_bridging_2021}
\bibfield{author}{\bibinfo{person}{Naomi Jacobs} {and} \bibinfo{person}{{Wijnand IJsselsteijn}}.} \bibinfo{year}{2021}\natexlab{}.
\newblock \showarticletitle{Bridging the {Theory}-{Practice} {Gap}: {Design}-{Experts} on {Capability} {Sensitive} {Design}}.
\newblock \bibinfo{journal}{\emph{International Journal of Technoethics}} \bibinfo{volume}{12}, \bibinfo{number}{2} (\bibinfo{date}{July} \bibinfo{year}{2021}), \bibinfo{pages}{1--16}.
\newblock
\showISSN{1947-3451, 1947-346X}
\href{https://doi.org/10.4018/IJT.2021070101}{doi:\nolinkurl{10.4018/IJT.2021070101}}


\bibitem[Jang et~al\mbox{.}(2024)]%
        {jang_its_2024}
\bibfield{author}{\bibinfo{person}{JiWoong Jang}, \bibinfo{person}{Sanika Moharana}, \bibinfo{person}{Patrick Carrington}, {and} \bibinfo{person}{Andrew Begel}.} \bibinfo{year}{2024}\natexlab{}.
\newblock \showarticletitle{“{It}’s the only thing {I} can trust”: {Envisioning} {Large} {Language} {Model} {Use} by {Autistic} {Workers} for {Communication} {Assistance}}. In \bibinfo{booktitle}{\emph{Proceedings of the {CHI} {Conference} on {Human} {Factors} in {Computing} {Systems}}}. \bibinfo{publisher}{ACM}, \bibinfo{address}{Honolulu HI USA}, \bibinfo{pages}{1--18}.
\newblock
\showISBNx{9798400703300}
\href{https://doi.org/10.1145/3613904.3642894}{doi:\nolinkurl{10.1145/3613904.3642894}}


\bibitem[Jeung and Huang(2024)]%
        {jeung_unlocking_2024}
\bibfield{author}{\bibinfo{person}{Jun~Li Jeung} {and} \bibinfo{person}{Janet Yi-Ching Huang}.} \bibinfo{year}{2024}\natexlab{}.
\newblock \showarticletitle{Unlocking {Memories} with {AI}: {Exploring} the {Role} of {AI}-{Generated} {Cues} in {Personal} {Reminiscing}}. In \bibinfo{booktitle}{\emph{Extended {Abstracts} of the {CHI} {Conference} on {Human} {Factors} in {Computing} {Systems}}}. \bibinfo{publisher}{ACM}, \bibinfo{address}{Honolulu HI USA}, \bibinfo{pages}{1--6}.
\newblock
\showISBNx{9798400703317}
\href{https://doi.org/10.1145/3613905.3650979}{doi:\nolinkurl{10.1145/3613905.3650979}}


\bibitem[Jin et~al\mbox{.}(2024)]%
        {jin_exploring_2024}
\bibfield{author}{\bibinfo{person}{Yucheng Jin}, \bibinfo{person}{Wanling Cai}, \bibinfo{person}{Li Chen}, \bibinfo{person}{Yizhe Zhang}, \bibinfo{person}{Gavin Doherty}, {and} \bibinfo{person}{Tonglin Jiang}.} \bibinfo{year}{2024}\natexlab{}.
\newblock \showarticletitle{Exploring the {Design} of {Generative} {AI} in {Supporting} {Music}-based {Reminiscence} for {Older} {Adults}}. In \bibinfo{booktitle}{\emph{Proceedings of the {CHI} {Conference} on {Human} {Factors} in {Computing} {Systems}}}. \bibinfo{publisher}{ACM}, \bibinfo{address}{Honolulu HI USA}, \bibinfo{pages}{1--17}.
\newblock
\showISBNx{9798400703300}
\href{https://doi.org/10.1145/3613904.3642800}{doi:\nolinkurl{10.1145/3613904.3642800}}


\bibitem[Jose et~al\mbox{.}(2012)]%
        {jose_does_2012}
\bibfield{author}{\bibinfo{person}{Paul~E. Jose}, \bibinfo{person}{Nicholas Ryan}, {and} \bibinfo{person}{Jan Pryor}.} \bibinfo{year}{2012}\natexlab{}.
\newblock \showarticletitle{Does {Social} {Connectedness} {Promote} a {Greater} {Sense} of {Well}‐{Being} in {Adolescence} {Over} {Time}?}
\newblock \bibinfo{journal}{\emph{Journal of Research on Adolescence}} \bibinfo{volume}{22}, \bibinfo{number}{2} (\bibinfo{date}{June} \bibinfo{year}{2012}), \bibinfo{pages}{235--251}.
\newblock
\showISSN{1050-8392, 1532-7795}
\href{https://doi.org/10.1111/j.1532-7795.2012.00783.x}{doi:\nolinkurl{10.1111/j.1532-7795.2012.00783.x}}


\bibitem[Joshi(2025)]%
        {joshi_review_2025}
\bibfield{author}{\bibinfo{person}{Satyadhar Joshi}.} \bibinfo{year}{2025}\natexlab{}.
\newblock \bibinfo{title}{Review of {Data} {Pipelines} and {Streaming} for {Generative} {AI} {Integration}: {Challenges}, {Solutions}, and {Future} {Directions}}.
\newblock
\href{https://doi.org/10.2139/ssrn.5268508}{doi:\nolinkurl{10.2139/ssrn.5268508}}


\bibitem[Khabarov and Samsonovich(2024)]%
        {samsonovich_registrar_2024}
\bibfield{author}{\bibinfo{person}{Dmitry Khabarov} {and} \bibinfo{person}{Alexei~V. Samsonovich}.} \bibinfo{year}{2024}\natexlab{}.
\newblock \showarticletitle{Registrar: {A} {Social} {Conversational} {Agent} {Based} on {Cognitive} and {Statistical} {Models} for a {Limited} {Paradigm}}.
\newblock In \bibinfo{booktitle}{\emph{Biologically {Inspired} {Cognitive} {Architectures} 2023}}, \bibfield{editor}{\bibinfo{person}{Alexei~V. Samsonovich} {and} \bibinfo{person}{Tingting Liu}} (Eds.). Vol.~\bibinfo{volume}{1130}. \bibinfo{publisher}{Springer Nature Switzerland}, \bibinfo{address}{Cham}, \bibinfo{pages}{444--452}.
\newblock
\showISBNx{978-3-031-50380-1 978-3-031-50381-8}
\href{https://doi.org/10.1007/978-3-031-50381-8_46}{doi:\nolinkurl{10.1007/978-3-031-50381-8_46}}
\newblock
\shownote{Series Title: Studies in Computational Intelligence}.


\bibitem[Lamblin et~al\mbox{.}(2017)]%
        {lamblin_social_2017}
\bibfield{author}{\bibinfo{person}{M. Lamblin}, \bibinfo{person}{C. Murawski}, \bibinfo{person}{S. Whittle}, {and} \bibinfo{person}{A. Fornito}.} \bibinfo{year}{2017}\natexlab{}.
\newblock \showarticletitle{Social connectedness, mental health and the adolescent brain}.
\newblock \bibinfo{journal}{\emph{Neuroscience \& Biobehavioral Reviews}}  \bibinfo{volume}{80} (\bibinfo{date}{Sept.} \bibinfo{year}{2017}), \bibinfo{pages}{57--68}.
\newblock
\showISSN{01497634}
\href{https://doi.org/10.1016/j.neubiorev.2017.05.010}{doi:\nolinkurl{10.1016/j.neubiorev.2017.05.010}}


\bibitem[Li et~al\mbox{.}(2025)]%
        {li_always_2025}
\bibfield{author}{\bibinfo{person}{Jiachen Li}, \bibinfo{person}{Elizabeth~D. Mynatt}, \bibinfo{person}{Varun Mishra}, {and} \bibinfo{person}{Jonathan Bell}.} \bibinfo{year}{2025}\natexlab{}.
\newblock \showarticletitle{'{Always} {Nice} and {Confident}, {Sometimes} {Wrong}': {Developer}'s {Experiences} {Engaging} {Generative} {AI} {Chatbots} {Versus} {Human}-{Powered} {Q}\&{A} {Platforms}}.
\newblock \bibinfo{journal}{\emph{Proceedings of the ACM on Human-Computer Interaction}} \bibinfo{volume}{9}, \bibinfo{number}{2} (\bibinfo{date}{May} \bibinfo{year}{2025}), \bibinfo{pages}{1--22}.
\newblock
\showISSN{2573-0142}
\href{https://doi.org/10.1145/3710927}{doi:\nolinkurl{10.1145/3710927}}


\bibitem[Li et~al\mbox{.}(2023)]%
        {li_blibug_2023}
\bibfield{author}{\bibinfo{person}{Yihua Li}, \bibinfo{person}{Yuqian Sun}, \bibinfo{person}{Ying Xu}, {and} \bibinfo{person}{Jihong Yu}.} \bibinfo{year}{2023}\natexlab{}.
\newblock \showarticletitle{Blibug: {AI} {Vtuber} {Based} on {Bilibili} {Danmuku} {Interaction}}. In \bibinfo{booktitle}{\emph{Creativity and {Cognition}}}. \bibinfo{publisher}{ACM}, \bibinfo{address}{Virtual Event USA}, \bibinfo{pages}{387--390}.
\newblock
\showISBNx{9798400701801}
\href{https://doi.org/10.1145/3591196.3596618}{doi:\nolinkurl{10.1145/3591196.3596618}}


\bibitem[Li et~al\mbox{.}(2024)]%
        {li_exploring_2024}
\bibfield{author}{\bibinfo{person}{Ziming Li}, \bibinfo{person}{Pinaki~Prasanna Babar}, \bibinfo{person}{Mike Barry}, {and} \bibinfo{person}{Roshan~L Peiris}.} \bibinfo{year}{2024}\natexlab{}.
\newblock \showarticletitle{Exploring the {Use} of {Large} {Language} {Model}-{Driven} {Chatbots} in {Virtual} {Reality} to {Train} {Autistic} {Individuals} in {Job} {Communication} {Skills}}. In \bibinfo{booktitle}{\emph{Extended {Abstracts} of the {CHI} {Conference} on {Human} {Factors} in {Computing} {Systems}}}. \bibinfo{publisher}{ACM}, \bibinfo{address}{Honolulu HI USA}, \bibinfo{pages}{1--7}.
\newblock
\showISBNx{9798400703317}
\href{https://doi.org/10.1145/3613905.3651996}{doi:\nolinkurl{10.1145/3613905.3651996}}


\bibitem[Liu et~al\mbox{.}(2024b)]%
        {liu_when_2024}
\bibfield{author}{\bibinfo{person}{Di Liu}, \bibinfo{person}{Hanqing Zhou}, {and} \bibinfo{person}{Pengcheng An}.} \bibinfo{year}{2024}\natexlab{b}.
\newblock \showarticletitle{"{When} {He} {Feels} {Cold}, {He} {Goes} to the {Seahorse}"—{Blending} {Generative} {AI} into {Multimaterial} {Storymaking} for {Family} {Expressive} {Arts} {Therapy}}. In \bibinfo{booktitle}{\emph{Proceedings of the {CHI} {Conference} on {Human} {Factors} in {Computing} {Systems}}}. \bibinfo{publisher}{ACM}, \bibinfo{address}{Honolulu HI USA}, \bibinfo{pages}{1--21}.
\newblock
\showISBNx{9798400703300}
\href{https://doi.org/10.1145/3613904.3642852}{doi:\nolinkurl{10.1145/3613904.3642852}}


\bibitem[Liu et~al\mbox{.}(2024a)]%
        {liu_peergpt_2024}
\bibfield{author}{\bibinfo{person}{Jiawen Liu}, \bibinfo{person}{Yuanyuan Yao}, \bibinfo{person}{Pengcheng An}, {and} \bibinfo{person}{Qi Wang}.} \bibinfo{year}{2024}\natexlab{a}.
\newblock \showarticletitle{{PeerGPT}: {Probing} the {Roles} of {LLM}-based {Peer} {Agents} as {Team} {Moderators} and {Participants} in {Children}'s {Collaborative} {Learning}}. In \bibinfo{booktitle}{\emph{Extended {Abstracts} of the {CHI} {Conference} on {Human} {Factors} in {Computing} {Systems}}}. \bibinfo{publisher}{ACM}, \bibinfo{address}{Honolulu HI USA}, \bibinfo{pages}{1--6}.
\newblock
\showISBNx{9798400703317}
\href{https://doi.org/10.1145/3613905.3651008}{doi:\nolinkurl{10.1145/3613905.3651008}}


\bibitem[Liu et~al\mbox{.}(2021)]%
        {liu_effectiveness_2021}
\bibfield{author}{\bibinfo{person}{Zhuo Liu}, \bibinfo{person}{Fan Yang}, \bibinfo{person}{Yifan Lou}, \bibinfo{person}{Wei Zhou}, {and} \bibinfo{person}{Feng Tong}.} \bibinfo{year}{2021}\natexlab{}.
\newblock \showarticletitle{The {Effectiveness} of {Reminiscence} {Therapy} on {Alleviating} {Depressive} {Symptoms} in {Older} {Adults}: {A} {Systematic} {Review}}.
\newblock \bibinfo{journal}{\emph{Frontiers in Psychology}}  \bibinfo{volume}{12} (\bibinfo{date}{Aug.} \bibinfo{year}{2021}), \bibinfo{pages}{709853}.
\newblock
\showISSN{1664-1078}
\href{https://doi.org/10.3389/fpsyg.2021.709853}{doi:\nolinkurl{10.3389/fpsyg.2021.709853}}


\bibitem[Louie et~al\mbox{.}(2020)]%
        {louie_novice-ai_2020}
\bibfield{author}{\bibinfo{person}{Ryan Louie}, \bibinfo{person}{Andy Coenen}, \bibinfo{person}{Cheng~Zhi Huang}, \bibinfo{person}{Michael Terry}, {and} \bibinfo{person}{Carrie~J. Cai}.} \bibinfo{year}{2020}\natexlab{}.
\newblock \showarticletitle{Novice-{AI} {Music} {Co}-{Creation} via {AI}-{Steering} {Tools} for {Deep} {Generative} {Models}}. In \bibinfo{booktitle}{\emph{Proceedings of the 2020 {CHI} {Conference} on {Human} {Factors} in {Computing} {Systems}}}. \bibinfo{publisher}{ACM}, \bibinfo{address}{Honolulu HI USA}, \bibinfo{pages}{1--13}.
\newblock
\showISBNx{978-1-4503-6708-0}
\href{https://doi.org/10.1145/3313831.3376739}{doi:\nolinkurl{10.1145/3313831.3376739}}


\bibitem[Maeda and Quan-Haase(2024)]%
        {maeda_when_2024}
\bibfield{author}{\bibinfo{person}{Takuya Maeda} {and} \bibinfo{person}{Anabel Quan-Haase}.} \bibinfo{year}{2024}\natexlab{}.
\newblock \showarticletitle{When {Human}-{AI} {Interactions} {Become} {Parasocial}: {Agency} and {Anthropomorphism} in {Affective} {Design}}. In \bibinfo{booktitle}{\emph{The 2024 {ACM} {Conference} on {Fairness}, {Accountability}, and {Transparency}}}. \bibinfo{publisher}{ACM}, \bibinfo{address}{Rio de Janeiro Brazil}, \bibinfo{pages}{1068--1077}.
\newblock
\showISBNx{9798400704505}
\href{https://doi.org/10.1145/3630106.3658956}{doi:\nolinkurl{10.1145/3630106.3658956}}


\bibitem[Mannheim et~al\mbox{.}(2023)]%
        {mannheim_ageism_2023}
\bibfield{author}{\bibinfo{person}{Ittay Mannheim}, \bibinfo{person}{Eveline J~M Wouters}, \bibinfo{person}{Hanna Köttl}, \bibinfo{person}{Leonieke~C Van~Boekel}, \bibinfo{person}{Rens Brankaert}, {and} \bibinfo{person}{Yvonne Van~Zaalen}.} \bibinfo{year}{2023}\natexlab{}.
\newblock \showarticletitle{Ageism in the {Discourse} and {Practice} of {Designing} {Digital} {Technology} for {Older} {Persons}: {A} {Scoping} {Review}}.
\newblock \bibinfo{journal}{\emph{The Gerontologist}} \bibinfo{volume}{63}, \bibinfo{number}{7} (\bibinfo{date}{Aug.} \bibinfo{year}{2023}), \bibinfo{pages}{1188--1200}.
\newblock
\showISSN{0016-9013, 1758-5341}
\href{https://doi.org/10.1093/geront/gnac144}{doi:\nolinkurl{10.1093/geront/gnac144}}


\bibitem[Moll et~al\mbox{.}(2020)]%
        {moll_are_2020}
\bibfield{author}{\bibinfo{person}{Sandra Moll}, \bibinfo{person}{Michelle Wyndham-West}, \bibinfo{person}{Gillian Mulvale}, \bibinfo{person}{Sean Park}, \bibinfo{person}{Alexis Buettgen}, \bibinfo{person}{Michelle Phoenix}, \bibinfo{person}{Robert Fleisig}, {and} \bibinfo{person}{Emma Bruce}.} \bibinfo{year}{2020}\natexlab{}.
\newblock \showarticletitle{Are you really doing ‘codesign’? {Critical} reflections when working with vulnerable populations}.
\newblock \bibinfo{journal}{\emph{BMJ Open}} \bibinfo{volume}{10}, \bibinfo{number}{11} (\bibinfo{date}{Nov.} \bibinfo{year}{2020}), \bibinfo{pages}{e038339}.
\newblock
\showISSN{2044-6055, 2044-6055}
\href{https://doi.org/10.1136/bmjopen-2020-038339}{doi:\nolinkurl{10.1136/bmjopen-2020-038339}}


\bibitem[Muniesa(2015)]%
        {muniesa_actor-network_2015}
\bibfield{author}{\bibinfo{person}{Fabian Muniesa}.} \bibinfo{year}{2015}\natexlab{}.
\newblock \showarticletitle{Actor-{Network} {Theory}}.
\newblock In \bibinfo{booktitle}{\emph{International {Encyclopedia} of the {Social} \& {Behavioral} {Sciences}}}. \bibinfo{publisher}{Elsevier}, \bibinfo{pages}{80--84}.
\newblock
\showISBNx{978-0-08-097087-5}
\href{https://doi.org/10.1016/B978-0-08-097086-8.85001-1}{doi:\nolinkurl{10.1016/B978-0-08-097086-8.85001-1}}


\bibitem[Munn et~al\mbox{.}(2018)]%
        {munn_systematic_2018}
\bibfield{author}{\bibinfo{person}{Zachary Munn}, \bibinfo{person}{Micah D.~J. Peters}, \bibinfo{person}{Cindy Stern}, \bibinfo{person}{Catalin Tufanaru}, \bibinfo{person}{Alexa McArthur}, {and} \bibinfo{person}{Edoardo Aromataris}.} \bibinfo{year}{2018}\natexlab{}.
\newblock \showarticletitle{Systematic review or scoping review? {Guidance} for authors when choosing between a systematic or scoping review approach}.
\newblock \bibinfo{journal}{\emph{BMC Medical Research Methodology}} \bibinfo{volume}{18}, \bibinfo{number}{1} (\bibinfo{date}{Dec.} \bibinfo{year}{2018}), \bibinfo{pages}{143}.
\newblock
\showISSN{1471-2288}
\href{https://doi.org/10.1186/s12874-018-0611-x}{doi:\nolinkurl{10.1186/s12874-018-0611-x}}


\bibitem[Na et~al\mbox{.}(2023)]%
        {na_social_2023}
\bibfield{author}{\bibinfo{person}{Peter~Jongho Na}, \bibinfo{person}{Dilip~V. Jeste}, {and} \bibinfo{person}{Robert~H. Pietrzak}.} \bibinfo{year}{2023}\natexlab{}.
\newblock \showarticletitle{Social {Disconnection} as a {Global} {Behavioral} {Epidemic}—{A} {Call} to {Action} {About} a {Major} {Health} {Risk} {Factor}}.
\newblock \bibinfo{journal}{\emph{JAMA Psychiatry}} \bibinfo{volume}{80}, \bibinfo{number}{2} (\bibinfo{date}{Feb.} \bibinfo{year}{2023}), \bibinfo{pages}{101}.
\newblock
\showISSN{2168-622X}
\href{https://doi.org/10.1001/jamapsychiatry.2022.4162}{doi:\nolinkurl{10.1001/jamapsychiatry.2022.4162}}


\bibitem[Nass et~al\mbox{.}(1994)]%
        {nass_computers_1994}
\bibfield{author}{\bibinfo{person}{Clifford Nass}, \bibinfo{person}{Jonathan Steuer}, {and} \bibinfo{person}{Ellen~R. Tauber}.} \bibinfo{year}{1994}\natexlab{}.
\newblock \showarticletitle{Computers are social actors}. In \bibinfo{booktitle}{\emph{Proceedings of the {SIGCHI} {Conference} on {Human} {Factors} in {Computing} {Systems}}}. \bibinfo{publisher}{ACM}, \bibinfo{address}{Boston Massachusetts USA}, \bibinfo{pages}{72--78}.
\newblock
\showISBNx{978-0-89791-650-9}
\href{https://doi.org/10.1145/191666.191703}{doi:\nolinkurl{10.1145/191666.191703}}


\bibitem[Numans et~al\mbox{.}(2021)]%
        {numans_vulnerable_2021}
\bibfield{author}{\bibinfo{person}{Wilma Numans}, \bibinfo{person}{Tine~Van Regenmortel}, \bibinfo{person}{René Schalk}, {and} \bibinfo{person}{Juliette Boog}.} \bibinfo{year}{2021}\natexlab{}.
\newblock \showarticletitle{Vulnerable persons in society: an insider’s perspective}.
\newblock \bibinfo{journal}{\emph{International Journal of Qualitative Studies on Health and Well-being}} \bibinfo{volume}{16}, \bibinfo{number}{1} (\bibinfo{date}{Jan.} \bibinfo{year}{2021}), \bibinfo{pages}{1863598}.
\newblock
\showISSN{1748-2631}
\href{https://doi.org/10.1080/17482631.2020.1863598}{doi:\nolinkurl{10.1080/17482631.2020.1863598}}


\bibitem[Oh et~al\mbox{.}(2018)]%
        {oh_systematic_2018}
\bibfield{author}{\bibinfo{person}{Catherine~S. Oh}, \bibinfo{person}{Jeremy~N. Bailenson}, {and} \bibinfo{person}{Gregory~F. Welch}.} \bibinfo{year}{2018}\natexlab{}.
\newblock \showarticletitle{A {Systematic} {Review} of {Social} {Presence}: {Definition}, {Antecedents}, and {Implications}}.
\newblock \bibinfo{journal}{\emph{Frontiers in Robotics and AI}}  \bibinfo{volume}{5} (\bibinfo{date}{Oct.} \bibinfo{year}{2018}), \bibinfo{pages}{114}.
\newblock
\showISSN{2296-9144}
\href{https://doi.org/10.3389/frobt.2018.00114}{doi:\nolinkurl{10.3389/frobt.2018.00114}}


\bibitem[Onorati et~al\mbox{.}(2023)]%
        {bravo_creating_2023}
\bibfield{author}{\bibinfo{person}{Teresa Onorati}, \bibinfo{person}{Álvaro Castro-González}, \bibinfo{person}{Javier~Cruz Del~Valle}, \bibinfo{person}{Paloma Díaz}, {and} \bibinfo{person}{José~Carlos Castillo}.} \bibinfo{year}{2023}\natexlab{}.
\newblock \showarticletitle{Creating {Personalized} {Verbal} {Human}-{Robot} {Interactions} {Using} {LLM} with the {Robot} {Mini}}.
\newblock In \bibinfo{booktitle}{\emph{Proceedings of the 15th {International} {Conference} on {Ubiquitous} {Computing} \& {Ambient} {Intelligence} ({UCAmI} 2023)}}, \bibfield{editor}{\bibinfo{person}{José Bravo} {and} \bibinfo{person}{Gabriel Urzáiz}} (Eds.). Vol.~\bibinfo{volume}{835}. \bibinfo{publisher}{Springer Nature Switzerland}, \bibinfo{address}{Cham}, \bibinfo{pages}{148--159}.
\newblock
\showISBNx{978-3-031-48305-9 978-3-031-48306-6}
\href{https://doi.org/10.1007/978-3-031-48306-6_15}{doi:\nolinkurl{10.1007/978-3-031-48306-6_15}}
\newblock
\shownote{Series Title: Lecture Notes in Networks and Systems}.


\bibitem[Ooi et~al\mbox{.}(2023)]%
        {ooi_potential_2023}
\bibfield{author}{\bibinfo{person}{Keng-Boon Ooi}, \bibinfo{person}{Garry Wei-Han Tan}, \bibinfo{person}{Mostafa Al-Emran}, \bibinfo{person}{Mohammed~A. Al-Sharafi}, \bibinfo{person}{Alexandru Capatina}, \bibinfo{person}{Amrita Chakraborty}, \bibinfo{person}{Yogesh~K. Dwivedi}, \bibinfo{person}{Tzu-Ling Huang}, \bibinfo{person}{Arpan~Kumar Kar}, \bibinfo{person}{Voon-Hsien Lee}, \bibinfo{person}{Xiu-Ming Loh}, \bibinfo{person}{Adrian Micu}, \bibinfo{person}{Patrick Mikalef}, \bibinfo{person}{Emmanuel Mogaji}, \bibinfo{person}{Neeraj Pandey}, \bibinfo{person}{Ramakrishnan Raman}, \bibinfo{person}{Nripendra~P. Rana}, \bibinfo{person}{Prianka Sarker}, \bibinfo{person}{Anshuman Sharma}, \bibinfo{person}{Ching-I Teng}, \bibinfo{person}{Samuel~Fosso Wamba}, {and} \bibinfo{person}{Lai-Wan Wong}.} \bibinfo{year}{2023}\natexlab{}.
\newblock \showarticletitle{The {Potential} of {Generative} {Artificial} {Intelligence} {Across} {Disciplines}: {Perspectives} and {Future} {Directions}}.
\newblock \bibinfo{journal}{\emph{Journal of Computer Information Systems}} (\bibinfo{date}{Oct.} \bibinfo{year}{2023}), \bibinfo{pages}{1--32}.
\newblock
\showISSN{0887-4417, 2380-2057}
\href{https://doi.org/10.1080/08874417.2023.2261010}{doi:\nolinkurl{10.1080/08874417.2023.2261010}}


\bibitem[{OpenAI}({[n.\,d.]})]%
        {openai_vision_nodate}
\bibfield{author}{\bibinfo{person}{{OpenAI}}.} \bibinfo{year}{[n.\,d.]}\natexlab{}.
\newblock \bibinfo{title}{Vision}.
\newblock
\urldef\tempurl%
\url{https://platform.openai.com/docs/guides/vision}
\showURL{%
\tempurl}


\bibitem[{OpenAI}(2023)]%
        {openai_gpt-4_2023}
\bibfield{author}{\bibinfo{person}{{OpenAI}}.} \bibinfo{year}{2023}\natexlab{}.
\newblock \bibinfo{title}{{GPT}-4 [{Large} {Language} {Model}]}.
\newblock
\urldef\tempurl%
\url{https://openai.com/gpt-4}
\showURL{%
\tempurl}


\bibitem[Pani et~al\mbox{.}(2024)]%
        {lyu_can_2024}
\bibfield{author}{\bibinfo{person}{Bianca Pani}, \bibinfo{person}{Joseph Crawford}, {and} \bibinfo{person}{Kelly-Ann Allen}.} \bibinfo{year}{2024}\natexlab{}.
\newblock \showarticletitle{Can {Generative} {Artificial} {Intelligence} {Foster} {Belongingness}, {Social} {Support}, and {Reduce} {Loneliness}? {A} {Conceptual} {Analysis}}.
\newblock In \bibinfo{booktitle}{\emph{Applications of {Generative} {AI}}}, \bibfield{editor}{\bibinfo{person}{Zhihan Lyu}} (Ed.). \bibinfo{publisher}{Springer International Publishing}, \bibinfo{address}{Cham}, \bibinfo{pages}{261--276}.
\newblock
\showISBNx{978-3-031-46237-5 978-3-031-46238-2}
\href{https://doi.org/10.1007/978-3-031-46238-2_13}{doi:\nolinkurl{10.1007/978-3-031-46238-2_13}}


\bibitem[Petersen et~al\mbox{.}(2023)]%
        {petersen_association_2023}
\bibfield{author}{\bibinfo{person}{Berkley Petersen}, \bibinfo{person}{Najmeh Khalili-Mahani}, \bibinfo{person}{Caitlin Murphy}, \bibinfo{person}{Kim Sawchuk}, \bibinfo{person}{Natalie Phillips}, \bibinfo{person}{Karen Z.~H. Li}, {and} \bibinfo{person}{Shannon Hebblethwaite}.} \bibinfo{year}{2023}\natexlab{}.
\newblock \showarticletitle{The association between information and communication technologies, loneliness and social connectedness: {A} scoping review}.
\newblock \bibinfo{journal}{\emph{Frontiers in Psychology}}  \bibinfo{volume}{14} (\bibinfo{date}{March} \bibinfo{year}{2023}), \bibinfo{pages}{1063146}.
\newblock
\showISSN{1664-1078}
\href{https://doi.org/10.3389/fpsyg.2023.1063146}{doi:\nolinkurl{10.3389/fpsyg.2023.1063146}}


\bibitem[Popay et~al\mbox{.}(2006)]%
        {popay_guidance_2006}
\bibfield{author}{\bibinfo{person}{Jennie Popay}, \bibinfo{person}{Helen Roberts}, \bibinfo{person}{Amanda Sowden}, \bibinfo{person}{Mark Petticrew}, \bibinfo{person}{Lisa Arai}, \bibinfo{person}{Mark Rodgers}, \bibinfo{person}{Nicky Britten}, \bibinfo{person}{Katrina Roen}, {and} \bibinfo{person}{Steven Duffy}.} \bibinfo{year}{2006}\natexlab{}.
\newblock \bibinfo{booktitle}{\emph{Guidance on the conduct of narrative synthesis in systematic reviews: {A} product from the {ESRC} {Methods} {Programme}}}.
\newblock \bibinfo{publisher}{Lancaster University}.
\newblock
\href{https://doi.org/10.13140/2.1.1018.4643}{doi:\nolinkurl{10.13140/2.1.1018.4643}}


\bibitem[Rebholz et~al\mbox{.}(2024)]%
        {rebholz_conversational_2024}
\bibfield{author}{\bibinfo{person}{Tobias~R. Rebholz}, \bibinfo{person}{Alena Koop}, {and} \bibinfo{person}{Mandy Hütter}.} \bibinfo{year}{2024}\natexlab{}.
\newblock \showarticletitle{Conversational user interfaces: {Explanations} and interactivity positively influence advice taking from generative artificial intelligence.}
\newblock \bibinfo{journal}{\emph{Technology, Mind, and Behavior}} \bibinfo{volume}{5}, \bibinfo{number}{4} (\bibinfo{date}{Dec.} \bibinfo{year}{2024}).
\newblock
\showISSN{2689-0208}
\href{https://doi.org/10.1037/tmb0000136}{doi:\nolinkurl{10.1037/tmb0000136}}


\bibitem[Reeves and Nass(1996)]%
        {reeves_media_1996}
\bibfield{author}{\bibinfo{person}{Byron Reeves} {and} \bibinfo{person}{Clifford Nass}.} \bibinfo{year}{1996}\natexlab{}.
\newblock \bibinfo{booktitle}{\emph{The media equation: how people treat computers, television, and new media like real people and places} (\bibinfo{edition}{1. paperback ed., [reprint.]} ed.)}.
\newblock \bibinfo{publisher}{Cambridge University Press}, \bibinfo{address}{Stanford, Calif}.
\newblock
\showISBNx{978-1-57586-053-4}


\bibitem[{Ronit} et~al\mbox{.}(2024)]%
        {ronit_enhancing_2024}
\bibfield{author}{\bibinfo{person}{{Ronit}}, \bibinfo{person}{Parth Agrawal}, {and} \bibinfo{person}{Manni Kumar}.} \bibinfo{year}{2024}\natexlab{}.
\newblock \showarticletitle{Enhancing {Human}-{Computer} {Interaction} with {Generative} {AI}}. In \bibinfo{booktitle}{\emph{2024 {IEEE} 4th {International} {Conference} on {ICT} in {Business} {Industry} \&amp; {Government} ({ICTBIG})}}. \bibinfo{publisher}{IEEE}, \bibinfo{address}{Indore, India}, \bibinfo{pages}{1--6}.
\newblock
\showISBNx{9798331518981}
\href{https://doi.org/10.1109/ICTBIG64922.2024.10911352}{doi:\nolinkurl{10.1109/ICTBIG64922.2024.10911352}}


\bibitem[{Runway ML}(2025)]%
        {runway_ml_runway_2025}
\bibfield{author}{\bibinfo{person}{{Runway ML}}.} \bibinfo{year}{2025}\natexlab{}.
\newblock \bibinfo{title}{Runway {ML} [{AI} video generator]}.
\newblock
\urldef\tempurl%
\url{https://runwayml.com}
\showURL{%
\tempurl}


\bibitem[Said et~al\mbox{.}(2024)]%
        {said_design_2024}
\bibfield{author}{\bibinfo{person}{Sherif Said}, \bibinfo{person}{Karim Youssef}, \bibinfo{person}{Benrose Prasad}, \bibinfo{person}{Ghaneemah Alasfour}, \bibinfo{person}{Samer Alkork}, {and} \bibinfo{person}{Taha Beyrouthy}.} \bibinfo{year}{2024}\natexlab{}.
\newblock \showarticletitle{Design and {Implementation} of {Adam}: {A} {Humanoid} {Robotic} {Head} with {Social} {Interaction} {Capabilities}}.
\newblock \bibinfo{journal}{\emph{Applied System Innovation}} \bibinfo{volume}{7}, \bibinfo{number}{3} (\bibinfo{date}{May} \bibinfo{year}{2024}), \bibinfo{pages}{42}.
\newblock
\showISSN{2571-5577}
\href{https://doi.org/10.3390/asi7030042}{doi:\nolinkurl{10.3390/asi7030042}}


\bibitem[Sanders and Stappers(2008)]%
        {sanders_co-creation_2008}
\bibfield{author}{\bibinfo{person}{Elizabeth B.-N. Sanders} {and} \bibinfo{person}{Pieter~Jan Stappers}.} \bibinfo{year}{2008}\natexlab{}.
\newblock \showarticletitle{Co-creation and the new landscapes of design}.
\newblock \bibinfo{journal}{\emph{CoDesign}} \bibinfo{volume}{4}, \bibinfo{number}{1} (\bibinfo{date}{March} \bibinfo{year}{2008}), \bibinfo{pages}{5--18}.
\newblock
\showISSN{1571-0882, 1745-3755}
\href{https://doi.org/10.1080/15710880701875068}{doi:\nolinkurl{10.1080/15710880701875068}}


\bibitem[Schnitzer et~al\mbox{.}(2024)]%
        {schnitzer_prototyping_2024}
\bibfield{author}{\bibinfo{person}{Benjamin Schnitzer}, \bibinfo{person}{Umut~Can Vural}, \bibinfo{person}{Bastian Schnitzer}, \bibinfo{person}{Muhammad~Usman Sardar}, \bibinfo{person}{Oren Fuerst}, {and} \bibinfo{person}{Oliver Korn}.} \bibinfo{year}{2024}\natexlab{}.
\newblock \showarticletitle{Prototyping a {Zoomorphic} {Interactive} {Robot} {Companion} with {Emotion} {Recognition} and {Affective} {Voice} {Interaction} for {Elderly} {People}}.
\newblock \bibinfo{journal}{\emph{Proceedings of the ACM on Human-Computer Interaction}} \bibinfo{volume}{8}, \bibinfo{number}{EICS} (\bibinfo{date}{June} \bibinfo{year}{2024}), \bibinfo{pages}{1--32}.
\newblock
\showISSN{2573-0142}
\href{https://doi.org/10.1145/3660244}{doi:\nolinkurl{10.1145/3660244}}


\bibitem[Sengar et~al\mbox{.}(2024)]%
        {sengar_generative_2024}
\bibfield{author}{\bibinfo{person}{Sandeep~Singh Sengar}, \bibinfo{person}{Affan~Bin Hasan}, \bibinfo{person}{Sanjay Kumar}, {and} \bibinfo{person}{Fiona Carroll}.} \bibinfo{year}{2024}\natexlab{}.
\newblock \showarticletitle{Generative artificial intelligence: a systematic review and applications}.
\newblock \bibinfo{journal}{\emph{Multimedia Tools and Applications}} (\bibinfo{date}{Aug.} \bibinfo{year}{2024}).
\newblock
\showISSN{1573-7721}
\href{https://doi.org/10.1007/s11042-024-20016-1}{doi:\nolinkurl{10.1007/s11042-024-20016-1}}


\bibitem[Seo et~al\mbox{.}(2024)]%
        {seo_chacha_2024}
\bibfield{author}{\bibinfo{person}{Woosuk Seo}, \bibinfo{person}{Chanmo Yang}, {and} \bibinfo{person}{Young-Ho Kim}.} \bibinfo{year}{2024}\natexlab{}.
\newblock \showarticletitle{{ChaCha}: {Leveraging} {Large} {Language} {Models} to {Prompt} {Children} to {Share} {Their} {Emotions} about {Personal} {Events}}. In \bibinfo{booktitle}{\emph{Proceedings of the {CHI} {Conference} on {Human} {Factors} in {Computing} {Systems}}}. \bibinfo{publisher}{ACM}, \bibinfo{address}{Honolulu HI USA}, \bibinfo{pages}{1--20}.
\newblock
\showISBNx{9798400703300}
\href{https://doi.org/10.1145/3613904.3642152}{doi:\nolinkurl{10.1145/3613904.3642152}}


\bibitem[Shakeri et~al\mbox{.}(2021)]%
        {shakeri_saga_2021}
\bibfield{author}{\bibinfo{person}{Hanieh Shakeri}, \bibinfo{person}{Carman Neustaedter}, {and} \bibinfo{person}{Steve DiPaola}.} \bibinfo{year}{2021}\natexlab{}.
\newblock \showarticletitle{{SAGA}: {Collaborative} {Storytelling} with {GPT}-3}. In \bibinfo{booktitle}{\emph{Companion {Publication} of the 2021 {Conference} on {Computer} {Supported} {Cooperative} {Work} and {Social} {Computing}}}. \bibinfo{publisher}{ACM}, \bibinfo{address}{Virtual Event USA}, \bibinfo{pages}{163--166}.
\newblock
\showISBNx{978-1-4503-8479-7}
\href{https://doi.org/10.1145/3462204.3481771}{doi:\nolinkurl{10.1145/3462204.3481771}}


\bibitem[Shelby et~al\mbox{.}(2023)]%
        {shelby_sociotechnical_2023}
\bibfield{author}{\bibinfo{person}{Renee Shelby}, \bibinfo{person}{Shalaleh Rismani}, \bibinfo{person}{Kathryn Henne}, \bibinfo{person}{AJung Moon}, \bibinfo{person}{Negar Rostamzadeh}, \bibinfo{person}{Paul Nicholas}, \bibinfo{person}{N'Mah Yilla-Akbari}, \bibinfo{person}{Jess Gallegos}, \bibinfo{person}{Andrew Smart}, \bibinfo{person}{Emilio Garcia}, {and} \bibinfo{person}{Gurleen Virk}.} \bibinfo{year}{2023}\natexlab{}.
\newblock \showarticletitle{Sociotechnical {Harms} of {Algorithmic} {Systems}: {Scoping} a {Taxonomy} for {Harm} {Reduction}}. In \bibinfo{booktitle}{\emph{Proceedings of the 2023 {AAAI}/{ACM} {Conference} on {AI}, {Ethics}, and {Society}}}. \bibinfo{publisher}{ACM}, \bibinfo{address}{Montréal QC Canada}, \bibinfo{pages}{723--741}.
\newblock
\showISBNx{9798400702310}
\href{https://doi.org/10.1145/3600211.3604673}{doi:\nolinkurl{10.1145/3600211.3604673}}


\bibitem[Shenk et~al\mbox{.}(2002)]%
        {shenk_narratives_2002}
\bibfield{author}{\bibinfo{person}{Dena Shenk}, \bibinfo{person}{Boyd Davis}, \bibinfo{person}{James~R. Peacock}, {and} \bibinfo{person}{Linda Moore}.} \bibinfo{year}{2002}\natexlab{}.
\newblock \showarticletitle{Narratives and self-identity in later life}.
\newblock \bibinfo{journal}{\emph{Journal of Aging Studies}} \bibinfo{volume}{16}, \bibinfo{number}{4} (\bibinfo{date}{Nov.} \bibinfo{year}{2002}), \bibinfo{pages}{401--413}.
\newblock
\showISSN{08904065}
\href{https://doi.org/10.1016/S0890-4065(02)00073-7}{doi:\nolinkurl{10.1016/S0890-4065(02)00073-7}}


\bibitem[Short et~al\mbox{.}(1976)]%
        {short_social_1976}
\bibfield{author}{\bibinfo{person}{John Short}, \bibinfo{person}{Ederyn Williams}, {and} \bibinfo{person}{Bruce Christie}.} \bibinfo{year}{1976}\natexlab{}.
\newblock \bibinfo{booktitle}{\emph{The social psychology of telecommunications}}.
\newblock \bibinfo{publisher}{Wiley}, \bibinfo{address}{London ; New York}.
\newblock
\showISBNx{978-0-471-01581-9}


\bibitem[Steen et~al\mbox{.}(2011)]%
        {steen_benefits_2011}
\bibfield{author}{\bibinfo{person}{Marc Steen}, \bibinfo{person}{Menno Manschot}, {and} \bibinfo{person}{Nicole~De Koning}.} \bibinfo{year}{2011}\natexlab{}.
\newblock \showarticletitle{Benefits of {Co}-design in {Service} {Design} {Projects}}.
\newblock \bibinfo{journal}{\emph{International Journal of Design}} \bibinfo{volume}{5}, \bibinfo{number}{2} (\bibinfo{year}{2011}), \bibinfo{pages}{53--60}.
\newblock


\bibitem[Suh et~al\mbox{.}(2021)]%
        {suh_ai_2021}
\bibfield{author}{\bibinfo{person}{Minhyang~(Mia) Suh}, \bibinfo{person}{Emily Youngblom}, \bibinfo{person}{Michael Terry}, {and} \bibinfo{person}{Carrie~J Cai}.} \bibinfo{year}{2021}\natexlab{}.
\newblock \showarticletitle{{AI} as {Social} {Glue}: {Uncovering} the {Roles} of {Deep} {Generative} {AI} during {Social} {Music} {Composition}}. In \bibinfo{booktitle}{\emph{Proceedings of the 2021 {CHI} {Conference} on {Human} {Factors} in {Computing} {Systems}}}. \bibinfo{publisher}{ACM}, \bibinfo{address}{Yokohama Japan}, \bibinfo{pages}{1--11}.
\newblock
\showISBNx{978-1-4503-8096-6}
\href{https://doi.org/10.1145/3411764.3445219}{doi:\nolinkurl{10.1145/3411764.3445219}}


\bibitem[Suijkerbuijk et~al\mbox{.}(2019)]%
        {suijkerbuijk_active_2019}
\bibfield{author}{\bibinfo{person}{Sandra Suijkerbuijk}, \bibinfo{person}{Henk~Herman Nap}, \bibinfo{person}{Lotte Cornelisse}, \bibinfo{person}{Wijnand~A. IJsselsteijn}, \bibinfo{person}{Yvonne~A.W. De~Kort}, {and} \bibinfo{person}{Mirella~M.N. Minkman}.} \bibinfo{year}{2019}\natexlab{}.
\newblock \showarticletitle{Active {Involvement} of {People} with {Dementia}: {A} {Systematic} {Review} of {Studies} {Developing} {Supportive} {Technologies}}.
\newblock \bibinfo{journal}{\emph{Journal of Alzheimer's Disease}} \bibinfo{volume}{69}, \bibinfo{number}{4} (\bibinfo{date}{June} \bibinfo{year}{2019}), \bibinfo{pages}{1041--1065}.
\newblock
\showISSN{13872877, 18758908}
\href{https://doi.org/10.3233/JAD-190050}{doi:\nolinkurl{10.3233/JAD-190050}}


\bibitem[{Suno}(2023)]%
        {suno_suno_2023}
\bibfield{author}{\bibinfo{person}{{Suno}}.} \bibinfo{year}{2023}\natexlab{}.
\newblock \bibinfo{title}{Suno [{AI} audio generator]}.
\newblock
\urldef\tempurl%
\url{https://www.suno.ai/}
\showURL{%
\tempurl}


\bibitem[Tang et~al\mbox{.}(2024)]%
        {tang_emoeden_2024}
\bibfield{author}{\bibinfo{person}{Yilin Tang}, \bibinfo{person}{Liuqing Chen}, \bibinfo{person}{Ziyu Chen}, \bibinfo{person}{Wenkai Chen}, \bibinfo{person}{Yu Cai}, \bibinfo{person}{Yao Du}, \bibinfo{person}{Fan Yang}, {and} \bibinfo{person}{Lingyun Sun}.} \bibinfo{year}{2024}\natexlab{}.
\newblock \showarticletitle{{EmoEden}: {Applying} {Generative} {Artificial} {Intelligence} to {Emotional} {Learning} for {Children} with {High}-{Function} {Autism}}. In \bibinfo{booktitle}{\emph{Proceedings of the {CHI} {Conference} on {Human} {Factors} in {Computing} {Systems}}}. \bibinfo{publisher}{ACM}, \bibinfo{address}{Honolulu HI USA}, \bibinfo{pages}{1--20}.
\newblock
\showISBNx{9798400703300}
\href{https://doi.org/10.1145/3613904.3642899}{doi:\nolinkurl{10.1145/3613904.3642899}}


\bibitem[Tricco et~al\mbox{.}(2018)]%
        {tricco_prisma_2018}
\bibfield{author}{\bibinfo{person}{Andrea~C. Tricco}, \bibinfo{person}{Erin Lillie}, \bibinfo{person}{Wasifa Zarin}, \bibinfo{person}{Kelly~K. O'Brien}, \bibinfo{person}{Heather Colquhoun}, \bibinfo{person}{Danielle Levac}, \bibinfo{person}{David Moher}, \bibinfo{person}{Micah~D.J. Peters}, \bibinfo{person}{Tanya Horsley}, \bibinfo{person}{Laura Weeks}, \bibinfo{person}{Susanne Hempel}, \bibinfo{person}{Elie~A. Akl}, \bibinfo{person}{Christine Chang}, \bibinfo{person}{Jessie McGowan}, \bibinfo{person}{Lesley Stewart}, \bibinfo{person}{Lisa Hartling}, \bibinfo{person}{Adrian Aldcroft}, \bibinfo{person}{Michael~G. Wilson}, \bibinfo{person}{Chantelle Garritty}, \bibinfo{person}{Simon Lewin}, \bibinfo{person}{Christina~M. Godfrey}, \bibinfo{person}{Marilyn~T. Macdonald}, \bibinfo{person}{Etienne~V. Langlois}, \bibinfo{person}{Karla Soares-Weiser}, \bibinfo{person}{Jo Moriarty}, \bibinfo{person}{Tammy Clifford}, \bibinfo{person}{Özge Tunçalp}, {and} \bibinfo{person}{Sharon~E. Straus}.}
  \bibinfo{year}{2018}\natexlab{}.
\newblock \showarticletitle{{PRISMA} {Extension} for {Scoping} {Reviews} ({PRISMA}-{ScR}): {Checklist} and {Explanation}}.
\newblock \bibinfo{journal}{\emph{Annals of Internal Medicine}} \bibinfo{volume}{169}, \bibinfo{number}{7} (\bibinfo{date}{Oct.} \bibinfo{year}{2018}), \bibinfo{pages}{467--473}.
\newblock
\showISSN{0003-4819, 1539-3704}
\href{https://doi.org/10.7326/M18-0850}{doi:\nolinkurl{10.7326/M18-0850}}


\bibitem[Tucci et~al\mbox{.}(2024)]%
        {degen_retromind_2024}
\bibfield{author}{\bibinfo{person}{Cesare Tucci}, \bibinfo{person}{Ilaria Amaro}, \bibinfo{person}{Attilio Della~Greca}, {and} \bibinfo{person}{Genoveffa Tortora}.} \bibinfo{year}{2024}\natexlab{}.
\newblock \showarticletitle{{RetroMind} and the {Image} of {Memories}: {A} {Preliminary} {Study} of a {Support} {Tool} for {Reminiscence} {Therapy}}.
\newblock In \bibinfo{booktitle}{\emph{Artificial {Intelligence} in {HCI}}}, \bibfield{editor}{\bibinfo{person}{Helmut Degen} {and} \bibinfo{person}{Stavroula Ntoa}} (Eds.). Vol.~\bibinfo{volume}{14736}. \bibinfo{publisher}{Springer Nature Switzerland}, \bibinfo{address}{Cham}, \bibinfo{pages}{456--468}.
\newblock
\showISBNx{978-3-031-60614-4 978-3-031-60615-1}
\href{https://doi.org/10.1007/978-3-031-60615-1_31}{doi:\nolinkurl{10.1007/978-3-031-60615-1_31}}
\newblock
\shownote{Series Title: Lecture Notes in Computer Science}.


\bibitem[Van~Bel et~al\mbox{.}(2009)]%
        {van_bel_social_2009}
\bibfield{author}{\bibinfo{person}{Daniel~T Van~Bel}, \bibinfo{person}{Karin C H~J Smolders}, \bibinfo{person}{Wijnand~A IJsselsteijn}, {and} \bibinfo{person}{Yvonne A.~W. De~Kort}.} \bibinfo{year}{2009}\natexlab{}.
\newblock \showarticletitle{Social connectedness: concept and measurement}.
\newblock In \bibinfo{booktitle}{\emph{Intelligent environments 2009}}. \bibinfo{publisher}{IOS Press}, \bibinfo{pages}{67--74}.
\newblock


\bibitem[Van~Grunsven and IJsselsteijn(2022)]%
        {dennis_confronting_2022}
\bibfield{author}{\bibinfo{person}{Janna Van~Grunsven} {and} \bibinfo{person}{Wijnand IJsselsteijn}.} \bibinfo{year}{2022}\natexlab{}.
\newblock \showarticletitle{Confronting {Ableism} in a {Post}-{COVID} {World}: {Designing} for {World}-{Familiarity} {Through} {Acts} of {Defamiliarization}}.
\newblock In \bibinfo{booktitle}{\emph{Values for a {Post}-{Pandemic} {Future}}}, \bibfield{editor}{\bibinfo{person}{Matthew~J. Dennis}, \bibinfo{person}{Georgy Ishmaev}, \bibinfo{person}{Steven Umbrello}, {and} \bibinfo{person}{Jeroen Van Den~Hoven}} (Eds.). Vol.~\bibinfo{volume}{40}. \bibinfo{publisher}{Springer International Publishing}, \bibinfo{address}{Cham}, \bibinfo{pages}{185--200}.
\newblock
\showISBNx{978-3-031-08423-2 978-3-031-08424-9}
\href{https://doi.org/10.1007/978-3-031-08424-9_10}{doi:\nolinkurl{10.1007/978-3-031-08424-9_10}}
\newblock
\shownote{Series Title: Philosophy of Engineering and Technology}.


\bibitem[Wan et~al\mbox{.}(2024)]%
        {wan_building_2024}
\bibfield{author}{\bibinfo{person}{Hongyu Wan}, \bibinfo{person}{Jinda Zhang}, \bibinfo{person}{Abdulaziz~Arif Suria}, \bibinfo{person}{Bingsheng Yao}, \bibinfo{person}{Dakuo Wang}, \bibinfo{person}{Yvonne Coady}, {and} \bibinfo{person}{Mirjana Prpa}.} \bibinfo{year}{2024}\natexlab{}.
\newblock \showarticletitle{Building {LLM}-based {AI} {Agents} in {Social} {Virtual} {Reality}}. In \bibinfo{booktitle}{\emph{Extended {Abstracts} of the {CHI} {Conference} on {Human} {Factors} in {Computing} {Systems}}}. \bibinfo{publisher}{ACM}, \bibinfo{address}{Honolulu HI USA}, \bibinfo{pages}{1--7}.
\newblock
\showISBNx{9798400703317}
\href{https://doi.org/10.1145/3613905.3651026}{doi:\nolinkurl{10.1145/3613905.3651026}}


\bibitem[Wang et~al\mbox{.}(2024)]%
        {wang_aint_2024}
\bibfield{author}{\bibinfo{person}{Zining Wang}, \bibinfo{person}{Paul Reisert}, \bibinfo{person}{Eric Nichols}, {and} \bibinfo{person}{Randy Gomez}.} \bibinfo{year}{2024}\natexlab{}.
\newblock \showarticletitle{Ain't {Misbehavin}' - {Using} {LLMs} to {Generate} {Expressive} {Robot} {Behavior} in {Conversations} with the {Tabletop} {Robot} {Haru}}. In \bibinfo{booktitle}{\emph{Companion of the 2024 {ACM}/{IEEE} {International} {Conference} on {Human}-{Robot} {Interaction}}}. \bibinfo{publisher}{ACM}, \bibinfo{address}{Boulder CO USA}, \bibinfo{pages}{1105--1109}.
\newblock
\showISBNx{9798400703232}
\href{https://doi.org/10.1145/3610978.3640562}{doi:\nolinkurl{10.1145/3610978.3640562}}


\bibitem[Wei et~al\mbox{.}(2024)]%
        {wei_improving_2024}
\bibfield{author}{\bibinfo{person}{Rongxuan Wei}, \bibinfo{person}{Kangkang Li}, {and} \bibinfo{person}{Jiaming Lan}.} \bibinfo{year}{2024}\natexlab{}.
\newblock \showarticletitle{Improving {Collaborative} {Learning} {Performance} {Based} on {LLM} {Virtual} {Assistant}}. In \bibinfo{booktitle}{\emph{2024 13th {International} {Conference} on {Educational} and {Information} {Technology} ({ICEIT})}}. \bibinfo{publisher}{IEEE}, \bibinfo{address}{Chengdu, China}, \bibinfo{pages}{1--6}.
\newblock
\showISBNx{9798350372663}
\href{https://doi.org/10.1109/ICEIT61397.2024.10540942}{doi:\nolinkurl{10.1109/ICEIT61397.2024.10540942}}


\bibitem[Weidinger et~al\mbox{.}(2021)]%
        {weidinger_ethical_2021}
\bibfield{author}{\bibinfo{person}{Laura Weidinger}, \bibinfo{person}{John Mellor}, \bibinfo{person}{Maribeth Rauh}, \bibinfo{person}{Conor Griffin}, \bibinfo{person}{Jonathan Uesato}, \bibinfo{person}{Po-Sen Huang}, \bibinfo{person}{Myra Cheng}, \bibinfo{person}{Mia Glaese}, \bibinfo{person}{Borja Balle}, \bibinfo{person}{Atoosa Kasirzadeh}, \bibinfo{person}{Zac Kenton}, \bibinfo{person}{Sasha Brown}, \bibinfo{person}{Will Hawkins}, \bibinfo{person}{Tom Stepleton}, \bibinfo{person}{Courtney Biles}, \bibinfo{person}{Abeba Birhane}, \bibinfo{person}{Julia Haas}, \bibinfo{person}{Laura Rimell}, \bibinfo{person}{Lisa~Anne Hendricks}, \bibinfo{person}{William Isaac}, \bibinfo{person}{Sean Legassick}, \bibinfo{person}{Geoffrey Irving}, {and} \bibinfo{person}{Iason Gabriel}.} \bibinfo{year}{2021}\natexlab{}.
\newblock \bibinfo{title}{Ethical and social risks of harm from {Language} {Models}}.
\newblock
\href{https://doi.org/10.48550/ARXIV.2112.04359}{doi:\nolinkurl{10.48550/ARXIV.2112.04359}}
\newblock
\shownote{Version Number: 1}.


\bibitem[Weidinger et~al\mbox{.}(2022)]%
        {weidinger_taxonomy_2022}
\bibfield{author}{\bibinfo{person}{Laura Weidinger}, \bibinfo{person}{Jonathan Uesato}, \bibinfo{person}{Maribeth Rauh}, \bibinfo{person}{Conor Griffin}, \bibinfo{person}{Po-Sen Huang}, \bibinfo{person}{John Mellor}, \bibinfo{person}{Amelia Glaese}, \bibinfo{person}{Myra Cheng}, \bibinfo{person}{Borja Balle}, \bibinfo{person}{Atoosa Kasirzadeh}, \bibinfo{person}{Courtney Biles}, \bibinfo{person}{Sasha Brown}, \bibinfo{person}{Zac Kenton}, \bibinfo{person}{Will Hawkins}, \bibinfo{person}{Tom Stepleton}, \bibinfo{person}{Abeba Birhane}, \bibinfo{person}{Lisa~Anne Hendricks}, \bibinfo{person}{Laura Rimell}, \bibinfo{person}{William Isaac}, \bibinfo{person}{Julia Haas}, \bibinfo{person}{Sean Legassick}, \bibinfo{person}{Geoffrey Irving}, {and} \bibinfo{person}{Iason Gabriel}.} \bibinfo{year}{2022}\natexlab{}.
\newblock \showarticletitle{Taxonomy of {Risks} posed by {Language} {Models}}. In \bibinfo{booktitle}{\emph{2022 {ACM} {Conference} on {Fairness}, {Accountability}, and {Transparency}}}. \bibinfo{publisher}{ACM}, \bibinfo{address}{Seoul Republic of Korea}, \bibinfo{pages}{214--229}.
\newblock
\showISBNx{978-1-4503-9352-2}
\href{https://doi.org/10.1145/3531146.3533088}{doi:\nolinkurl{10.1145/3531146.3533088}}


\bibitem[{World Health Organization}(2021)]%
        {world_health_organization_social_2021}
\bibfield{author}{\bibinfo{person}{{World Health Organization}}.} \bibinfo{year}{2021}\natexlab{}.
\newblock \bibinfo{booktitle}{\emph{Social {Isolation} and {Loneliness} among {Older} {People}: {Advocacy} {Brief}} (\bibinfo{edition}{1st ed} ed.)}.
\newblock \bibinfo{address}{Geneva}.
\newblock
\showISBNx{978-92-4-003074-9}


\bibitem[Xygkou et~al\mbox{.}(2024)]%
        {xygkou_mindtalker_2024}
\bibfield{author}{\bibinfo{person}{Anna Xygkou}, \bibinfo{person}{Chee~Siang Ang}, \bibinfo{person}{Panote Siriaraya}, \bibinfo{person}{Jonasz~Piotr Kopecki}, \bibinfo{person}{Alexandra Covaci}, \bibinfo{person}{Eiman Kanjo}, {and} \bibinfo{person}{Wan-Jou She}.} \bibinfo{year}{2024}\natexlab{}.
\newblock \showarticletitle{{MindTalker}: {Navigating} the {Complexities} of {AI}-{Enhanced} {Social} {Engagement} for {People} with {Early}-{Stage} {Dementia}}. In \bibinfo{booktitle}{\emph{Proceedings of the {CHI} {Conference} on {Human} {Factors} in {Computing} {Systems}}}. \bibinfo{publisher}{ACM}, \bibinfo{address}{Honolulu HI USA}, \bibinfo{pages}{1--15}.
\newblock
\showISBNx{9798400703300}
\href{https://doi.org/10.1145/3613904.3642538}{doi:\nolinkurl{10.1145/3613904.3642538}}


\bibitem[You and Choi(2024)]%
        {you_sociocultural_2024}
\bibfield{author}{\bibinfo{person}{Soobin You} {and} \bibinfo{person}{Heejeong Choi}.} \bibinfo{year}{2024}\natexlab{}.
\newblock \showarticletitle{A sociocultural perspective on {\textless}span style="font-variant:small-caps;"{\textgreater}{AI}{\textless}/span{\textgreater} assistive technology for older adults' social connectedness: {A} scoping review}.
\newblock \bibinfo{journal}{\emph{Family Relations}} (\bibinfo{date}{Dec.} \bibinfo{year}{2024}), \bibinfo{pages}{fare.13128}.
\newblock
\showISSN{0197-6664, 1741-3729}
\href{https://doi.org/10.1111/fare.13128}{doi:\nolinkurl{10.1111/fare.13128}}


\bibitem[Zhang et~al\mbox{.}(2023)]%
        {zhang_reconfiguring_2023}
\bibfield{author}{\bibinfo{person}{Dan Zhang}, \bibinfo{person}{Zhiyong Lin}, \bibinfo{person}{Feinian Chen}, {and} \bibinfo{person}{Shuzhuo Li}.} \bibinfo{year}{2023}\natexlab{}.
\newblock \showarticletitle{Reconfiguring {Social} {Disconnectedness} and {Its} {Link} to {Psychological} {Well}-{Being} among {Older} {Adults} in {Rural} {China}}.
\newblock \bibinfo{journal}{\emph{Journal of Applied Gerontology}} \bibinfo{volume}{42}, \bibinfo{number}{1} (\bibinfo{date}{Jan.} \bibinfo{year}{2023}), \bibinfo{pages}{99--110}.
\newblock
\showISSN{0733-4648, 1552-4523}
\href{https://doi.org/10.1177/07334648221124915}{doi:\nolinkurl{10.1177/07334648221124915}}


\bibitem[Zhou et~al\mbox{.}(2024)]%
        {zhou_my_2024}
\bibfield{author}{\bibinfo{person}{Hanqing Zhou}, \bibinfo{person}{Anastasia Nikolova}, {and} \bibinfo{person}{Pengcheng An}.} \bibinfo{year}{2024}\natexlab{}.
\newblock \showarticletitle{'{My} lollipop dropped...'–{Probing} {Design} {Opportunities} for {SEL} {Agents} through {Children}'s {Peer} {Co}-{Creation} of {Social}-{Emotional} {Stories}}. In \bibinfo{booktitle}{\emph{Extended {Abstracts} of the {CHI} {Conference} on {Human} {Factors} in {Computing} {Systems}}}. \bibinfo{publisher}{ACM}, \bibinfo{address}{Honolulu HI USA}, \bibinfo{pages}{1--8}.
\newblock
\showISBNx{9798400703317}
\href{https://doi.org/10.1145/3613905.3651867}{doi:\nolinkurl{10.1145/3613905.3651867}}


\bibitem[Zhuo et~al\mbox{.}(2023)]%
        {zhuo_red_2023}
\bibfield{author}{\bibinfo{person}{Terry~Yue Zhuo}, \bibinfo{person}{Yujin Huang}, \bibinfo{person}{Chunyang Chen}, {and} \bibinfo{person}{Zhenchang Xing}.} \bibinfo{year}{2023}\natexlab{}.
\newblock \bibinfo{title}{Red teaming {ChatGPT} via {Jailbreaking}: {Bias}, {Robustness}, {Reliability} and {Toxicity}}.
\newblock
\href{https://doi.org/10.48550/ARXIV.2301.12867}{doi:\nolinkurl{10.48550/ARXIV.2301.12867}}
\newblock
\shownote{Version Number: 4}.


\end{thebibliography}

\appendix
\section{Search queries}\label{appendixA}

\begin{longtable}[c]{>{\raggedright\arraybackslash}p{.12\textwidth} >{\raggedright\arraybackslash}p{.84\textwidth}}
\caption{The exact search queries per database.}
\label{tab:appendixA_queries}\\
\hline
\textbf{Database} & \textbf{Search string} \\ \hline
\endfirsthead
\hline
\textbf{Database} & \textbf{Search string} \\ \hline
\endhead
\hline
\endfoot
\endlastfoot

Scopus & ( TITLE-ABS-KEY ( "generative AI" OR "generative-AI" OR "generative artificial intelligence" OR "ChatGPT" OR gpt* OR "large language model" OR llm* ) AND TITLE-ABS-KEY ( social* OR "conversation*" OR "interaction" ) AND TITLE-ABS-KEY ( design* OR creat* ) ) AND PUBYEAR > 2019 AND PUBYEAR < 2026 AND ( LIMIT-TO ( SUBJAREA , "COMP" ) OR LIMIT-TO ( SUBJAREA , "SOCI" ) OR LIMIT-TO ( SUBJAREA , "ENGI" ) OR LIMIT-TO ( SUBJAREA , "MATH" ) OR LIMIT-TO ( SUBJAREA , "MEDI" ) OR LIMIT-TO ( SUBJAREA , "DECI" ) OR LIMIT-TO ( SUBJAREA , "ARTS" ) OR LIMIT-TO ( SUBJAREA , "PSYC" ) OR LIMIT-TO ( SUBJAREA , "MULT" ) OR LIMIT-TO ( SUBJAREA , "HEAL" ) OR LIMIT-TO ( SUBJAREA , "NURS" ) ) AND ( LIMIT-TO ( DOCTYPE , "cp" ) OR LIMIT-TO ( DOCTYPE , "ar" ) OR LIMIT-TO ( DOCTYPE , "ch" ) ) \\
ACM Library & (title:("generative AI" OR "generative-AI" OR "generative artificial intelligence" OR "ChatGPT" OR GPT* OR "large language model" OR LLM*) OR abstract:("generative AI" OR "generative-AI" OR "generative artificial intelligence" OR "ChatGPT" OR GPT* OR "large language model" OR LLM*) OR Keyword:("generative AI" OR "generative-AI" OR "generative artificial intelligence" OR "ChatGPT" OR GPT* OR "large language model" OR LLM*)) AND (title:(social* OR "conversation" OR "interaction") OR abstract:(social* OR "conversation" OR "interaction") OR Keyword:(social* OR "conversation" OR "interaction")) AND (title:(design* OR creat*) OR abstract:(design* OR creat*) OR Keyword:(design* OR creat*)) \\
IEEE Xplore & ("All Metadata":"generative AI" OR "All Metadata":"generative-AI" OR "All Metadata":"generative artificial intelligence" OR "All Metadata":"ChatGPT" OR "All Metadata":GPT* OR "All Metadata":"large language model" OR "All Metadata":LLM*) AND ("All Metadata":social* OR "All Metadata":conversation* OR "All Metadata":"interaction") AND ("All Metadata":design* OR "All Metadata":creat*) \\
PubMed & (((("generative AI"[Title/Abstract] OR "generative-AI"[Title/Abstract] OR "generative artificial intelligence"[Title/Abstract] OR "ChatGPT"[Title/Abstract] OR "GPT*"[Title/Abstract] OR "large language model"[Title/Abstract] OR "LLM*"[Title/Abstract])) AND ("social*"[Title/Abstract])) AND ("design*"[Title/Abstract] OR "creat*"[Title/Abstract])) \\ \hline
\end{longtable}

\section{Screening pipeline}\label{appendixB}
\begin{figure}[H]
    \centering
    \includegraphics[width=\textwidth]{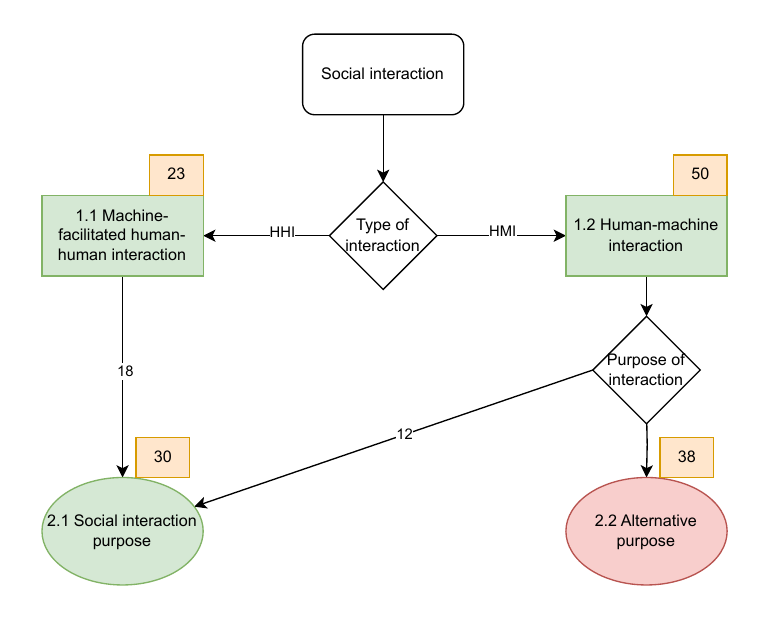}
    \caption{The pipeline for screening studies based on the nature of the social interaction that they stimulate. When a design creates an opportunity for machine-facilitated human-human interaction, it was included by default. If the interaction had a human-machine nature only, the purpose of the interaction was assessed to determine inclusion.}
    \label{fig:appendixB_interaction_inclusion_pipeline}
\end{figure}

\section{Overview of GAI-based applications}\label{appendixC}

\begin{longtable}[c]{>{\raggedright\arraybackslash}p{.06\textwidth} >{\raggedright\arraybackslash}p{.14\textwidth} >{\raggedright\arraybackslash}p{.1\textwidth} >{\raggedright\arraybackslash}p{.36\textwidth} >{\centering\arraybackslash}p{.1\textwidth} >{\raggedright\arraybackslash}p{.1\textwidth}}
\caption{Overview of the technical GAI-details of each technology described in each of the papers.}
\label{tab:appendixC_tech_details}\\
\hline
\textbf{Output} & \textbf{User perception} & \textbf{Software} & \textbf{Prompt} & \textbf{Open source?} & \textbf{Reference} \\ \hline
\endfirsthead
\hline
\textbf{Output} & \textbf{User perception} & \textbf{Software} & \textbf{Prompt} & \textbf{Open source?} & \textbf{Reference} \\ \hline
\endhead
\hline
\endfoot
\hline
\endlastfoot

Text & Written text \& Speech & GPT-3.5 & `\textit{You are a conversational companion for an elderly person. You should be polite, helpful, empathetic, sociable, friendly, and factually correct.}' & $\square$ & \citet{alessa_towards_2023} \\
Text & Written text \& Speech & ChatGPT & N/A & $\square$ & \citet{bhati_bookmate_2023} \\
Text & Speech & GPT-3 (text-davinci-002) & Example from GitHub repository: `\textit{You are the Pepper robot, social assistive robot located at the Interaction Lab, University of Skövde, Sweden.}' & \ding{110} & \citet{billing_language_2023} \\
Text & Written text & GPT-3.5 & Paper presents five prompts. One prompt to illustrate: `\textit{Your name is {{botname}}. Your job is to listen to a conversation between students discussing a topic and determine when is the best time to chime in. The chat messages will be sent with the role `user' and will include the student’s name and message. These students will be having a conversation about a certain topic, and they will be sharing their opinions and asking questions amongst themselves. When you notice that the conversation is going smoothly and the students have much to discuss, do not chime in. Respond only with: `. . .' If you notice that the conversation is repetitive, not productive, or worth discussing further, respond only with:`CHIME.' If a student asks you a question using your name, respond only with: `CHIME.'}' & $\square$ & \citet{cai_advancing_2024} \\
Text & Written text & ChatGPT & Paper presents six prompts. One prompt to illustrate: `\textit{Please rewrite the story to make it suitable for 15-year-old children.}' & $\square$ & \citet{chen_chatgpt_2023} \\
Text & Written text & GPT-3 & N/A & $\square$ & \citet{elgarf_creativebot_2022} \\
Text & Written text & GPT-4 & Paper presents four prompt components. One prompt to illustrate: `\textit{I envision a hierarchy of positive response levels for topics as follows: Level 1 — Intimate and interested. It means· · ··Level 4 — Distant and disgusted· · · · · Please refine the current expressions, generate two intermediate levels, and provide the level divisions and reasons.}' & $\square$ & \citet{fang_socializechat_2023} \\
Text & Written text & GPT-3.5 Turbo & Paper presents two prompts. One prompt to illustrate: `\textit{You are a Speech Language Pathologist specialized in Augmentative and Alternative Communication. Your task is to provide vocabulary related to a situation to help a person with communication disability to formulate messages about the situation. This vocabulary must contain words that people would often use to talk about that situation, either to describe it or to discuss it in a general context. The vocabulary must contain 20 verbs, 20 descriptors (adjectives and adverbs not terminating with LY), 20 objects, and 20 prepositions. All words must be in the first person singular, infinitive form without `to'. Provide your answer as a JSON object as in the following example: EXAMPLE ANSWER: `verbs': [VERBS], `prepositions': [PREPOSITIONS] `descriptors': [ADJECTIVES AND ADVERBS NOT TERMINATING WITH LY], `objects': [OBJECTS], SITUATION: [Photo caption from the computer vision model is inserted here]}' & $\square$ & \citet{fontana_de_vargas_co-designing_2024} \\
Text & Written text & GPT-4 & `\textit{(1) You are a helpful assistant named Paprika. Provide clear and thorough answers but be concise. (2) Use a more conversational but still workplace appropriate style. Make sure your answers are short, make sure your responses are around two paragraphs. (3) Also, if I am not asking a question that is workplace-communication related, let me know that I am off-topic and steer the conversation back on-topic to workplace communication. Do not attempt to answer the question if it is off-topic.}' & $\square$ & \citet{jang_its_2024} \\
Text & Written text \& Speech & GPT-3.5 & `\textit{First, find relevant information that could spark thoughts about their past. Then, based on the relevant information, ask one question to reflect on the past. Do it from the following text:}' & $\square$ & \citet{jeung_unlocking_2024} \\
Text & Written text \& Speech & GPT-3.5 & N/A & $\square$ & \citet{samsonovich_registrar_2024} \\
Text & Written text & GPT-3 & N/A & $\square$ & \citet{li_blibug_2023} \\
Text & Speech & GPT-3.5 Turbo & Paper presents two prompts. One prompt to illustrate: `\textit{Act as a cafe customer. You are ordering something from a coffee shop. Make your sentence as oral as possible. Do not exceed 180 characters. Now let’s start acting.}' & $\square$ & \citet{li_exploring_2024} \\
Text & Speech & GPT-3.5 & N/A & $\square$ & \citet{liu_peergpt_2024} \\
Text & Speech & GPT-3.5 (text-davinci-003) & Paper presents six prompts. One prompt to illustrate: `\textit{Your task is to summarize in Spanish the block of text that is placed between <>}' & $\square$ & \citet{bravo_creating_2023} \\
Text & Speech & ChatGPT & N/A & $\square$ & \citet{said_design_2024} \\
Text & Speech & GPT-3.5 & Paper presents two example prompts. One prompt to illustrate: `\textit{Your task is to summarize in Spanish the block of text that is placed between <>}' & $\square$ & \citet{schnitzer_prototyping_2024} \\
Text & Written text & GPT-4 \& GPT-3.5 & Paper presents six different prompts. One prompt to illustrate: `\textit{• Ask the user about potential solutions to the problem of the episode. • Ask only one question each conversation turn. • If the episode involves other people, such as friends or parents, ask the user how they would feel. • Help the user to find an `actionable' solution. • Do not overly suggest a specific solution.}' & \ding{110} & \citet{seo_chacha_2024} \\
Text & Written text & GPT-3 & N/A & $\square$ & \citet{shakeri_saga_2021} \\
Text & Written text \& Speech & GPT-4 & N/A & $\square$ & \citet{wan_building_2024} \\
Text & Written text \& Speech & Llama 2 (Llama-2-70B-chat) & N/A & $\square$ & \citet{wang_aint_2024} \\
Text & Speech & GPT-4 & N/A & $\square$ & \citet{xygkou_mindtalker_2024} \\
Image & Images & DALL-E2 & N/A & \ding{110} & \citet{chen_closer_2023} \\
Image & Images & Midjourney & Paper presents two prompts. One prompt to illustrate: `\textit{image URLs + subjects + descriptors of subjects + style descriptors + white background + parameters (aspect ratio: 3:2)}' & $\square$ & \citet{liu_when_2024} \\
Image & Images & Midjourney & `\textit{Image URL + Subject + Subject Descriptor + Style Descriptor + White Background + Parameters (Aspect Ratio: 16:9)}' & $\square$ & \citet{zhou_my_2024} \\
Music & Music & Cococo & N/A & $\square$ & \citet{suh_ai_2021} \\
Text \& Image & Written text \& Images & GPT-4 (BingChat1) \& Midjourney & N/A & $\square$ & \citet{jin_exploring_2024} \\
Text \& Image & Written text \& Images & GPT-4 \& Midjourney & Paper presents more than ten prompts for both GPT-4 and Midjourney. One prompt to illustrate: `\textit{You are {the character chosen by children}, I am {children’s name}, you need to share an experience..., your language should be simple and vivid.}' & $\square$ & \citet{tang_emoeden_2024} \\
Text \& Image & Written text \& Images & GPT-4 \& DALL-E3 & N/A & $\square$ & \citet{degen_retromind_2024} \\
Text \& Image & Written text \& Images & ChatGPT & N/A & $\square$ & \citet{wei_improving_2024} \\ \hline

\end{longtable}

\end{document}